\documentclass{aa}
\usepackage{graphicx}
\usepackage{txfonts}
\usepackage{float}
\usepackage{hyperref}
\hypersetup{
     colorlinks = true,
     linkcolor = blue,
     anchorcolor = blue,
     citecolor = blue,
     filecolor = blue,
     urlcolor = blue
     }
\usepackage{pdflscape} 
\usepackage{afterpage} 

\usepackage{tabularx}
\usepackage{breqn}

\makeatletter
\renewcommand*\aa@pageof{, page \thepage{} of \pageref*{LastPage}}

\begin{document} 

   \title{Exploring the nature of dark matter with the extreme galaxy AGC~114905}
   \titlerunning{Exploring the nature of dark matter with the extreme galaxy AGC~114905}
\authorrunning{Mancera Pi\~na et al.}

   \author{Pavel E. Mancera Pi\~na\inst{1}\fnmsep\thanks{\email{pavel@strw.leidenuniv.nl}},
        Giulia Golini\inst{2,3}, Ignacio Trujillo\inst{2,3}, and Mireia Montes\inst{2,3}.
          }
   \institute{Leiden Observatory, Leiden University, P.O. Box 9513, 2300 RA, Leiden, The Netherlands 
         \and Instituto de Astrofísica de Canarias, c/ Vía Láctea s/n, E-38205, La Laguna, Tenerife, Spain
         \and Departamento de Astrofísica, Universidad de La Laguna, E-38206, La Laguna, Tenerife, Spain
          }


 
  \abstract{AGC~114905 is a dwarf gas-rich ultra-diffuse galaxy seemingly in tension with the cold dark matter (CDM) model. Specifically, the galaxy appears to have an extremely low-density halo and a high baryon fraction, while CDM predicts dwarfs to have very dense and dominant dark haloes.
  The alleged tension relies on the galaxy's rotation curve decomposition, which depends heavily on its inclination. This inclination, estimated from the gas (neutral atomic hydrogen, H\,{\sc i}) morphology, remains somewhat uncertain.  
  We present unmatched ultra-deep optical imaging of AGC~114905 reaching surface brightness limits $\mu_{\rm r,lim} \approx 32$~mag/arcsec$^2$ ($3\sigma$; 10 arcsec $\times$ 10 arcsec) obtained with the 10.4--m Gran Telescopio Canarias. 
  With the new imaging, we characterise the galaxy's optical morphology, surface brightness, colours, and stellar mass profiles in great detail. The stellar disc has a similar extent as the H\,{\sc i} disc, presents spiral arms-like features, and shows a well-defined edge. Stars and gas share similar morphology, and crucially, we find an inclination of $31\pm2^\circ$, in agreement with the previous determinations.
  We revisit the rotation curve decomposition of the galaxy, and we explore different mass models in the context of CDM, self-interacting dark matter (SIDM), fuzzy dark matter (FDM) or Modified Newtonian Dynamics (MOND). We find that the latter does not fit the circular speed of the galaxy, while CDM only does so with dark halo parameters rarely seen in cosmological simulations. Within the uncertainties, SIDM and FDM remain feasible candidates to explain the observed kinematics of AGC~114905.}

   \keywords{galaxies: kinematics and dynamics – galaxies: formation – galaxies: evolution – galaxies: fundamental parameters – galaxies: dwarfs }

   \maketitle
%

\newcommand{\gsblim}{$31.6$}
\newcommand{\rsblim}{$31.9$}
\newcommand{\isblim}{$30.6$}

\section{Introduction}
\label{sec:intro}
Dark matter dominates the mass budget of the Universe, providing the gravitational seed for baryons to form structures, and it is an essential ingredient in our theories of galaxy formation and evolution (e.g. \citealt{press1974,binney1977,white1978,fall1980,nfw}).
The coupling between dark matter (or, more generally, missing mass) and baryons is evident on galactic scales. The shape of spatially resolved rotation curves correlates strongly with the surface brightness profile of the galaxy. While high surface brightness galaxies show steeply rising rotation curves and the baryons typically dominate their inner regions, dwarf low surface brightness galaxies have slowly rising rotation curves and usually are dominated by the dark matter at all radii (e.g. \citealt{bosma1978,begeman,deblok1996, marc_phd,swatersPhD,renzo2004,noordermeer,iorio}). On a more global scale, the coupling is also seen in scaling relations such as the baryonic Tully-Fisher relation (BTFR, \citealt{mcgaugh2000,anastasia_SED}), linking the baryonic mass (stars and cold interstellar medium) of galaxies with their total dynamical mass via the circular speed of the underlying gravitational potential, or the radial acceleration relation (e.g. \citealt{lelli_rar, stiskalek2023}) connecting the baryonic and total dynamical accelerations. Whether attributed to dark matter or alternative theories (e.g. MOdified Newtonian Dynamics, MOND, see \citealt{mond,mond_famaey_review}), these phenomena are the foundation of our understanding of galaxy dynamics and the missing mass problem.

Because the observations described above seem to be the rule, finding objects that do not conform to them is puzzling. For example, in recent years, several gas-rich dwarf galaxies, some of them within the category of the so-called ultra-diffuse galaxies (UDGs\footnote{The term UDG was coined by \citet{vandokkum2015} to refer to low surface brightness galaxies with central surface brightness fainter than $\mu_{0,g} \sim 24~\rm{mag/arcsec^2}$ and effective radius $R_{\rm e} \geq 1.5~\rm{kpc}$. For historical context, see also, e.g. \citet{sandage1984,conselice2018,chamba_udgs}.}), have been found to present a set of unexpected dynamical properties. First, six isolated gas-rich UDGs with circular speeds around $25-40~\rm{km/s}$ appear to shift off the BTFR, having a baryonic mass $10-100$ times larger than other galaxies with similar speeds, or equivalently, they have circular speeds $2-5$ times lower than galaxies of the same baryonic mass \citep{huds2019,huds2020}. This result is based on the kinematic modelling of resolved neutral atomic hydrogen (H\,{\sc i}) interferometric data, but unresolved studies seem to find a similar behaviour (e.g. \citealt{leisman2017,karunakaran2020,hu2023}). The offset from the BTFR also implies that the galaxies can have baryon fractions as high as the cosmological average\footnote{$f_{\rm {bar,cosmic}} = \Omega_{\rm b} / \Omega_{\rm m} \approx 0.16$ is the average cosmological baryon fraction, e.g. \citet{komatsu2011,planck2020_fbar}.}.

Second, the rotation curve decomposition of those galaxies suggests that their dark matter properties are very atypical (\citealt{agc114905,demao}, see also \citealt{shi2021}). Perhaps the clearest manifestation of this is the fact that the circular speed of the galaxies can be largely explained by the gravitational potential provided by the baryons, with less room for dark matter than in typical dwarfs. One potential interpretation is that the galaxies violate the cosmological baryon fraction limit. Unless exotic channels for acquiring baryons are in place, this would be at odds with our understanding of galaxies forming in the centre of more massive dark matter haloes. A second interpretation, potentially more in line with our ideas of galaxy formation, is that these gas-rich UDGs do not exceed the cosmological baryon limit, but their dark haloes are structurally different from those of other galaxies. Assuming these objects do not violate the cosmological baryon fraction, their dark matter densities in scales as large as 10~kpc must be extremely low \citep{agc114905,demao}. This can be seen from the fact that the systems shift down the so-called dark matter concentration--mass relation (i.e. they have low concentration parameters) and equivalently shift up the $R_{\rm max}-V_{\rm max}$ relation (with $R_{\max}$ the radius at which the circular speed of the haloes reaches its maximum value $V_{\rm max}$), both cornerstones of N-body cosmological simulations in the context of cold dark matter (e.g. \citealt{duttonmaccio2014,diemer2019})

The deviations from the expectations of the cold dark matter (CDM) model are so large that it has been argued that gas-rich UDGs may suggest that the CDM model should be revised. Interestingly, different works have started to explore whether alternative dark matter models can better explain the observed kinematics (e.g. \citealt{khalifeh2021,demao,roshan2022,varieschi2023,nadler2023}). In particular, \citet{nadler2023} has shown that dark-matter-only simulations considering self-interacting dark matter (SIDM, e.g. \citealt{tulin_sidm} and references therein) produce systems that lie in a similar region of the concentration--mass and $R_{\rm max}-V_{\rm max}$ relations as observed gas-rich UDGs.\\

Among the galaxies that lead to the results highlighted above, AGC~114905 \citep{agc114905} is particularly interesting and important. The gas kinematics of this UDG have been derived using state-of-the-art forward-modelling techniques \citep{barolo}, but a critical assumption made is that of the inclination angle. The inclination is crucial, as it converts the line-of-sight rotation velocities into deprojected values. In particular, if the inclination of the disc has been underestimated, the correction would lead to higher rotational speeds, allowing for a more typical dark matter contribution. 

The inclination of AGC~114905 (and the other five UDGs) has been measured from the morphology of the gas in the outer regions of the galaxy \citep{huds2020,agc114905}, completely independent of the kinematics (although constraining the inclination from the kinematic modelling gives similar results, see \citealt{agc114905}). While this has been carefully modelled, the constraining power of the data is somewhat limited. For instance, it has been argued that the H\,{\sc i} isocontours (which are not completely smooth) may not be an appropriate tracer of the disc inclination \citep{banik2022}. Adding to this debate, \citet{sellwood_agc114905} performed a set of N-body simulations to analyse the disc stability of AGC~114905 under a shallow dark matter halo as proposed by \citet{agc114905}. \citet{sellwood_agc114905} reported that the system develops strong instabilities unless the gas velocity dispersion is higher than previously reported or the dark matter halo mass is increased, possibly by invoking a lower inclination angle.\\


Obtaining a new measure of the inclination of AGC~114905 independent from the gas morphology is imperative. Similar estimates are traditionally obtained through broad-band photometric observations of the stellar emission, but given the low surface brightness nature of AGC~114905 ($\mu_{0,g} \approx 24~\mathrm{mag/arcsec^2}$), existing data were too shallow to trace the outer parts of the stellar disc \citep{lexi,agc114905}. To overcome this, we have used the 10.4-m Gran Telescopio Canarias, the largest optical telescope in the world, to obtain ultra-deep imaging of the galaxy and constrain the inclination of the stellar disc. Besides, we revisit and refine the mass model of AGC~114905. The rest of this paper is organised as follows. In Sec.~\ref{sec:data}, we present the optical and H\,{\sc i} data used in this work. In Sec.~\ref{sec:optical}, we delve into the properties of the stellar disc of AGC~114905, including the derivation of its inclination. In Sec.~\ref{sec:kinematics}, we model the H\,{\sc i} kinematics of the galaxy, finding its circular speed and gas velocity dispersion. In Sec.~\ref{sec:dm_content}, we investigate the dark matter content of AGC~114905, discussing whether CDM or alternative dark matter theories can explain the dynamical properties of this galaxy. In Sec.~\ref{sec:stability}, we present a short discussion on stability diagnostics and the star formation state given the observed kinematics and mass distribution of AGC~114905. Finally, in Sec.~\ref{sec:conclusions}, we summarise our results and present our main conclusions.

Throughout this work, magnitudes are reported in the AB system, and a $\Lambda$CDM cosmology is adopted with $\Omega_{\rm m} = 0.3$, $\Omega_{\Lambda} = 0.7$ and $H_0 = 70~\rm{km\,s^{-1}\,Mpc^{-1}}$.

\section{Data}
\label{sec:data}
\subsection{Optical data}
Ultra-deep images of AGC~114905 were obtained using the Optical System for Imaging and low-Intermediate-Resolution Integrated Spectroscopy \citep[OSIRIS;][]{2000SPIE.4008..623C} imager of the 10.4--m Gran Telescopio Canarias (GTC). The field of view of OSIRIS spans 7.8~arcmin by 8.5~arcmin, and the camera pixel scale is 0.254 arcsec. We observed AGC~114905 using the $g$, $r$ and $i$ Sloan filters during eight nights between October and December 2022. The total time on source was 3h12min, 3h06min and 3h45min, in the $g$, $r$ and $i$ filters, respectively, with an average seeing of 1.12 arcsec, 1.09 arcsec, and 1.24 arcsec. 

Accurate data reduction is required to handle different and complex observational biases such as scattered light, saturation, and ghosts to reach the magnitude depths needed to characterise the properties of the low surface brightness disc of AGC~114905. An accurate flat field estimation and careful treatment of the sky background are also crucial. This last task can be very challenging, as observational conditions change throughout the night due to factors like air mass variation and cloud passage, which can degrade the quality of the sky background. To tackle these issues, we have designed a specific observational strategy involving a dithering pattern and developed an optimised data reduction pipeline \citep[see, e.g.][]{2016ApJ...823..123T,LIGHTSs,golini2024}. 
Our pipeline's main steps involve deriving flat field images, background removal, astrometric and photometric calibrations, and mosaic co-addition. Appendix~\ref{appendix:data_reduction} gives detailed information on each step.

\begin{figure*}
    \centering
    \includegraphics[width=\textwidth]{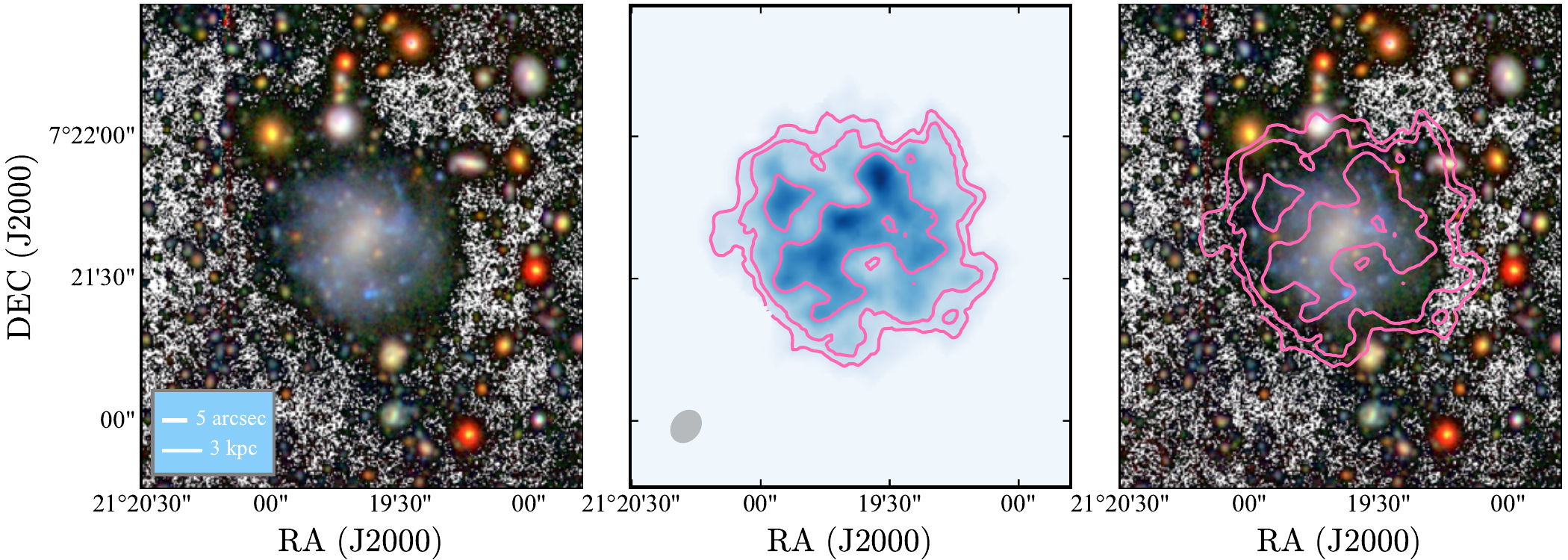}
    \caption{\emph{Left:} Optical colour image of the stellar disc of AGC~114905 generated using the filters $g$, $r$ and $i$. The black and white background was created using the $g-$band image. For reference, scales of 5~arcsec and 3~kpc (assuming a distance of 78.7~Mpc) are shown.
    \emph{Middle:} H\,{\sc i} total intensity map with H\,{\sc i} column density contours overlaid. The contours are at 1, 2, 4 $\times 10^{20}\,\mathrm{cm}^{-2}$, and the noise of the H\,{\sc i} column density map is $4.1 \times 10^{19}\,\mathrm{cm}^{-2}$. The grey ellipse shows the beam of the observations. \emph{Right:} H\,{\sc i} column density map overlaid on the optical emission.}
    \label{fig:images}
\end{figure*}

Our final mosaics have surface brightness limits of $\gsblim{}$ mag/arcsec$^2$ ($g$ band), $\rsblim{}$
mag/arcsec$^2$ ($r$ band), and $\isblim{}$ ($i$ band), corresponding to a 3$\sigma$ fluctuation of the background in areas equivalent to $10\,\rm{arcsec}\, \times\, 10\, \rm{arcsec}$ \citep[see, e.g.][]{roman2020}. Our data allow us to reach $r-$band surface brightness limits about 3 mag/arcsec$^{2}$ (2 mag/arcsec$^{2}$ for $g$) deeper than the previous imaging of this particular source \citep{lexi} and also with respect to typical works imaging large numbers of UDGs (e.g. \citealt{koda2015,vanderburg2016,udgs_kiwics, roman_outsideclusters,smudges,ultra-puffy_1}). 
Fig.~\ref{fig:images} shows a close-up of the colour image centred on AGC~114905 created with \texttt{astscript-color-faint-gray} \citep{scriptcolorimagesRaul} by combining our $g$, $r$ and $i$ filters and using the $g$ band as a grey-scaled background. We discuss the properties of the stellar disc galaxy in Sec.~\ref{sec:optical}.

\subsection{H\,{\sc i} data}

The H\,{\sc i} observations, data reduction, and final data products of AGC~114905 are fully described in \citet{agc114905}. Summarising, the galaxy was observed with the Karl G. Jansky Very Large Array (VLA) using the D-, C-, and B-array configurations (see \citealt{leisman2017,lexi,agc114905}). The different observations combined resulted in a data cube with a beam size of $7.88$ arcsec $\times$ 6.36~arcsec. The rms noise per channel is 0.26~mJy/beam, and the spectral resolution (FWHM) is 3.4~km~s$^{-1}$. 

The global profile of the H\,{\sc i} data and our kinematic modelling shown below (Sec.~\ref{sec:kinematics}) indicate that the systemic velocity of the galaxy is $5433\pm3~\rm{km\,s^{-1}}$. We convert this velocity to redshift and then to luminosity distance using \texttt{Astropy} \citep{astropy}. For the uncertainties, we consider the isolation of AGC~114905 and assume a potential peculiar velocity of $150~\rm{km\,s^{-1}}$ (e.g. \citealt{davis1997,karachentsev2003}). All this results in a distance of $78.7\pm1.5$~Mpc. We check that our distance range covers the expected value from modelled large-scale peculiar velocity fields \citep{graziani2019}.

The H\,{\sc i} column density peaks at a value of 8.4$\times 10^{20}$ atoms~cm$^{-2}$, and the noise level is 4.1$\times 10^{19}$~cm$^{-2}$. The flux of the source is 0.73$\pm 0.07$~Jy~km~s$^{-1}$, which at the distance of 78.7~Mpc implies (e.g. \citealt{bookFilippo}) a neutral atomic gas mass $M_{\rm HI} = (1.04\pm 0.11)\times 10^9~M_\odot$. Once corrected for helium, we obtain $M_{\rm gas} = 1.33\, M_{\rm HI} = (1.38 \pm 0.14)\times 10^{9}~M_\odot$.

The gas morphology can be seen in the middle panel of Fig.~\ref{fig:images}, and the azimuthally-averaged H\,{\sc i} surface density profile is shown in the following section. As we show below, the optical and H\,{\sc i} morphologies largely resemble each other, aside from some gas excess on the west side of the galaxy. It is unclear whether this slight asymmetry is real or a consequence of it being close to a noisy region, but it is worth mentioning that small asymmetries in the H\,{\sc i} distribution of galaxies are more often present than not, as noticed first by \citet{richter1994}.

\section{Stellar properties of AGC~114905}
\label{sec:optical}
\subsection{The inclination and position angle of the stellar disc }

The multi-band ultra-deep imaging we have performed on AGC~114905 allows us to appreciate its optical morphology in detail for the very first time, as previous imaging (see \citealt{dey2019,lexi}) was not deep enough ($2-3$ mag/arcsec$^{-2}$ shallower than our new data) to uncover the low surface brightness features. The new data reveal a regular galaxy with a well-defined nucleus, a bluer outer disc, bright blue knots (arguably ongoing star formation), and striking structures that resemble the spiral arms of more massive disc galaxies. Our observations also show that the stellar disc is essentially as extended as the H\,{\sc i} disc, which opens the possibility of constraining the geometry of the galaxy from its stellar emission. 
We measure the inclination and position angle of the stellar disc using a two-pronged approach, combining isophotal fitting and a custom version of the Modified Hausdorff Distance (MHD, \citealt{mhd,montes2019}) method, which we denote as modified MHD or $\mathrm{M^2HD}$. 

\begin{figure*}
    \centering
    \includegraphics[width=\textwidth]{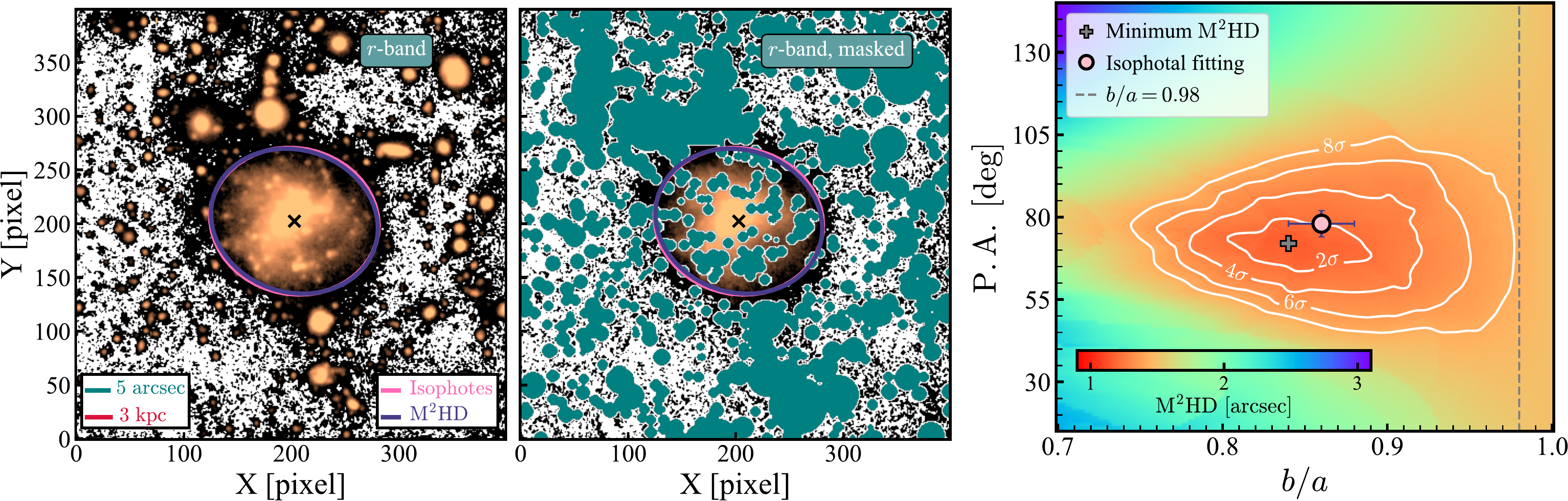}
    \caption{\emph{Left} and \emph{middle} show the $r-$band stellar emission of AGC~114905 without and with our mask, respectively. Both panels include two ellipses, one obtained through isophotal fitting (pink) and one through the $\mathrm{M^2HD}$ method (purple). The black crosses represent the centre of the galaxy. \emph{Right:} $\mathrm{M^2HD}$ map. The grey marker highlights the minimum of the map, and the white contours represent 2, 4, 6, and 8 standard deviations. The parameters obtained through isophotal fitting (pink marker) are shown to be in $2\sigma$ agreement. Very high axis ratios (i.e. nearly face-on inclinations) are in significant tension with the data. 
    }
    \label{fig:isophotes}
\end{figure*}

As a first step, we produce a mask. Creating an accurate mask to avoid contamination from background and foreground sources is crucial when fitting ellipses to the low surface brightness stellar emission of AGC~114905. We inspect our deep $r-$band image together with the colour image displayed in Fig.~\ref{fig:images} to mask the extended diffuse light of neighbouring objects and more localised clumps with different colours from AGC~114905. In the spiral-arm region, some knots could be part of the galaxy and are being masked, but we checked, and less strict masking in these regions does not affect the results below. With the masked image, we proceed to the preliminary isophotal fitting. For this, we use the Astropy package \texttt{photutils} \citep{photutils}. We sample isophotes every 5 pixels (1 arcsec), approximately the seeing of our data. Since we are predominantly interested in the outer disc (that will give the constraints on the inclination of the stellar and H\,{\sc i} discs), we focus on radii beyond $\sim17$ arcsec. This choice is purely practical, as beyond $\sim17$ arcsec the emission stops being dominated by bright and irregular regions. We use our $r-$band image, as it is our deepest band.

We perform the fitting in two steps. First, we treat as free parameters the centre $(\alpha, \delta)$, position angle (P.A.), and axis ratio ($b/a$). Then, we perform a second iteration, fixing the centre to the median value of the first iteration and fitting the remaining parameters. The parameters show no significant variation with radius and the global (median) results of our analysis yield $\alpha = 01\mathrm{h}25\mathrm{m}18.55\mathrm{s}$, $\delta = +07\mathrm{d}21\mathrm{m}37.53\mathrm{s}$, $\rm{P.A.} = (78\pm5)^\circ$, and $b/a=0.86 \pm 0.02$. The uncertainty in position angle and axis ratio is estimated as the sum in quadrature of the median error found by \texttt{photutils} and the standard deviation of the collection of measurements.
Fig.~\ref{fig:isophotes} illustrates the results of this exercise by showing an ellipse (pink) with our best-fitting geometry, sampled at the truncation or break radius of the stellar disc (see below) on top of the $r-$band image. The galaxy's unmasked emission can be seen in the left panel, while the middle panel shows the masked image. 

However, we note that the isophotal results can have some moderate variations depending on the initial guess of the geometric parameters and the exact mask being used. This is where our second prong comes into play, as we use the $\mathrm{M^2HD}$ metric to obtain an independent constraint on the ellipse parameters better representing our data. The advantages of this method are that it allows us to explore the effects that our mask introduces to the data, does not depend on initial guesses, and gives us a statistically meaningful uncertainty. Our procedure is fully described in Appendix~\ref{appendix:MMHD}. In summary, our method builds thousands of mock elliptical isophotes sampling the parameter space $(\mathrm{P.A.}, b/a)$ and finds the model ellipse, which, point by point, is the closest to our data, i.e. it finds the geometrical parameters of the ellipse that minimise the distance ($\mathrm{M^2HD}$) between model and data. We evaluate the $\mathrm{M^2HD}$ at 19 arcsec, immediately before the edge of the stellar disc ($\sim 20~\rm{arcsec}$, see Sec.~\ref{sec:stellar_mass}) of the galaxy. This is a meaningful radius which not only marks the optical edge of the galaxy $R_{\rm edge}$ \citep{trujillo2020,chamba2022}, but it is also structurally relevant: it contains most of the stellar (and gas) mass of the system, and both stars and gas show a break at this radius (see Sec.~\ref{sec:sbprofs}). The right panel of Fig.~\ref{fig:isophotes} shows the resulting $\mathrm{M^2HD}$ map, highlighting the minimum value and its uncertainty contours. The minimum $\mathrm{M^2HD}$ occurs $\mathrm{P.A.} = 72^\circ$ and $b/a=0.84$, but as shown in the figure, there is a $2\sigma$ agreement with the values obtained from the isophotal fitting. Crucially, the $\mathrm{M^2HD}$ method shows that we can safely reject very high $b/a$ values.

With all the information of our two methods, we decide to adopt as final values those from the isophotal analysis, i.e. $\rm{P.A.} = (78\pm5)^\circ$, and $b/a=0.86 \pm 0.02$. While the $\mathrm{M^2HD}$ values would be as good, exploration of the kinematics (see Sec~\ref{sec:kinematics}) show that $\rm{P.A.} = 78^\circ$ is favoured over $\rm{P.A.} = 72^\circ$, and using $b/a=0.86$ over $b/a=0.84$ is a conservative approach (as it minimises the inclination of the galaxy), as we discuss later. We have checked that all the results shown below (surface brightness profiles, rotation curves, mass modelling) are robust against these minor differences in $\mathrm{P.A.}$ and $b/a$.

We convert the axis ratio $b/a$ to inclination using the standard equation from \citet{hubble1926}:
\begin{equation}
    \cos^2(i) = \dfrac{(b/a)^2 - q_0^2}{1 - q_0^2}~,
\end{equation}
with $q_0$ the intrinsic axis ratio, with typical values between 0 and 0.4 (e.g. \citealt{fouque1990,binggeli1995,roychowdhur2013}). This means we assume we are seeing a circular disc at a given inclination rather than a non-circular disc (cf. \citealt{banik2022}). We find $i = (31 \pm 2)^\circ$, where the uncertainty is estimated by Monte Carlo sampling considering the measurement $b/a=0.86 \pm 0.02$ and the range $0 < q_0 \lesssim 0.4$. 

Reassuringly, the values obtained in this Section agree with previous determinations based on H\,{\sc i} data. By measuring the morphology of the H\,{\sc i} total intensity map, \citet{agc114905} inferred a position angle of $(89 \pm 5)^\circ$ and an inclination of $(32 \pm 3)^\circ$. Earlier work on lower-resolution data \citep{huds2019} reported similar values, a position angle of $(85 \pm 8)^\circ$ and inclination of $(33 \pm 5)^\circ$. 

\subsection{A cautionary comment on optical inclinations}

\begin{figure}
    \centering
    \includegraphics[width=0.48\textwidth]{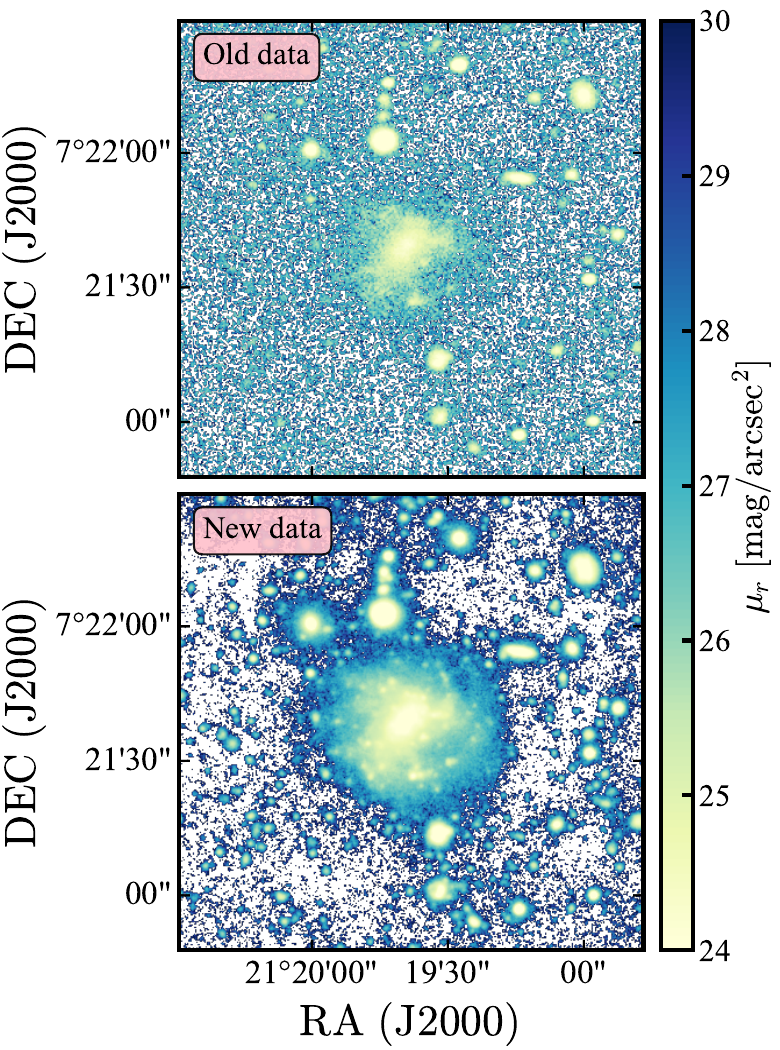}
    \caption{Comparison between the previous $r-$band imaging of AGC~114905 (\emph{top}) and the new data (\emph{bottom}). The colours represent different surface brightness levels. The previous data only traced the brighter central regions, while the new data, going 3 mag/arcsec$^2$ deeper, are able to capture the underlying fainter extended disc. }
    \label{fig:comparison}
\end{figure}

We take this opportunity to emphasise how crucial ultra-deep imaging is if attempting to constrain the inclination from optical imaging for galaxies of very low surface brightness galaxies like AGC~114905. In the case of AGC~114905, the previous imaging from \citep{lexi} or from the Dark Energy Camera Legacy Survey (DECaLs, DR10, \citealt{dey2019}) or the Sloan Digital Sky Survey (SDSS, DR14 \citealt{2018sdss}) would not have been good enough. Those data only revealed the redder central part of the galaxy and were not able to trace well the outer disc. This is illustrated in Fig.~\ref{fig:comparison}, where we show the comparison between the depth of the old photometric observations \citep{agc114905,lexi} and the new data.
Estimates of the inclination and position angle using our previous imaging would have been incorrect, as the redder region of the galaxy has a different orientation and axis ratio from the overall disc. As reported in \citet{agc114905,lexi}, the previous imaging would have suggested a position angle and inclination of about $50^\circ$ and $45^\circ$, respectively.

Studies using large samples of gas-rich UDGs with unresolved gas kinematics and optical inclinations from relatively shallow imaging (e.g. \citealt{hu2023,rong2024}) should, therefore, be taken with caution. While they might be correct in a statistical sense, they can be highly uncertain on a galaxy-by-galaxy basis. Constraining the optical inclination with deep photometry such as that presented in this work is desirable, ideally complementing it with resolved measurements of the H\,{\sc i} morphology (see also e.g. \citealt{barbara_udgs}).

\subsection{Surface brightness and colour profiles}
\label{sec:sbprofs}

Once the inclination and position angle of the stellar disc are known, we proceed to derive surface brightness and colour profiles using our three-band (masked) imaging. Before extracting the surface brightness profiles, we correct our images for any residual background contribution close to the galaxy. To do this, we compute the value of the background of the data in masked circular regions with a radius of 10~arcsec further than 40~arcsec from the southern part of the galaxy, where the scattered light contribution from a nearby star is lower (see Appendix~\ref{sec:scatteredlight}).
Once the local background value is subtracted from each band, we use the task \texttt{astscript-radial-profile} in Gnuastro to derive the azimuthally-averaged radial surface brightness profiles of AGC~114905, using the centre, position angle, and axis ratio derived in the previous section. The uncertainty in the profiles incorporates the rms of the flux inside each ring and the standard deviation of the image's background. 

\begin{figure*}
    \centering
    \includegraphics[width=\textwidth]{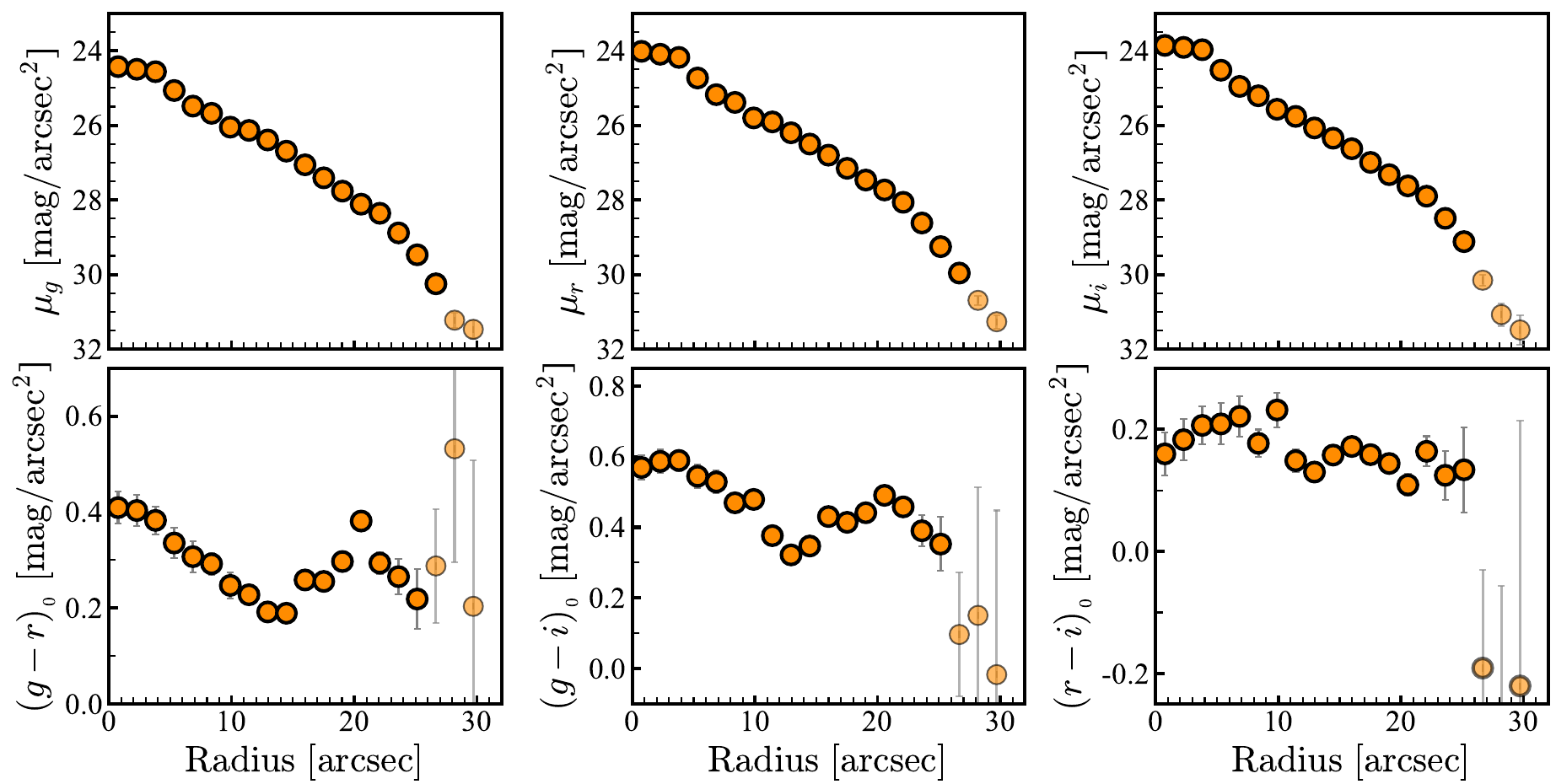}
   \caption{Surface brightness and colour profiles of AGC~114905. Corrections for Galactic extinction and inclination have been included. Error bars that are not visible are smaller than the size of the markers. We have marked those values with uncertainties larger than 0.1~mag/arcsec$^2$ with higher transparency. At our fiducial distance of 78.7~Mpc, the physical scale is 0.368~kpc/arcsec.}
    \label{fig:sb_profs}
\end{figure*}

We correct the surface brightness profiles (and colours) for Galactic extinction using the values from \citet{galactic_extinction} for the sky position of AGC~114905. Specifically, we use the corrections 0.119 mag, 0.082 mag, and 0.061 mag for the $g$, $r$, and $i$ filters, respectively. Then, we correct the surface brightness for inclination effects, assuming an optically thin disc using the expression $\mu = \mu_{\rm obs} -2.5\log(b/a)$. Given the axis ratio of AGC~114905, this correction is robust for any sensible thickness of the stellar disc, as shown in \citet{trujillo2020}. We do not correct for internal dust extinction as the calibrations in the low-mass regime are uncertain, but typical literature calibrations (e.g. \citealt{tully1998,sorce2012}) yield negligible corrections for all our bands of the order of 0.02~mag, as expected for a low-mass galaxy with little dust content.

The final surface brightness profiles are shown in the top panels of Fig~\ref{fig:sb_profs}. The ultra-deep data from GTC reveals that at a distance of 20~arcsec from the centre, the nearly exponential profile of AGC~114905 presents a break or truncation of `Type II', as seen in many nearby disc galaxies (e.g. \citealt{vanderkruit1979,pohlen2006,erwin2008,vanderkruit2011} and references therein), where the surface brightness profiles decay more steeply and the colour profiles bend. The truncation is particularly clear in the redder bands and in the stellar mass surface density profile of the galaxy (Fig.~\ref{fig:mass_profs}), as we discuss in Sec.~\ref{sec:stellar_mass}. 
As discussed in \cite{fiteni2024} and references therein, truncations can be related to small instabilities generating spiral-arm-like features, just as we observe in AGC~114905.
The location of the truncation in disc galaxies has also been found to be a good proxy for the gas density threshold for star formation \citep{trujillo2020,chamba2022}. In the case of AGC 114905, the position of the edge ($R_{\rm edge}$) at $\sim$7.5 kpc and the stellar surface mass density at this location ($\sim$0.1 M$_{\odot}$/pc$^2$) places the galaxy in the upper envelope of the stellar mass--size relation when using $R_{\rm edge}$ as a proxy for the size, in line with its blue colour \citep{chamba2022}.

Truncated surface brightness profiles of disc galaxies typically give rise to U-shaped colour gradients \citep[see, e.g.][]{azzollini2008,bakos2008}. While the median $(g-r)_{_0}$, $(g-i)_{_0}$, and $(r-i)_{_0}$ colours of AGC~114905 are 0.29~mag, 0.44~mag, and 0.16~mag, respectively, the galaxy shows clear colour gradients, as seen in the bottom panels of Fig.~\ref{fig:sb_profs}. The innermost part of the galaxy is significantly redder than the outer, bluer disc, as also visible in the colour image of Fig.~\ref{fig:images}. According to stellar population tracks (e.g. \citealt{vazdekis2015}), for typical dwarf-galaxy metallicities, the $(g-r)_{_0}$ vs. $(r-i)_{_0}$ colours of AGC~114905 suggest that the central redder region has an age of about 1--2~Gyr, and the outer bluer disc of about 0.5--1~Gyr.

\subsection{Stellar mass surface density}
\label{sec:stellar_mass}
To derive the stellar mass surface density $\Sigma_\ast(R)$, we must convert our surface brightness profile into physical units and adopt a stellar mass-to-light ratio $M/L$. The former is done with the equation \citep{bookFilippo}:
\begin{equation}
    \log(I_{r}) = \frac{\mu_{r} - M_{\odot,r} - 21.572}{-2.5}~,
\end{equation}
with $I_r$ luminosity profile in units of $L_\odot/\rm{pc}^2$ and $M_{\odot,r}=4.65$ mag \citep{magnitude_sun} the absolute magnitude of the Sun, both in the $r$ band. 

For $M/L$, we use the  $M/L-$colour relations from \citet{du2020}, which were calibrated specifically for low surface brightness galaxies assuming a Chabrier initial mass function \citep{chabrier2003}. Taking advantage of our three photometric filters, we use their equations:
\begin{equation}
    \log(M/L_r) = 1.252\times(g-r)_{_0} - 0.700~,
\end{equation}
\begin{equation}
        \log(M/L_r) = 1.088\times(g-i)_{_0} - 0.947~,
\end{equation}
for which we adopt an uncertainty of 0.15~dex based on the typical scatter observed in the calibrations (e.g. \citealt{roediger2015,herrmann_m2l,du2020}). Our final radial $M/L$ profile is the average of the two individual calibrations. We checked that the resulting $M/L$ profile is in good agreement with the values expected from the E-MILES stellar population models \citep{vazdekiz2010} for sub-solar metallicities $-1.3 \lesssim Z \lesssim -0.4$ (as expected given the typical luminosity and mass of UDGs, e.g. \citealt{kirby_metallicity, buzzo2024}; although no measurements of this galaxy are available).

\begin{figure}
    \includegraphics[width=0.47\textwidth]{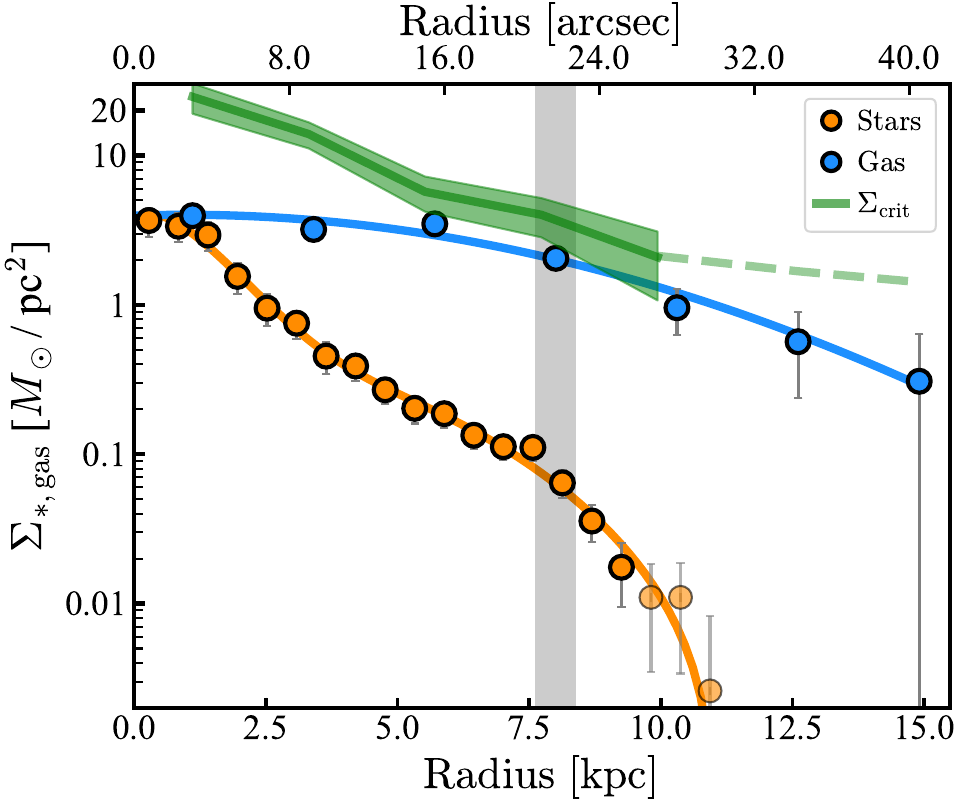}
    \caption{Stellar (orange) and gas (blue) mass surface density profiles. The stellar profile shows a clear truncation at about 7.5~kpc ($\sim 20$~arcsec) (grey band), which coincides with the radius at which the gas profile also decays more strongly. We have marked with higher transparency the stellar mass density data points derived from those values in the colour and surface brightness profiles with uncertainties larger than 0.1 mag/arcsec$^2$ (see Fig.~\ref{fig:sb_profs}). The gas profile lies below $\Sigma_{\rm crit}$ (green), a theoretical threshold to trigger star formation (see Sec.~\ref{sec:stability} for details). The dashed part of $\Sigma_{\rm crit}$ has been extrapolated from the data, assuming that the rotation curve and velocity dispersion profiles stay flat after the outermost observed radius (see Sec.~\ref{sec:kinematics}).
    The blue and orange curves represent functional forms fitted to the surface densities and are used during our mass modelling (see Sec.~\ref{sec:dm_content}).}
    \label{fig:mass_profs}
\end{figure}

The stellar mass surface density profile ($\Sigma_\ast = I\times (M/L)$) of AGC~114905 is shown in Fig~\ref{fig:mass_profs}, together with the gas surface density, both assuming a distance of 78.7~Mpc. The enclosed stellar mass within 10.9~kpc is $M_\ast = (9 \pm 1)\times 10^{7}~M_\odot$. The estimation of $M_\ast$ implies a baryonic mass of $M_{\rm bar} = M_\ast + 1.33M_{\rm HI} = (1.47\pm0.14)\times 10^{9}~M_\odot$, and a gas fraction $f_{\rm gas} = M_{\rm gas}/M_{\rm bar} = 0.94 \pm 0.01$, highlighting the extremely gas-rich nature of this UDG. Considering the stellar mass beyond 10.9~kpc negligible, we measure a stellar half-\emph{mass} radius of $R_{\rm e} = (2.80 \pm 0.04)~\rm{kpc}$. 

\section{Kinematic modelling}
\label{sec:kinematics}
\citet{agc114905} inferred the gas kinematics of AGC~114905 through forward modelling of the H\,{\sc i} data cube using the software $^{\rm 3D}$Barolo \citep{barolo}. $^{\rm 3D}$Barolo performs a fit to the H\,{\sc i} emission by building 3D realisations of the galaxy data cube using the tilted-ring method (see, e.g. \citealt{rogstad1974,begeman}). Importantly, the models are convolved to the observational beam and then compared against the data on a channel-by-channel basis, in contrast to traditional methods that fit the (beam-smeared) velocity field. This allows for reliable recovery of the intrinsic kinematics of the galaxy regardless of the spatial resolution (see, e.g. \citealt{begeman,swatersPhD,barolo}).

We revisit the kinematic modelling of the galaxy considering the new position angle and inclination. During the modelling with $^{\rm 3D}$Barolo, we consider tilted-ring models with five rings separated by 6~arcsec (similar to the beam size, with a minor oversampling of 15\%). In a first iteration, we fix the position angle\footnote{We have also explored leaving the position angle as a free parameter. Depending on the initial conditions, the preferred kinematic position angle ranges between $80^\circ$ and $90^\circ$, in fair agreement with the value found from the optical analysis.} and inclination of all the rings to the results of our optical analysis and leave as free parameters the rotation velocity ($V_{\rm rot}$) and the velocity dispersion ($\sigma_{_{\rm HI}}$). We also allow $^{\rm 3D}$Barolo to obtain the centre and systemic velocity of the galaxy based on the geometry of the H\,{\sc i} disc and the global HI profile, respectively; the resulting centre is consistent (at about 1 arcsec) with the optical centre within the uncertainties. In a subsequent iteration, we fix the systemic velocity and the kinematic centre and keep $V_{\rm rot}$ and $\sigma_{_{\rm HI}}$ free. 

\begin{figure*}
    \centering
    \includegraphics[width=\textwidth]{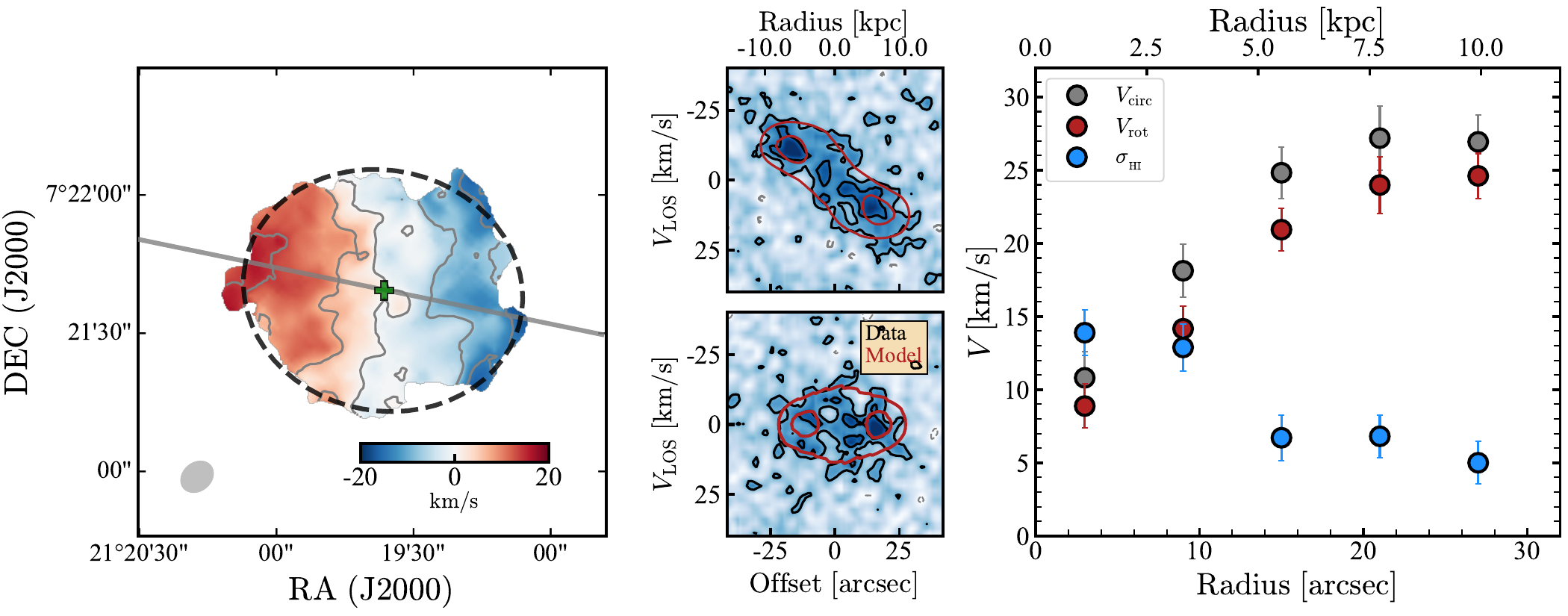}
    \caption{Gas kinematics of AGC~114905. \emph{Left:} H\,{\sc i} velocity field. The grey line and filled ellipse show the position angle and beam of the observations, respectively. The black dashed line shows an ellipse with the global position angle and inclination. Grey curves in the velocity field are iso-velocity contours ($V_{\rm sys}$ is in zero). \emph{Middle:} Major- (\emph{top}) and minor-axis (\emph{bottom}) position-velocity diagrams. Black (grey for negative values) and red contours represent the data and best-fitting kinematic model, respectively. Contours are at $-2, 2, 4\times$ the rms noise of the slice. \emph{Right:} Circular speed profile (grey), rotation velocity (red), and gas velocity dispersion (blue) of AGC~114905. To convert from $V_{\rm rot}$ to $V_{\rm circ}$, we apply the asymmetric drift correction (see the text for details). The kinematic model assumes an inclination of $31^\circ$.}
    \label{fig:kinematics}
\end{figure*}

$^{\rm 3D}$Barolo successfully finds a model that reproduces the overall kinematics of the system. In the left and middle panels of Fig.~\ref{fig:kinematics}, we show the observed velocity field of the galaxy (with the new position angle and the beam of the observations) and the position-velocity diagrams along the major and minor axes for both the data and the best-fitting model. 

The resulting $V_{\rm rot}$ (assuming our fiducial inclination of $31^\circ$) and $\sigma_{_{\rm HI}}$ profiles are shown in the right panel of Fig.~\ref{fig:kinematics}. The uncertainties in both $V_{\rm rot}$ and $\sigma_{\rm HI}$ are estimated as:
\begin{equation}
    \delta_{V^2} = \left(  \dfrac{ V_{\rm app} -  V_{\rm rec}}{4}   \right)^2 + \delta_{\rm spectral}^2 ~~ \rm{and}
\end{equation}
\begin{equation}
    \delta_{\sigma_{_{\rm HI}}^2} = \left(  \dfrac{ \sigma_{_{\rm HI, app}} -  \sigma_{_{\rm HI, rec}}}{4}   \right)^2 + \delta_{\rm spectral}^2 ~.
\end{equation}

\noindent
In these equations, the first term is traditionally used to account for deviations from kinematic symmetry (e.g. \citealt{noordermeer,swaters09,sparc}), with $V_{\rm app}$ and $\sigma_{_{\rm HI, app}}$ the rotation velocity and velocity dispersion obtained by fitting only the approaching side of the galaxy, and $V_{\rm rec}$ and $\sigma_{_{\rm HI, rec}}$ the receding sides, respectively. The second term captures the limitations of the H\,{\sc i} data, with $\delta_{\rm spectral}$ the standard deviation of the spectral axis of the data cube. Note that we do not include an uncertainty contribution from the inclination angle here, as that will be considered in Sec.~\ref{sec:dm_content} when performing the rotation curve decomposition of the galaxy.

While our findings generally agree with the previous analysis by \citet{agc114905}, there are some subtle but important differences. On average, our $\sigma_{_{\rm HI}}$ is about 40\% higher, mainly due to a less restricting kinematic masking that includes faint H\,{\sc i} emission. With a mean value of $9~\mathrm{km/s}$, our $\sigma_{_{\rm HI}}$ profiles match well the velocity dispersion of other dwarf galaxies \citep{iorio,paperIBFR}. $V_{\rm rot}$ is also higher by about 10\%, owing to the slightly lower inclination and the different position angle. We conclude this section by making the remark that our kinematic model, in agreement with previous findings \citep{huds2019,huds2020}, indicates that AGC~114905 is an outlier of the BTFR (e.g. \citealt{anastasia_SED,lelliBTFR}) and the stellar mass and luminosity Tully-Fisher relation (e.g. \citealt{reyes2011,anastasiaphotometry}) found for more massive galaxies\footnote{The distance required to put the galaxies on these relations is about 40-50 Mpc.}. 

\section{Testing dark matter models with AGC~114905}
\label{sec:dm_content}

Historically, the rotation curves of disc galaxies have played a major role in understanding the distribution of dark matter in galaxies (see references in, e.g. \citealt{paper_massmodels,lelli_review,bosma_review}). Under the assumptions of axisymmetry and dynamical equilibrium, the circular speed of a galaxy ($V_{\rm circ}$) is a direct tracer of the underlying gravitational potential ($\Phi$) and the total mass distribution (e.g. \citealt{binney,bookFilippo}, cf. \citealt{downing2023} for cases where this might not be the case). Therefore, $V_{\rm circ}$ is also linked with the observed kinematics of the galaxy through
\begin{equation}
    V_{\rm circ}^2 \equiv R \dfrac{\partial \Phi}{\partial R} \bigg |_{z=0} = V_{\rm rot}^2 + \dfrac{R}{\rho} \dfrac{\partial(\rho\, \sigma_{_{\rm HI}}^2 )}{\partial R}~,
    \label{eq:vcirc}
\end{equation}
with $R$ the cylindrical radius and $\rho$ the density of the gas, which under the assumption of a constant gas scale height becomes simply the observed gas surface density $\Sigma_{\rm gas}$\footnote{In reality, gas discs are flared (e.g. \citealt{romeo1992,nakanishi2003,yim2014,ceciVSFL,paper_massmodels}). However, the corrections for this to the right-hand side of Eq.~\ref{eq:vcirc} are typically minor. We have estimated the H\,{\sc i} flaring of AGC~114905 based on hydrostatic equilibrium, finding it more moderate than in other dwarfs and further justifying this simplification (see below and also \citealt{ceci_q3d}).}. The second term of the right-hand side of Eq.~\ref{eq:vcirc} corrects $V_{\rm rot}$ for pressure-supported motions, and it is often called the asymmetric drift correction (e.g. \citealt{binney, swatersPhD}). We compute this correction following the prescription of \citet{iorio}, finding that, on average, it increases the rotation curve by about $2-3$~km/s. The right panel of Fig.~\ref{fig:kinematics} shows both $V_{\rm rot}$ and $V_{\rm circ}$.

Since our deep H\,{\sc i} and optical data enable us to characterise the mass contribution from the baryons accurately, we can use rotation curve decomposition techniques to infer the distribution of dark matter in AGC~114905.
The technique consists of finding a dark matter distribution (typically parameterised with haloes of specific functional forms) whose circular speed ($V_{\rm DM}$) satisfies
\begin{equation}
    V_{\rm circ}^2 = V_\ast^2 + V_{\rm gas}^2 + V_{\rm DM}^2~,
    \label{eq:massmodel}
\end{equation}
with $V_{\rm circ}$ as described above, and $V_{\rm gas}$ and $V_\ast$ the components generated by the gravity of the gas and stellar distributions, respectively.

Our strategy to obtain a mass model closely follows the one detailed in \citet{agc114905} and \citet{paper_massmodels}. We use the software \textsc{galpynamics}\footnote{\url{https://gitlab.com/iogiul/galpynamics/}, see \citet{iorio_phd}} to compute $V_{\rm gas}$ and $V_\ast$. \textsc{galpynamics} computes the gravitational potential of a given mass distribution represented by fairly flexible functional forms describing its density profile through numerical integration (see \citealt{cuddeford1993}). Then, \textsc{galpynamics} computes the circular speed of the distribution from the derivative of the potential evaluated at the midplane of the mass component. 

Given the above, we fit our stellar and gas surface density profile with functional forms. For the stars, we use a `poly-exponential' disc of fourth-degree
\begin{equation}
\label{eq:polyexp}
    \Sigma_\ast(R) = \Sigma_{\rm 0,pex}\, e^{-R/R_{\rm pex}}~(1 + c_1\, R + c_2\, R^2 + c_3\, R^3 + c_4\, R^4)~,
\end{equation}
with $\Sigma_{\rm 0,pex} = 3.6703~M_\odot/\rm{pc^2}$, $R_{\rm pex} = 3.6155~\rm{arcsec}$, $c_1 = 0.5576~\rm{arcsec^{-1}}$, $c_2 = -0.1126~\rm{arcsec^{-2}}$, $c_3 = 0.0085~\rm{arcsec^{-3}}$, and $c_4 = -0.0002~\rm{arcsec^{-4}}$ the best-fitting values (for our inclination-corrected data).

In the case of the gas, we use the function (e.g. \citealt{Tom891})
\begin{equation}
\label{eq:Frat_disc}
    \Sigma_{\rm gas}(R) = \Sigma_{0,\rm gas}\, e^{-R/{\rm R_{1}}}\, (1+R/{\rm R_{2}})^\alpha~,
\end{equation}
with $\Sigma_{0,\rm gas} = 3.953~M_\odot/\rm{pc^2}$, $R_1 = 0.676~\rm{arcsec}$, $R_2 = 375.220~\rm{arcsec}$, and $\alpha = 559.231$ the best-fitting parameters\footnote{The gas profile has some small differences to that used in \citet{agc114905} owning to the different position angle, inclination, centre, and masking (see Sec.~\ref{sec:optical} and Sec.~\ref{sec:kinematics}). While the profiles are consistent within the uncertainties, the new one has moderately higher values in the centre but decays slightly faster at large radii.} for our measured inclination. Eqs.~\ref{eq:polyexp} and \ref{eq:Frat_disc} describe well the radial surface mass density profiles, as shown in Fig.~\ref{fig:mass_profs}. 

Besides the radial behaviour of the profiles, we need to consider the vertical direction. For the stars, we assume that the disc follows a sech$^2$ shape along the vertical direction \citep{vanderkruit2011} with a constant thickness $z_{\rm d} = 0.196\,(R_{\rm e,\ast}/1.678)^{0.633}~\rm{pc}$, as found in nearby disc galaxies \citep{bershady_thickness}.
For the gas, we assume a Gaussian profile for the vertical direction with a constant scale height of $z_{\rm d} = 1.3~\rm{kpc}$. This value of $z_{\rm d}$ for the gas is motivated by tests following \citet{paper_massmodels} estimating the scale height of the galaxy assuming vertical hydrostatic equilibrium. We checked that the difference of this assumption with respect to the approach of \cite{paper_massmodels} of constraining $z_{\rm d}$ simultaneously with the mass model is negligible.  With this, $V_\ast$ and $V_{\rm gas}$ are fully defined, and Eq.~\ref{eq:massmodel} can be used to find $V_{\rm DM}$. In the following sections, we explore different scenarios for $V_{\rm DM}$ under different dark matter theories.

\subsection{Models without dark matter}
\subsubsection{Only baryons}
\label{sec:onlybaryons}

Motivated by the low dynamical mass of AGC~114905 inferred by its low $V_{\rm circ}$ and a high baryonic mass, \cite{agc114905} showed that a mass model without dark matter was consistent with their observed gas kinematics within the uncertainties. We start by revisiting this scenario, i.e. we obtain a mass model using Eq.~\ref{eq:massmodel} but fixing $V_{\rm DM} = 0$.

To account for the uncertainties in the distance $D$ (which normalises the x-axis of the velocity profiles) and inclination $i$ (which normalises the value of $V_{\rm circ}$, $V_{\rm gas}$, and $V_\ast$), we consider them free parameters. We fit them using the Bayesian MCMC routine \texttt{emcee} \citep{emcee} and minimising a likelihood of the form $\exp(-0.5 \chi^2)$, with $\chi^2 =  (V_{\rm circ}-V_{\rm circ,mod})^2 / \delta_{V_{\rm circ}}^2$, where $V_{\rm circ}$ and $V_{\rm circ,mod}$ are the observed and model circular speed profiles, respectively, and $\delta_{V_{\rm c}}$ the uncertainty in $V_{\rm circ}$. For both $D$ and $i$, we consider Gaussians priors with centres and standard deviations given by our fiducial values, i.e. ($31\pm 2)^\circ$ and $(78.7\pm1.5)~\rm{Mpc}$, and bounded within $\pm 3\sigma$ limits, i.e. $25^\circ \leq i \leq 37^\circ$ and $74.2 \leq D/\rm{Mpc} \leq 83.2$.

\begin{figure}
    \centering
    \includegraphics[width=0.47\textwidth]{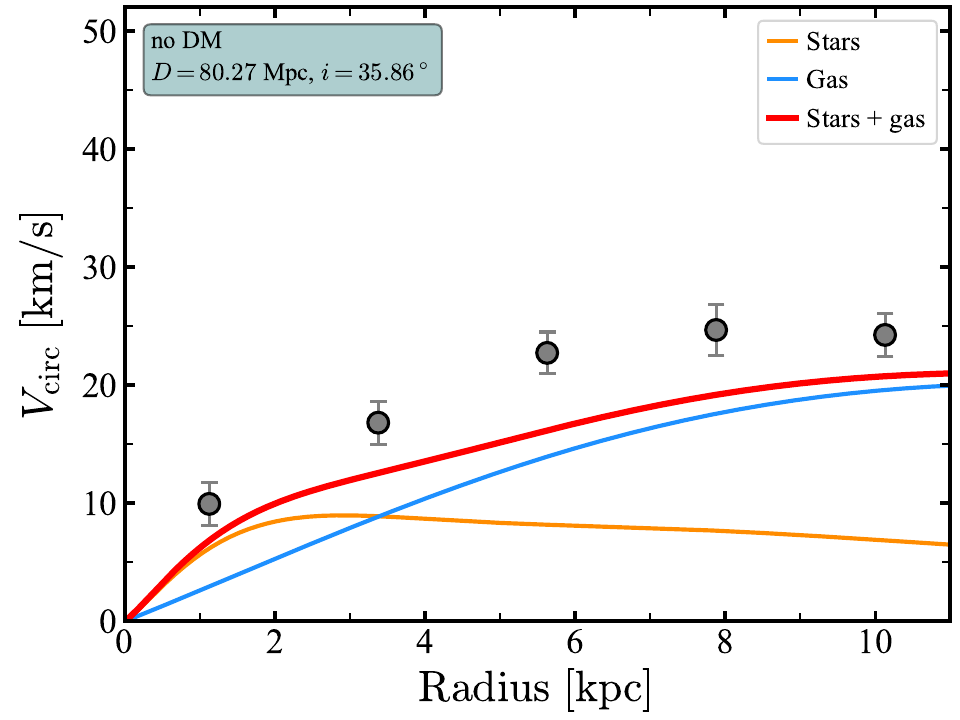}
    \caption{Only-baryons mass model. The orange and blue curves show the contribution from $V_\ast$ and $V_{\rm gas}$ to the gravitational potential, respectively. The red curve shows their sum in quadrature. The blue box lists the best-fitting distance and inclination. According to our priors on distance and inclination, the distribution of the baryons alone cannot explain the observed kinematics of AGC~114905. }
    \label{fig:noDM}
\end{figure}

The results are shown in Fig.~\ref{fig:noDM}, where we display the mass model; its posterior distribution is presented in Fig.~\ref{fig:post_noDM}. Clearly, the fit is very poor. To further quantify this, we compute
the Bayesian Information Criterion (BIC, \citealt{bic,bic2}), calculated as $\rm{BIC}$ = $\chi ^2 + N_{\rm param}\, \times\, \ln(N_{\rm data})$, with $N_{\rm param}$ and $N_{\rm data}$ the number of free parameters and the number of data points, respectively. The BIC penalises $\chi^2$ and $N_{\rm param}$; the lower the BIC, the better. For this mass model without dark matter, BIC = $39.98^{+4.56}_{-3.18}$, which is significantly higher (worse) than the mass models we discuss in the following sections. We explore broader priors as well, finding that the only way in which the baryons could reproduce the kinematics is if $i\sim 47^\circ$ and $D \sim 124~\rm{Mpc}$, both values prohibited by our measurements of $D$ and $i$. We also investigated whether a higher stellar $M/L$ can fit the data at our fiducial distance and inclinations, but the values needed for this are $\sim3-6$ times higher than expected in stellar population synthesis models \citep{vazdekiz2010} and also require a higher $M/L$ in the outer disc than in the central regions, which seems nonphysical. The disagreement between the mass model and the observations implies that the gravitational field from the baryons alone cannot explain the observed $V_{\rm circ}$ of AGC~114905. 

\subsubsection{Modified Newtonian Dynamics}

Modified Newtonian Dynamics (MOND) is an alternative no-dark-matter theory that postulates a modification to the law of gravity (see details in \citealt{mond,sanders_mond_review,mond_famaey_review}) and which has been proven to be very successful in reproducing galactic rotation curves, albeit some discrepancies have been reported in the literature (e.g. \citealt{gentile2004,frusciante2012,ren_SIDM_SPARC,mercadoMOND, khelashvili2024}). Provided that the baryonic mass distribution is known, MOND can predict the rotation curve of galaxies with only one additional parameter, $a_0 = 1.2\times10^{-10}\,\rm{m\, s^{-2}}$, suggested to be a universal constant. MOND also postulates the existence of a BTFR with a slope of 4 and zero intrinsic scatter.

As discussed in \citet{agc114905}, AGC~114905 and other gas-rich UDGs off the BTFR represent a challenge to MOND (see also \citealt{shi2021}). \citet{agc114905} showed explicitly that the rotation curve of AGC~114905 appears to be in tension with MOND expectations. The primary source of uncertainty was, however, the inclination of the galaxy: if the inclination turned out to be overestimated, the tension would be largely alleviated. 

Using our new dataset, we compare the observed circular speed of AGC~114905 against the MOND expectation, given by the expression of \citet{gentile2008} 
\begin{equation}
  V^2_{\rm MOND}(r) =  V_{\rm bar}^2  + \dfrac{V_{\rm bar}^2}{2} \left(\sqrt{1+\dfrac{4\, a_0\, r}{V_{\rm bar}^2}}-1 \right)~
\end{equation}
where $V_{\rm bar}^2 = V_{\rm gas}^2 + V_\ast^2$.\\

\begin{figure}
    \centering
    \includegraphics[width=0.47\textwidth]{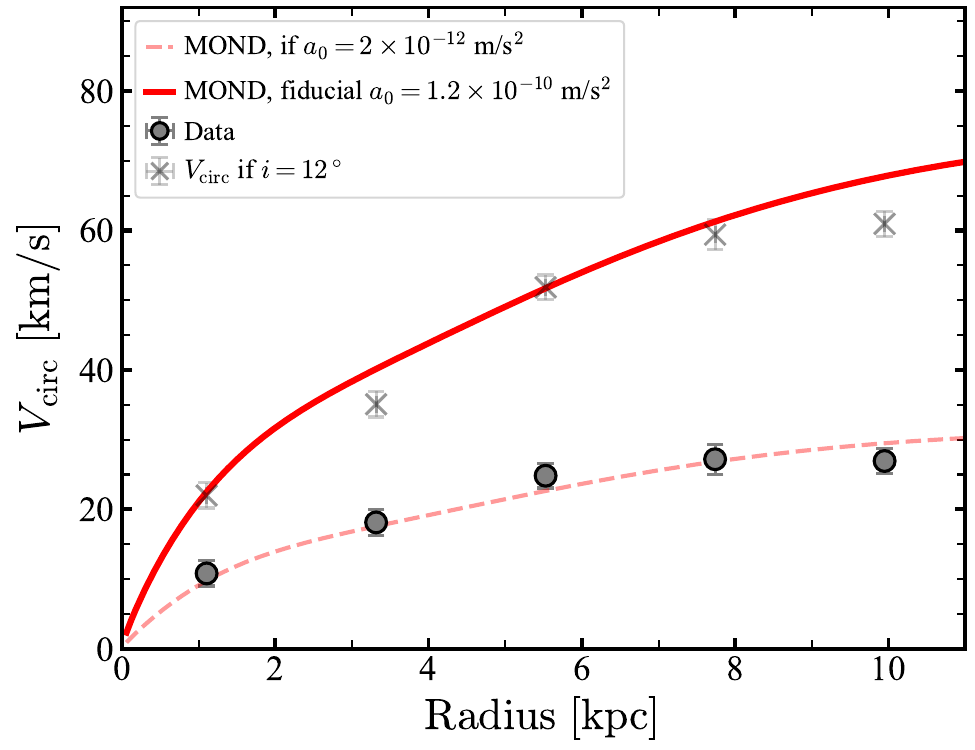}
    \caption{Comparison between the circular speed of AGC~114905 (solid grey symbols) against the MOND expectations (red solid curve). There is a marked disagreement between the data and the MOND prediction. The transparent grey symbols show a hypothetical rotation curve if the inclination correction is increased to match the MOND prediction; the needed inclination ($i \approx 12^\circ$) is inconsistent with the optical and gas morphologies. The transparent red dashed curve shows a MOND prediction if $a_0$ is 60 times smaller than the accepted value.}
    \label{fig:mond}
\end{figure}

The MOND comparison is shown in Fig.~\ref{fig:mond}. Assuming $a_0 = 1.2\times10^{-10}\,\rm{m\, s^{-2}}$ results in a strong overestimation of $V_{\rm circ}$ (red solid curve vs grey solid points), reinforcing the discrepancy with MOND postulated in \citet{agc114905}. To test the extent of this claim, we perform the additional exercise of finding which value of $a_0$ would bring the MOND prediction in agreement with the data and which value of the inclination would do the same. We find that a value for $a_0$ around $2\times10^{-12}~\rm{m/s^2}$, about 60 times lower than expected, would be needed to fit the kinematics of AGC~114905 (red dashed curve with transparency in Fig.~\ref{fig:mond}). Conversely, an inclination of $12^\circ$ would be required to make the data match the MOND expectation (small grey points with transparency in Fig.~\ref{fig:mond}) by renormalising the data. To obtain $i = 12^\circ$, the axis ratio should be $b/a \approx 0.98$, which is inconsistent with the data at a high significance level (see Fig.~\ref{fig:isophotes}). Under a scenario where the MOND phenomenology arises actually from dark matter dynamics and/or stellar feedback (see \citealt{ren_SIDM_SPARC,mercadoMOND}, and references therein), the low value of the fitted $a_0$ is expected if the halo has a low concentration (or a large $R_{\rm max}$ at fixed $V_{\rm max}$), since $a_0 \approx 1.59\,\pi\,V_{\rm max}^2 /R_{\rm max}$ (see, e.g. Sec. IV in \citealt{ren_SIDM_SPARC}).

Throughout the text, we have assumed that the stellar and gas discs of AGC~114905 are axisymmetric, follow circular orbits, and are seen at a given inclination $i$, generating a given apparent $b/a$. This is the same assumption always made when studying rotation curves, both in the context of dark matter or alternative theories and when studying both massive or dwarf galaxies. \citet{banik2022} have argued that face-on galaxies in MOND simulations may not always have fully circular isophotes and suggest that the inclination of AGC~114905 could still be overestimated because of this. We cannot directly test this interesting hypothesis with our data. However, if correct (and under the assumption that the stars also suffer from a similar potential effect as the gas), it is unclear why it would only manifest itself now for AGC~114905 and not for other dwarf galaxies, many of which have morphologies and kinematic patterns more disturbed and complex than AGC~114905, that appear to lie on the BTFR and conform to MOND \citep[e.g. ][]{lelli_starburst,mcnichols_shield,iorio}.

Certainly, it will be exciting to investigate this once large-volume MOND-based cosmological hydrodynamical simulations become available. For now, the available evidence presented here suggests that the challenge to MOND persists. In this context, it is essential to highlight that AGC~114905 may not be the only galaxy sustaining this claim. About ten other gas-rich UDGs with resolved H\,{\sc i} kinematics seem to deviate from MOND expectations (see \citealt{huds2019,shi2021,barbara_udgs}), all having different inclinations between $30^\circ-70^\circ$. Unresolved studies, albeit more uncertain, also show a whole population of gas-rich galaxies deviating from the BTFR \citep{karunakaran2020,hu2023,du2024} as predicted in \citet{huds2019}. Together with the aforementioned simulations, larger samples of gas-rich UDGs with high spatial resolution and ultra-deep photometric observations will be crucial to assessing this tension.\\

\noindent
Given the results of this section, we conclude that the no dark matter and MOND hypothesis cannot explain our data. In what follows, we explore mass models incorporating a dark matter halo component.

\subsection{Cold dark matter}
\label{sec:cdm}
We start with the CDM model. In this case, for $V_{\rm DM}$, we consider the functional form of the \textsc{coreNFW} \citep{coreNFW,readAD} profile. This type of halo has been shown to fit well the circular speed profiles of nearby dwarf galaxies (e.g. \citealt{read2017,paper_massmodels}). In Appendix~\ref{appendix:dpl}, we demonstrate that the results shown below also hold if using a double power-law halo, which has been used when studying some gas-rich UDGs \citep[e.g.][]{brook2021,demao}.

The \textsc{coreNFW} is a modification to the usual NFW halo \citep{nfw} proposed to adopt the classic profile to a context where supernovae feedback can create cores in the dark matter distribution (e.g. \citealt{navarro_cores, read2005, pontzen2012}), which helps to solve the `cusp-core' problem \citep{bullock2017,sales_review_dwarfs}. NFW haloes have a density profile as a function of their spherical radius in cylindrical coordinates $r$ ($r~=~\sqrt{R^2+z^2}$~) parameterised with the expression
\begin{equation}
    \rho_{\rm NFW}(r) = \dfrac{4\, \rho_{\rm s}}{(r/r_{\rm s})\,(1 + r/r_{\rm s})^2}~,
\end{equation}
with $r_{\rm s}$ a `scale radius' and $\rho_{\rm s}$ the density at $r_{\rm s}$. The corresponding mass profile $M_{\rm NFW}(r)$ is given by:
\begin{multline}
    M_{\rm NFW}(<r) = \dfrac{M_{200}}{\ln(1+c_{200}) - \dfrac{c_{200}}{1+c_{200}}} \\ 
    \times\, \left[\ln\left( 1 + \dfrac{r}{r_{\rm s}}\right) - \dfrac{r}{r_{\rm s}} \left( 1 + \dfrac{r}{r_{\rm s}}\right)^{-1} \right]~.
\end{multline}
The parameter $M_{200}$ is the mass within a sphere of radius $r_{200}$ within which the average density is 200 times the critical density of the Universe. The `concentration' parameter $c_{200}$ is defined as $c_{200} = r_{200}/r_{-2}$, where $r_{-2}$ is the radius at which the logarithmic slope of the density profile equals $-2$. For an NFW halo $r_{\rm s} = r_{-2}$, and thus $c_{200} = r_{200} / r_{\rm s}$.

Instead, the density profile of the \textsc{coreNFW} halo is given by \citep{coreNFW,readAD}
\begin{equation}
\label{eq:corenfw}
    \rho_{\rm \textsc{coreNFW}}(r) = f^n\, \rho_{\rm NFW}(r) + \dfrac{n\,f^{n-1}\,(1-f^2)}{4\,\pi\, r^2\, r_{\rm c}}\, M_{\rm NFW}(r)~.
\end{equation}
In this equation, $\rho_{\rm NFW}$ and $M_{\rm NFW}$ are the NFW parameters defined above, while $f~=~\tanh(r/r_{\rm c})$ is a function that generates a core of size $r_{\rm c}$. The degree of transformation from cusp to core is set by the parameter $n$, with $n=0$ defining a cuspy NFW profile and $n = 1$ a fully cored profile. In practice, we fix $n=1$ assuming a cored profile, but we have checked that the exact value of $n$ does not change the results shown in the following paragraphs.

Again, we perform our rotation curve decomposition fitting $V_{\rm DM}$ to minimise Eq.~\ref{eq:massmodel} using \texttt{emcee} and the same likelihood as in the previous section. We treat as free parameters $\log(M_{200})$, $\log(c_{200})$, $\log(r_{\rm c}/\rm{kpc})$, $i$ and $D$. Both $i$ and $D$ are nuisance parameters that account for their uncertainties in the fit. \citet{coreNFW,readAD} proposed specific calibrations for $r_{\rm c}$ and $n$ based on their simulations, but it is unclear whether those results can be extended to AGC~114905, so we decided not to use those calibrations. As mentioned above, since we want to explore explicitly whether AGC~114905 shows a core, we set $n=1$.

We start by adopting the following flat priors: $7 \leq \log(M_{200}/M_\odot) \leq 12 $, $\log(1) \leq \log(c_{200}) \leq \log(30)$, and $\log(0.01) \leq \log(r_{\rm c}/\rm{kpc}) \leq \log(3.75\,R_{\rm e,\ast})$. The lower bound in the prior of $\log(r_{\rm c})$ effectively means no core, and the upper bound comes from considerations on the energy injected by supernovae, which is not enough to create cores larger than $\sim 3.75\,R_{\rm e,\ast}$ (\citealt{coreNFW,read2017}, see also e.g. \citealt{benitezllambay19,coreEinasto}). 
Finally, we consider the same Gaussian priors for $i$ and $D$ as used in the only-baryons mass model (Sec.~\ref{sec:onlybaryons}). We will call this exploration Case 1.

The top panel of Fig.~\ref{fig:coreNFW} presents our resulting mass model for Case 1, and Fig.~\ref{fig:post_cdm_corenfw_nominM200} in Appendix~\ref{appendix:posteriors} the corresponding corner plot of the posterior distributions. Fig.~\ref{fig:coreNFW} shows the data (grey markers) as well as the contribution from the stars (orange), gas (blue) and dark halo (black) to the total circular speed (red). To illustrate the uncertainties in the model, the red band shows the $16^{\rm th}$ and $84^{\rm th}$ percentiles of our MCMC samples, \emph{at} the best-fitting distance and inclination. Considering that sometimes some of the parameters in these and other posterior distributions shown below are skewed, we also include mass models (see red and black dashed curves) derived from the maximum-likelihood parameters of the posterior distributions, obtained with the Python \texttt{scipy} \citep{scipy} Kernel Density Estimator. The blue box in Fig.~\ref{fig:coreNFW} quotes the median parameters of our fit, and in parenthesis, we give the maximum-likelihood values. The median and the maximum-likelihood $\log(M_{200})$ and $\log(c_{200})$ are also reported in Table~\ref{tab:v2kpc}. Besides, Table~\ref{tab:v2kpc} lists the maximum circular speed of the dark matter halo ($V_{\rm max}$), its circular speed at 2 kpc ($V_{\rm 2kpc}$), and the radius at which $V_{\rm DM} = V_{\rm max}$ ($R_{\rm max}$). Finally, we provide the BIC value.

The CDM mass model is reasonably well-defined and in excellent agreement with the data. Inspecting the posterior distributions (Fig.~\ref{fig:post_cdm_corenfw_nominM200}), it is clear that halo mass, concentration, distance, and inclination are well constrained, while the core radius presents large uncertainties and with a $1\sigma$ interval of between 0 and 3.5 kpc. However, the inferred halo mass of Case 1 is very low: $\log(M_{200}/M_\odot) = 9.36^{+0.37}_{-0.31}$. This is significantly lower than the minimum expectation within $\Lambda$CDM, i.e. $\log(M_{\rm 200,min}/M_\odot) = \log(M_{\rm bar}/0.16\,M_\odot) \approx 9.96$. In this case, the baryon fraction of AGC~114905 would be $\sim 4$ times higher than the cosmological average. Under this interpretation, our results would indicate that AGC~114905 misses a significant amount of dark matter relative to its baryonic mass. The concentration parameters are also lower than expected from the CDM $c_{200}-M_{200}$ relation \citep{duttonmaccio2014}.

\begin{figure}
    \centering
    \includegraphics[width=0.47\textwidth]{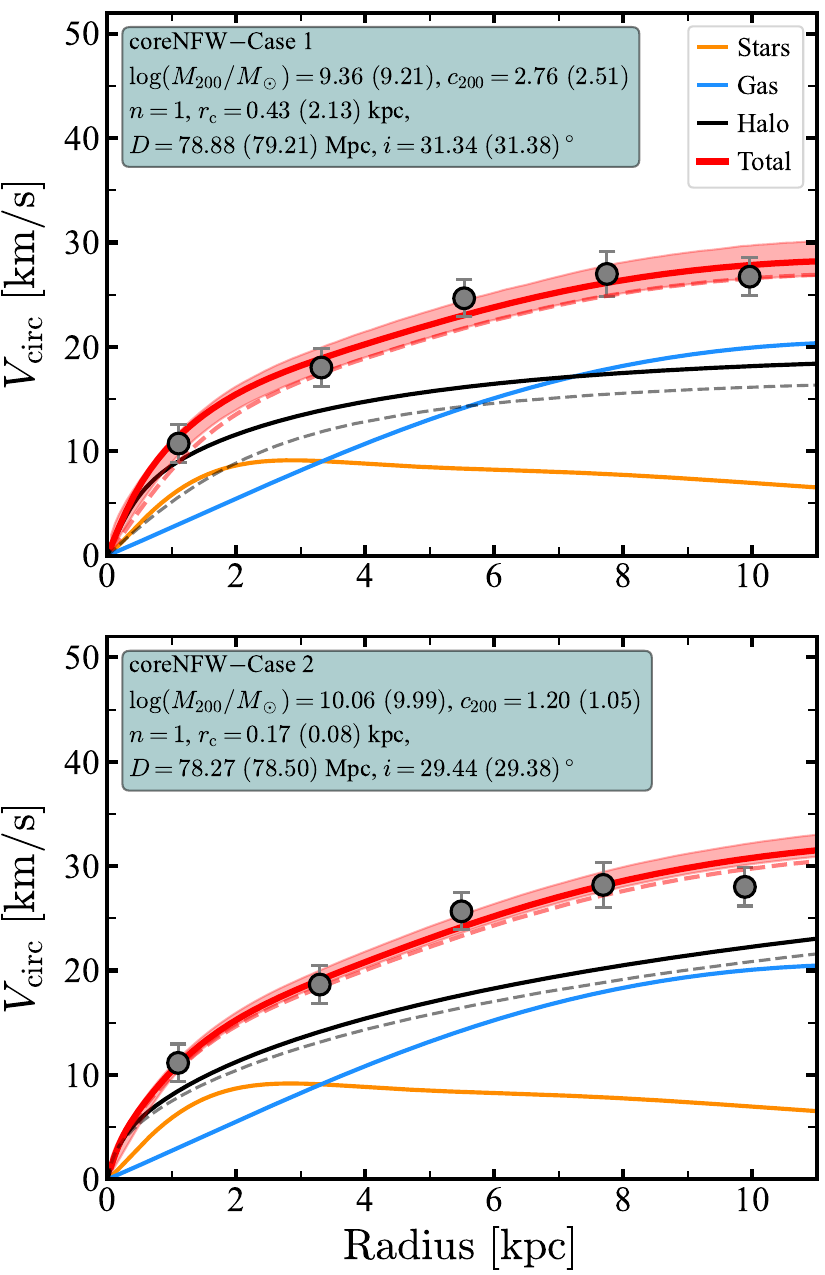}
    \caption{Mass models within the CDM framework. The \emph{top} panel shows the fit for Case 1 when $M_{200}$ can be as low as $10^7~M_\odot$, which results in a baryon fraction higher than the cosmological average. The \emph{bottom} corresponds to Case 2, when $M_{200}$ ensures a baryon fraction lower or equal to the cosmological baryon fraction. The median parameters of the best-fitting distributions are listed on each panel, together with the maximum-likelihood values (in parenthesis). In both panels, the orange, blue, and black curves represent the contribution to the total circular speed (red) from the stars, gas, and dark matter halo, respectively. The red band represents the $16^{\rm th}$ and $84^{\rm th}$ percentiles of our posteriors, \emph{at} the best-fitting distance and inclination.
    Note that the best-fitting distance and inclination are slightly different between both panels. The additional black and red dashed curves correspond to the maximum-likelihood mass model.
    }
    \label{fig:coreNFW}
\end{figure}

Following the discovery of a couple of gas-poor UDGs in high-density environments that appear to lack dark matter (e.g. \citealt{vandokkum_DF2,vandokkum_DF4,danieli_DF2}, cf. \citealt{trujillo_distanceDF2,laporte_udgs}), different mechanisms have been proposed in the last years to explain the existence of dwarf galaxies with less dark matter than expected (e.g. \citealt{duc_tdgs, jackson_tidalUDGs,silk_udgs,montes_df4,shin_udgs,trujillogomez_udgs,benavides2021,moreno2022,ivleva2024}). However, as discussed in detail in \citet{agc114905}, such mechanisms rely on the strong influence of the environment and typically produce gas-poor galaxies and, therefore, cannot directly explain the properties of AGC~114905, which is an isolated gas-rich system. 

Another intriguing possibility, albeit unclear if feasible, is that AGC~114905 and similar galaxies experienced an atypical accretion history that allowed them to have more baryons than primordially given. In this scenario, one could evoke strong feedback episodes at high-$z$ expelling a significant fraction of baryons out of early galaxies into the intergalactic medium (e.g. \citealt{hayward2017,veilleux2020}, see also \citealt{romano2023} at low-$z$), and those baryons falling back into different hosts; some dark haloes would end up accreting material from a big pool of gas previously ejected by other galaxies. In this process, some galaxies may end up with more baryons than they originally had, and therefore exceeding the cosmological baryon fraction. There are, however, some limitations to the above idea. First, it is unclear if such processes would actually be efficient at low halo masses (e.g. \citealt{white2015,mcquinn2019,romano_ultrafait,bassini2023}) and if there is at any point so much gas available in the intergalactic medium.
Moreover, the isolation of AGC~114905 might be at odds with the idea, as the system does not seem to have neighbouring galaxies that would have lost their baryons to it (unless there is a significant presence of dark galaxies around the galaxy, or if its location has dramatically changed from the past). Finally, modern cosmological hydrodynamical simulations do not report finding galaxies with baryon fractions higher than the cosmological average. However, the halo mass regime of UDGs is mainly unexplored as it requires large volumes at high resolution. Therefore, it is unclear if the little dark matter scenario is a viable option, considering also that such systems may develop global instabilities \citep{ostriker1973,sellwood_agc114905}.


Considering this, and following \citet{agc114905,demao}, we explored a second mass model where the condition of not exceeding the cosmological baryon fraction is reinforced. In this fit, which we call Case 2, $\log(M_{\rm 200,min})$ is imposed as the lower bound for the halo mass, with $\log(M_{\rm 200,min}/M_\odot)=\log(M_{\rm bar}/0.16\,M_\odot)$. The \emph{bottom} panel of Figure~\ref{fig:coreNFW} shows the best-fitting \textsc{coreNFW} halo, while its posterior distribution is shown in Fig.~\ref{fig:post_cdm_corenfw_minM200}. As before, Table~\ref{tab:v2kpc} gives the median and maximum-likelihood $\log(M_{200})$ and $c_{200}$, as well as $V_{\rm 2kpc}$, $V_{\rm max}$, and $R_{\rm max}$. 

Case 2 has $\log(M_{200}/M_\odot) = 10.06^{+0.13}_{-0.07}$, and it provides a good fit to the data; albeit it slightly overestimates the last value of the circular speed, it has a similar BIC as the previous mass model (see Table~\ref{tab:v2kpc}). As can be seen, the maximum-likelihood values slightly improve the mass model. As for the other parameters, we highlight that the concentration goes to its lower bound ($c_{200} \approx 1$), and our constrain in the core radius is essentially an upper limit on $r_{\rm c} \lesssim 1~\rm{kpc}$, which means that the profile is consistent with a classical NFW profile \citep[as also found by ][]{agc114905,shi2021} or at most with a modest core. We note that hydrodynamical simulations predict efficient dark matter core creation by stellar feedback for galaxies with $M_\ast/M_{200} \sim 0.01$ (e.g. \citealt{dicintio2014,coreEinasto,azartash2024}), similar to Case 2, with core sizes between $2-7$~kpc \citep{coreEinasto}. Our mass model disfavours a core of that size. We surmise that this could be due to star formation not having been sustained for long enough to drive a more significant dark matter expansion (see, e.g. \citealt{read2019,collins2021}). As we mention in Sec.~\ref{sec:sbprofs}, the older and younger stellar populations of AGC~114905 seem to have ages around $1-2$~Gyr and $0.5-1$~Gyr, respectively. The absence of a large core appears to be in line with this type of system having experienced relatively `weak' feedback through their evolution \citep{huds2020}.  \\

\begin{table*}
    \caption{Structural properties of the fitted dark matter haloes.}
    \label{tab:v2kpc}
\resizebox{1\textwidth}{!}{
    \centering
    \begin{tabular}{lcccccccc}
    \hline \noalign{\smallskip}
       Halo  & $\log(M_{200})$ & $\log(c_{200})$ & $\log(M_{200})_{\rm{MLE}}$ & $\log(c_{200})_{\rm{MLE}}$ & $V_{\rm 2kpc}$ & $V_{\rm max}$  & $R_{\rm max}$ & $\rm{BIC}$\\ \noalign{\smallskip}
       & $[\log(M_\odot)]$ & & $[\log(M_\odot)]$ & & [km/s] & [km/s]  & [kpc] & \\  
       \noalign{\smallskip}
   \hline \noalign{\smallskip}
      CDM, \textsc{coreNFW}, Case 1  & $9.36^{+0.37}_{-0.31}$ & $0.44^{+0.39}_{-0.27}$  & 9.21  & 0.40 & $11.1^{+1.6}_{-1.6}$ & $19.6^{+6.0}_{-3.5}$ & $21.4^{+29.4}_{-11.8}$ & $11.13^{+2.51}_{-1.05}$\\ \noalign{\smallskip}
      CDM, \textsc{coreNFW}, Case 2  & $10.06^{+0.13}_{0.07}$ & $0.08^{+0.10}_{-0.06}$  & 9.99  & 0.02 & $11.3^{+1.0}_{-0.8}$ & $33.6^{+3.7}_{-2.0}$ & $84.4^{+14.9}_{-18.2}$ & $11.32^{+1.67}_{-0.53}$\\ \noalign{\smallskip}
      CDM, DPL, Case 1$^\ast$        & $9.24^{+0.35}_{-0.29}$ & $0.57^{+0.15}_{-0.24}$  & 9.20  & 0.63 & $6.8^{+3.4}_{-2.9}$ & $13.0^{+4.4}_{-5.7}$ & $13.7^{+20.5}_{-5.8}$  & $12.99^{+3.17}_{-1.39}$ \\ \noalign{\smallskip}
      CDM, DPL, Case 2$^\ast$        & $10.16^{+0.39}_{-0.15}$ & $0.18^{+0.13}_{-0.12}$ & 10.03 & 0.19 & $4.3^{+1.5}_{-1.1}$ & $19.9^{+8.8}_{-5.2}$ & $71.5^{+40.0}_{-24.5}$ & $17.20^{+4.38}_{-3.48}$ \\ \noalign{\smallskip}
      SIDM, Case 1                   & $9.21^{+0.25}_{-0.22}$ & $0.61^{+0.16}_{-0.19}$  & 9.17  & 0.63 & $12.0^{+2.2}_{-2.4}$ & $19.0^{+3.5}_{-2.7}$ & $11.1^{+10.2}_{-4.5}$  & $9.81^{+2.97}_{-1.41}$ \\ \noalign{\smallskip}
      SIDM, Case 2                   & $10.08^{+0.21}_{-0.09}$ & $0.17^{+0.14}_{-0.11}$ & 10.01 & 0.10 & $7.5^{+1.4}_{-0.9}$ & $35.1^{+6.7}_{-2.5}$ & $65.6^{+26.4}_{-21.9}$ & $17.50^{+2.44}_{-1.49}$ \\ \noalign{\smallskip}
      FDM, Case 1                    & $8.99^{+0.41}_{-0.57}$ & $0.58^{0.60}_{-0.42}$   & 9.09  & 0.09 & $9.7^{+2.8}_{-2.1}$ & $18.8^{+2.6}_{-2.2}$ & $7.0^{+10.9}_{-1.8}$  & $16.26^{+2.03}_{-1.37}$ \\ \noalign{\smallskip}
      FDM, Case 2                   & $10.04^{+0.11}_{-0.06}$ & $0.08^{+0.10}_{-0.06}$  & 9.99 & 0.02 & $9.1^{+2.1}_{1.5}$ & $33.3^{+3.0}_{-1.7}$ & $84.1^{+12.8}_{-18.6}$ & $18.95^{+1.70}_{-1.08}$ \\ \noalign{\smallskip} \hline
    \end{tabular}} 
    \tablefoot{
The first column lists the corresponding mass model. Case 1 considers a broad $\log(M_{200})$ prior, while for Case 2 a lower limit on $\log(M_{200})$ was imposed. The second and third columns give the halo mass and concentration parameters. The values and uncertainties in $\log(M_{200})$ and $\log(c_{200})$ come from the $16^{\rm th}$, $50^{\rm th}$, and $84^{\rm th}$ percentiles of the posterior distributions shown in Appendix~\ref{appendix:posteriors}. The fourth and fifth columns give the maximum-likelihood estimation (MLE) of $\log(M_{200})$ and $\log(c_{200})$ (the remaining MLE parameters are given in the corresponding figures of each mass model). The sixth, seventh, and eight columns give $V_{\rm 2kpc}$, $V_{\rm max}$, and $R_{\rm max}$, respectively. These values are obtained by reconstructing and sampling the MCMC chains of each mass model and obtaining the distribution of those parameters. Finally, the last column gives the BIC of each model computed from the posterior distributions.
    ($^\ast$) The DPL (double power-law) fits are shown in Appendix~\ref{appendix:dpl} and included here for completeness.}
\end{table*}

\noindent
It is important to highlight two aspects implied by the structural parameters of the haloes. 
First, the values of $M_{200}$ found for Case 1 and Case 2 imply very high baryon fractions $f_{\rm bar} = M_{\rm bar}/(f_{\rm bar,cosmic}\times M_{200})$. For Case 1 $f_{\rm bar} \approx 4$, and for Case 2 $f_{\rm bar} \approx 0.8$. Even in Case 2, $f_{\rm bar}$ implies that AGC~114905 has virtually no `missing baryons' (see \citealt{huds2019,huds2020}) and is higher than in typical gas-rich dwarf irregular galaxies ($f_{\rm bar}\sim0.1-0.5$, although the scatter is large, see \citealt{paper_massmodels}) which have typically dynamically dominant dark haloes (e.g. \citealt{deblok_phd,read2017}). Instead, the high baryon fraction of  AGC~114905 is more similar to the values found in some massive spiral galaxies \citep{postinomissing,paper_massmodels}. This result sets important constraints for simulations aiming to produce gas-rich UDGs (e.g. \citealt{nihao_udgs}).

The remarkable second aspect concerns the concentration parameters of our best-fitting haloes. We find a similar result as \citet{agc114905} and \citet[][see also \citealt{shi2021}]{demao}, namely that the only way to fit these types of haloes (for both Case 1 and 2) is to consider shallow concentration parameters ($c_{200} = r_{200}/r_{-2}$). This is not typically expected in the context of CDM, where low-mass haloes have high concentrations (e.g. \citealt{duttonmaccio2014,ludlow2014,diemer2019}), which is generally seen in low-mass galaxies (e.g. \citealt{read2017,paper_massmodels}). 

For both Case 1 and 2, our best-fitting $c_{200}$ parameters are about $\sim 7-9\sigma$ below the mean value of the $c_{200}-M_{200}$ relation typically expected \citep{duttonmaccio2014}. Nevertheless, as shown by \citet{demao} in their Fig. 14, inspection of the $c_{200}-M_{200}$ relation in the TNG50 simulation \citep{illustris_tng50} shows that the nearly constant scatter of the $c_{200}-M_{200}$ relation seen at high masses breaks down at $M_{200} \lesssim 10^{10}~M_\odot$, where a strong tail of low-concentration (or high $R_{\rm max}$) simulated haloes emerges. Yet, while some TNG50 systems have low concentrations, \citet[][see also \citealt{nadler2023}]{demao} have shown that none is as extreme as AGC~114905 in our Case 2, and only a handful of outliers out of the $\sim 37500$ simulated haloes with similar halo masses are similar to our Case 1. 

Part of the reason why simulated systems with the dark matter properties of AGC~114905 are not found in abundance in TNG50 could be a consequence of the volume within which AGC~114905 is observed ($\sim 512000~\rm{Mpc}^3$) being about four times larger than the TNG50 volume \footnote{TNG100 cannot be explored in this case, as it lacks the resolution to simulate low-mass dark haloes.} ($\sim 125000~\rm{Mpc}^3$). While exploring simulations with enough resolution but large cosmological volumes would undoubtedly be desirable, the tension between AGC~114905 and TNG50 is exacerbated by the fact that even those simulated haloes with $c_{200}$ low enough to be nearly consistent with observed gas-rich UDGs are significantly denser on small scales, as their $V_{\rm 2kpc}$ are considerably higher \citep{demao,nadler2023}. This is likely because simulated haloes on the low-concentration tail might not be fully relaxed, and their density profiles can be somewhat steeper than an NFW \citep{demao}. Since stellar feedback does not modify strongly the density of the dark matter haloes in TNG50 \citep{lovell2018}, and even in simulations where it does \citep{nihao_udgs,fire_udgs} such extreme objects do not seem to be produced, it all suggest that the existence of these galaxies \citep{huds2020,shi2021,demao}, of which AGC~114905 might be the most extreme case, represents a challenge to our current understanding (or simulation implementation) of the CDM model.
 
Summarising, we have explored a set of CDM mass models for AGC~114905. In our models, the galaxy has a very high baryon fraction, potentially exceeding the cosmological average. Moreover, the inferred dark matter haloes have markedly low concentration parameters and are more extended than those halos of similar mass produced in CDM-based simulations. These results robustly confirm previous conclusions with more limited data \citep{agc114905,demao}. Either these systems are scarce, and they are informing us about the extreme, unexplored, and not well-understood conditions at which galaxy formation can take place within the CDM model, or they may be providing new clues on the nature of dark matter itself, favouring alternative models to CDM. In the following sections, we start exploring this second possibility.

\subsection{Self-interacting dark matter}
\label{sec:sidm}

An alternative way to overcome the small-scale problems of CDM other than baryonic physics (e.g. \citealt{bullock2017,sales_review_dwarfs,collins_feedback}) is to assume a different type of dark matter particle. A particularly promising candidate is self-interacting dark matter (SIDM, see \citealt{spergel2000,tulin_sidm} and references therein). Thanks to the self-interactions, dark matter particles thermalise the haloes and can generate cores without the need for stellar feedback (this does not mean there is no stellar feedback, see, e.g. \citealt{robles2017}). More generally, all SIDM haloes go through a gravothermal phase of expansion and subsequent collapse \citep{yang_gravothermal}, which can lead to a broader diversity in rotation curves than in CDM (e.g. \citealt{nadler2023,jiang2023_sidm}), solving the `diversity of rotation curves problem' \citep{oman2015,salucci2019,sales_review_dwarfs}.
Considering SIDM in the case of AGC~114905 is particularly appealing. While, as discussed above, large CDM-based cosmological simulations only rarely produce haloes with densities as low as observed in some gas-rich UDGs \citep{demao}, such low-density haloes appear to be a natural outcome of small-volume (216~Mpc$^3$) SIDM-based cosmological simulations, which consider a velocity-dependent cross-section with $\sigma/m \approx 120\, \rm{cm^2/g}$ for haloes with maximum circular speeds below $50~\rm{km/s}$ \citep{nadler2023}.

In this study, we build upon the work by \citet{yang_parametric} and \citet{nadler2023} to fit a SIDM halo to AGC~114905. \citet{yang_parametric} provide an analytic expression for the density profile of SIDM haloes, which captures their gravothermal evolution through time, as calibrated with SIDM simulations. Their profile uses the same functional form as the \textsc{coreNFW} halo, but the parameters defining it are different. Specifically, the SIDM halo considers $n=1$, and the parameters $\rho_{\rm s}$, $r_{\rm s}$, and $r_{\rm c}$ are given by the following set of equations
\begin{multline}
\dfrac{4\,\rho_{\rm s}}{\rho_{\rm s,NFW}} = 1.335 + 0.7746\,\tau + 8.042\,\tau^5 - 13.89\,\tau^7 + 10.18\,\tau^9 \\ 
+ \frac{(1 - 1.335)}{\ln(0.001)} \ln(\tau + 0.001)~, 
\end{multline}

\begin{multline}
\dfrac{r_{\rm s}}{r_{\rm s,NFW}} = 0.8771 - 0.2372\,\tau + 0.2216\,\tau^2 -0.3868\,\tau^3 \\ 
+ \frac{1 - 0.8771}{\ln(0.001)} \ln(\tau + 0.001)~, 
\end{multline}

\begin{multline}
\dfrac{r_{\rm c}}{r_{\rm s,NFW}} = 3.324\sqrt{\tau} - 4.897\,\tau + 3.367\,\tau^2 - 2.512\,\tau^3 + 0.8699\,\tau^4~, 
\end{multline}
where $\rho_{\rm s,NFW}$ and $r_{\rm s,NFW}$ are the usual NFW parameters and are modified by functions that depend on $\tau = 10$~Gyr $/t_{\rm c}$, with $t_{\rm c}$ a collapse timescale linked to the effective cross-section ($\sigma/m$) through the expression
\begin{equation}
    t_{\rm c} =  \dfrac{200}{(\sigma/m)\, \rho_{\rm s,NFW}\, r_{\rm s,NFW} } \dfrac{1}{\sqrt{4\, \pi\, G\, \rho_{\rm s,NFW}}}~.
\end{equation}

In this work, we test whether the cross-section proposed by \citet[][see also, e.g. \citealt{correa2022}]{nadler2023} is reasonable to fit the kinematics of AGC~114905. Specifically, we assume the functional form
\begin{equation}
    \sigma/m = \dfrac{\sigma_0/m}{ [ 1 + (V_{\rm max}/v_0)^{\delta} ]\,^\beta }~,
\end{equation}
with $\sigma_{0}/m = 147.1\, \rm{cm}^{2}\, \rm{g}^{-1}$, $v_0 = 80\, \rm{km/s}$, $\delta = 1.72$, $\beta=1.90$.

Using the above equations, one can start from a CDM halo and derive its SIDM counterpart for a given cross-section. In this way, the final SIDM halo depends on the `initial' NFW parameters (i.e. the parameters that the halo would have if it did not experience self-interactions, as in CDM) $M_{200}$ and $c_{200}$ and the cross-section $\sigma/m$. 
In practice, during our MCMC fitting, we consider the same priors on $\log(M_{200})$, $\log(c_{200})$, $D$, and $i$ as in the CDM fits.

\begin{figure}
    \centering
    \includegraphics[width=0.47\textwidth]{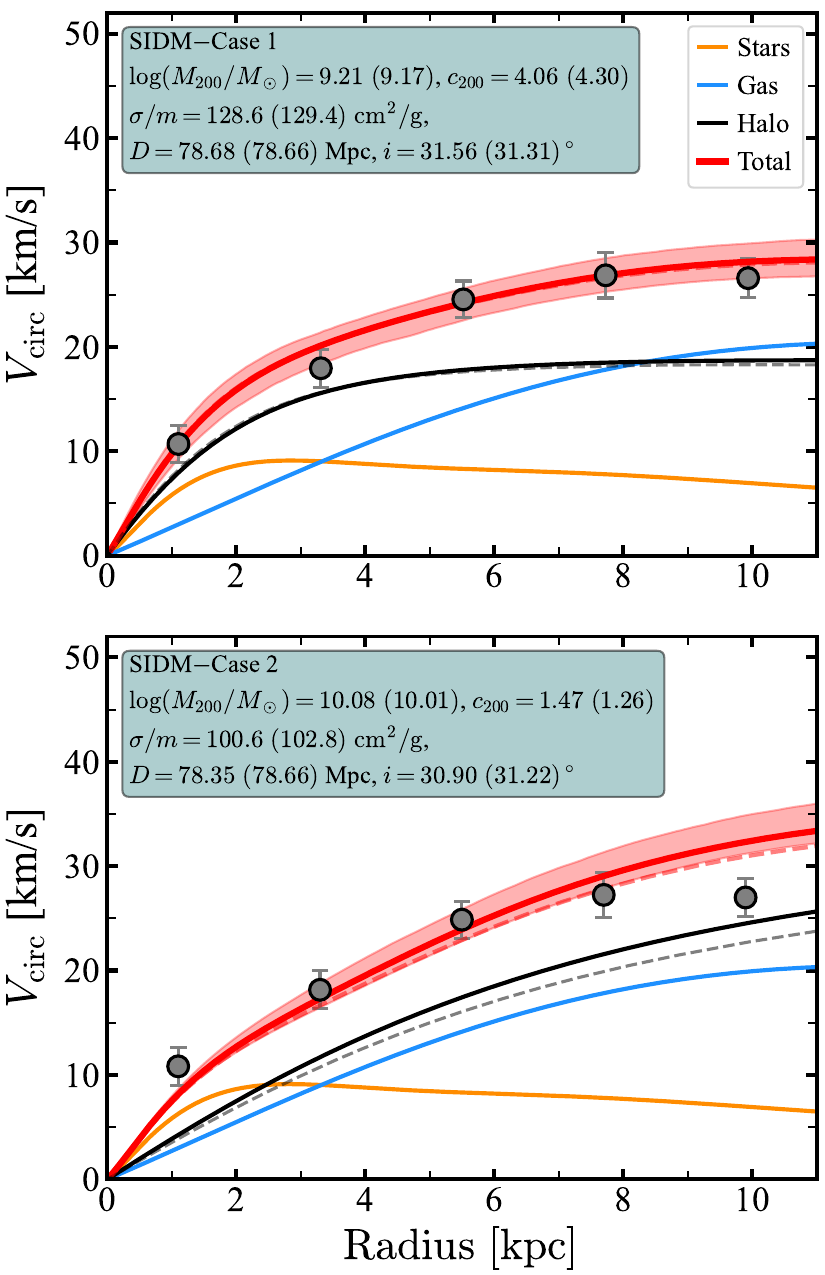}
    \caption{Mass models for a self-interacting dark matter halo. The top panel shows the results for Case 1 when $M_{200}$ is completely free (in which case the cosmological baryon fraction is surpassed). The bottom panel shows Case 2, imposing a lower bound in $M_{\rm 200,min}$. Curves are as in Fig.~\ref{fig:coreNFW}.}
    \label{fig:sidm}
\end{figure}

The top panel of Fig.~\ref{fig:sidm} shows our SIDM mass model when $M_{200}$ is allowed to be within the range $7 \leq \log(M_{200}/M_\odot) \leq 12 $ (Case 1); the posterior distribution is shown in Fig.~\ref{fig:post_sidm_nominM200}. The fit is very good, which shows that the cross-section proposed by \cite{nadler2023}, namely $\sigma/m \approx 129~\rm{cm^2/g}$ (with $\tau \approx 0.04$) works for Case 1. Similar to what happens with the CDM-Case 1, our SIDM-Case 1 requires a low halo mass: $\log(M_{200}/M_\odot) = 9.2$. This value implies a baryon fraction about 6 times larger than the cosmological baryon fraction. An advantage, with respect to the CDM-Case 1, is that SIDM-Case 1 has a higher $c_{200}$ parameter, albeit still low.

As before, we perform a second fit when the flat prior $\log(M_{\rm 200,min}) \leq \log(M_{200}/M_\odot) \leq 12 $ is used (Case 2). The resulting mass model is shown in the bottom panel of Fig.~\ref{fig:sidm}, and its posterior distribution in Fig.~\ref{fig:post_sidm_minM200}. Case 2 also provides a reasonable fit under the cross-section of \cite{nadler2023}, namely $\sigma/m \approx 101~\rm{cm^2/g}$ (with $\tau \approx 0.01$). The $c_{200}$ is as low as in the CDM case, and the halo mass is $\log(M_{200}) = 10.1$, with a baryon fraction of 0.76 relative to the cosmic mean. Table~\ref{tab:v2kpc} quotes the values for $V_{\rm 2kpc}$, $V_{\rm max}$, and $R_{\rm max}$, for both Case 1 and Case 2. We note that in both cases the $\tau$ parameters are low, meaning that the haloes are far from the collapse phase of the SIDM gravothermal evolution \citep{yang_gravothermal,yang_parametric}.

Overall, the SIDM mass models are of similar quality as their CDM counterparts (albeit the BIC of Case 2 is somewhat worse, see Table~\ref{tab:v2kpc}) and also have baryon fractions and concentration parameters as extreme as in CDM (somewhat less extreme for SIDM-Case 1). 
The SIDM haloes present other advantages over the CDM ones. First, the dark matter cores can form due to the dark matter self-interactions, without the need for stellar feedback processes to drive substantial dark matter expansion \citep{tulin_sidm,robles2017,kaplinghat2020}. As shown before, it remains uncertain whether AGC~114905 has a core or not, but in general, being able to create a core without stellar feedback is important as different simulations and models disagree on whether or not feedback-induced cores develop (e.g. \citealt{dicintio2014,coreNFW,lovell2018,freundlich2020,burger2021,orkney2021,sales_review_dwarfs,cuspcoreII}). And second, but perhaps more importantly, SIDM cosmological simulations appear to naturally produce haloes with very low densities (also in small $\sim 2\,\rm{kpc}$ scales) approaching that of AGC~114905, while CDM simulations struggle to do so \citep{demao,nadler2023}. Gas-rich UDGs like AGC~114905 will provide a crucial test for upcoming large-volume SIDM cosmological hydrodynamical simulations.


\subsection{Fuzzy dark matter}
\label{sec:fdm}
Another candidate is fuzzy dark matter (FDM, see \citealt{ferreira_review} for a recent review). In the framework of FDM, dark matter is made of axions of very low mass $-25 \lesssim \log(m_{\rm a}/\rm{eV}) \lesssim -21$. Due to quantum effects (see, e.g. \citealt{ferreira_review,banares2023} and references therein), the axions generate a sizeable dark matter core (referred to as soliton) and suppress the small scales of the power spectrum (e.g. \citealt{hui2017,ferreira_review}), alleviating two critical problems of CDM (the `cusp-core' and the `missing-satellites' problems) while keeping its success at large scales. FDM is, therefore, a tantalising type of dark matter to explain the kinematics of AGC~114905, where low dark matter densities in large scales are needed. In fact, \citet{dentler2022} has shown that the dark matter concentration-mass relation within FDM is suppressed compared to its CDM counterpart, reaching values as low as inferred from our CDM fits above. 

Following \citet{bar2018,bar2022}, and \citet{banares2023}, we use the following set of equations to define the FDM halo density profile \citep{schive2014}
\begin{equation}
        \rho_{\rm FDM}=
    \begin{cases}
     \, \rho_{\rm sol}, & \text{if}\ r\leq r_{\rm t} \\
     \, \rho_{\rm NFW}, & \text{if}\ r > r_{\rm t} ~,
    \end{cases}
\end{equation}
\begin{equation}
    \rho_{\rm sol} = \dfrac{\rho_0}{ \left[ 1 + 0.091\, (r/r_{\rm c})^2 \right]^8 }~,
\end{equation}
\begin{equation}
    \rho_0 = 0.083 \left( \dfrac{M_{\rm sol}}{M_\odot}  \right) \left( \dfrac{r_{\rm c}}{\rm{kpc}}\right)^{-3} \dfrac{M_\odot}{\rm{kpc}^3}~,
\end{equation}
\begin{equation}
    M_{\rm sol} = 2.28\times10^8~M_\odot \left(\dfrac{r_{\rm c}}{\rm{kpc}}\right)^{-1} \left( \dfrac{m_{\rm a}}{1\times10^{-22}~\rm{eV}}  \right)^{-2}
\end{equation}
where $m_{\rm a}$ is again the axion mass, $r_{\rm c}$ a core radius at which the central density $\rho_0$ drops by a factor of 2, $M_{\rm sol}$ the soliton mass, and $r_{\rm t}$ a transition radius between the density profile of the soliton ($\rho_{\rm sol}$) and that of the outer NFW halo ($\rho_{\rm NFW}$); $r_{\rm t}$ can be found as the first radius starting from $r = \infty$ at which $\rho_{\rm sol} \geq \rho_{\rm NFW}$. From this, it follows that $\rho_{\rm FDM}$ can be parameterised with four parameters: $M_{200}$, $c_{200}$, $m_{\rm a}$, and $r_{\rm c}$.

For $M_{200}$, $c_{200}$, $i$, and $D$, we adopt the same priors as the previous sections. For the core radius, we use the flat prior $-2 \leq \log(r_{\rm{c}}/\rm{kpc}) \leq$ $\log(r_{\rm t})$, while for the axion mass the prior is Gaussian, centred at $\log(m_{\rm a}/\rm{eV}) = \log(1.9\times10^{-23})$, with a scatter of 0.2~dex, and within the bounds $-25 \leq \log(m_{\rm a}/\rm{eV}) \leq -21$. These priors on $m_{\rm a}$ and $r_{\rm c}$ are motivated by the values found by \citet{banares2023} on a sample of 13 dwarf irregular galaxies with high-resolution gas kinematics \citep{iorio}, and consistent with the values found in \citet{2024A&A...681A..15M} based on the stellar distribution of a dwarf low surface brightness galaxy. We also explored fits under a flat prior on $\log(m_{\rm a})$, finding consistent results within the uncertainties as those shown below for our fiducial priors, but with a stronger degeneracy between $m_{\rm a}$ and $r_{\rm c}$.

\begin{figure}
    \centering
    \includegraphics[width=0.47\textwidth]{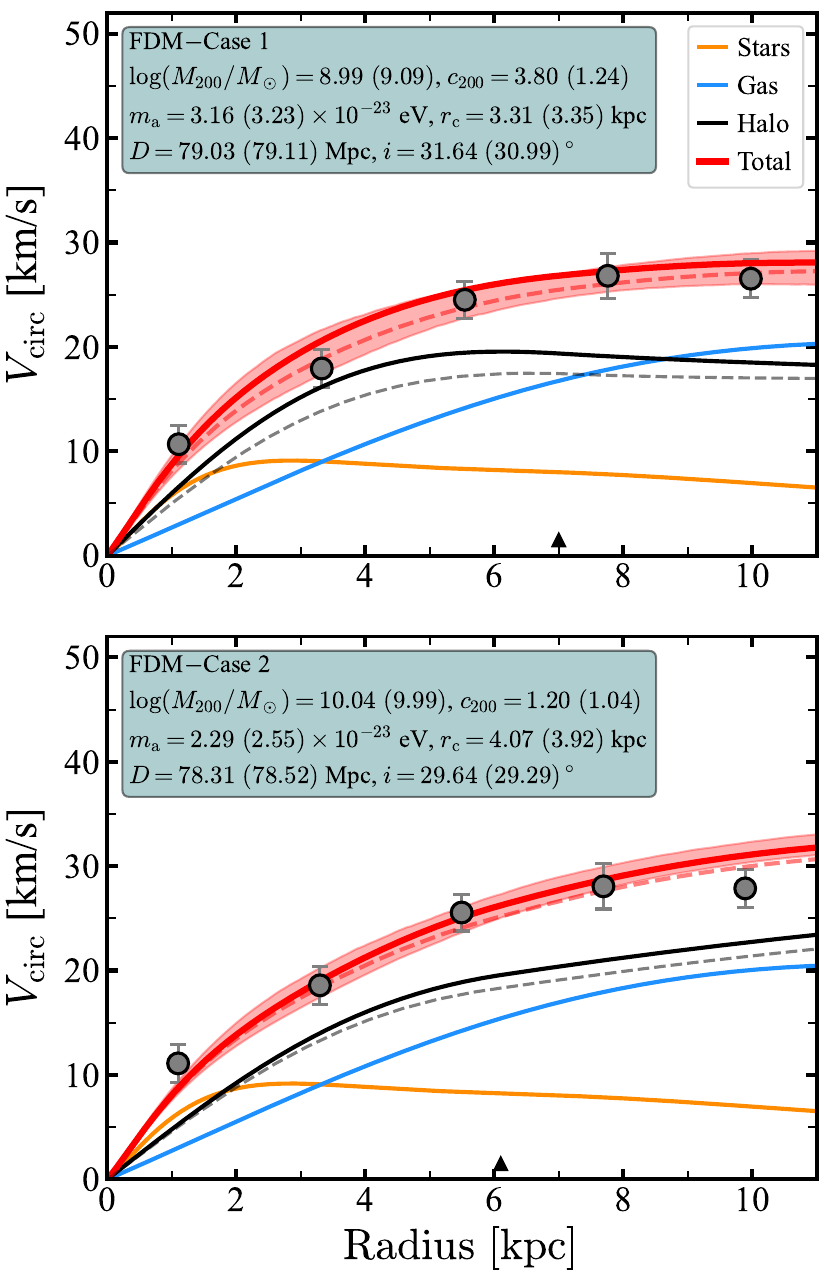}
    \caption{Mass models for a FDM halo. The top panel is Case 1, when $M_{200}$ is completely free (and the cosmological baryon fraction is not obeyed), while the bottom panel is Case 2, imposing a lower bound in $M_{\rm{200,min}}$. Curves are as in Fig.~\ref{fig:coreNFW}. The black triangles indicate the position of $r_{\rm t}$, the transition radius between the soliton and the NFW halo }.
    \label{fig:fuzzy}
\end{figure}

Fig.~\ref{fig:fuzzy} shows our two FDM mass models. The top panel shows the case where $M_{200}$ is allowed to be as low as preferred by the fit (Case 1), while the bottom panel is the mass model when $M_{\rm 200,min}$ is imposed as a lower bound (Case 2). The corresponding posterior distributions are shown in Figs.~\ref{fig:post_fuzzy_nominM200} and \ref{fig:post_fuzzy_minM200}. Both mass models fit the data well. The transition radius $r_{\rm t}$ is relatively large (7~kpc for Case 1 and 6.1~kpc for Case 2), highlighting the relevance of the solitonic component. 
In Case 1, the halo mass is low ($\sim$50\% smaller than $M_{\rm bar}$), resembling the CDM and SIDM cases. The nominal concentration is slightly higher than in the previous fits, but the posterior distribution shows barely any constraint on this parameter. The axion mass $m_{\rm a} = 3.16\times10^{-23}\,\rm{eV}$ is slightly larger than the mean value found by \citet{banares2023}, but within the scatter of their measurements, as expected from our priors.
Case 2 (bottom panel of Fig.~\ref{fig:fuzzy}) also reproduces the data at a similar level as the CDM and SIDM cases. We find a best-fitting $m_{\rm a} = 2.29\times10^{-23}\, \rm{eV}$. The structural parameters and BIC values for these models are given in Table~\ref{tab:v2kpc}. At face value, Fig.~\ref{fig:fuzzy} suggests that FDM can also explain the kinematics of AGC~114905, provided that the halo has a relatively low dark matter and a low concentration\footnote{For completeness, we have also obtained a mass model where we assume $\rho_{\rm FDM} = \rho_{\rm sol}$. The best-fitting soliton (derived with a flat prior) has $m_{\rm a} = 2.96\times10^{-23}\, \rm{eV}$ and $r_{\rm c} = 3.6\, \rm{kpc}$, and the mass model reproduces the data point by point. This soliton model, however, cannot be put into a cosmological context as it does not provide information on the host dark matter halo.}. Similarly to the SIDM case, the FDM profile does so without feedback-induced modifications.

Nevertheless, low values of the concentration parameter and halo mass are not as atypical in the FDM framework as in CDM. This is for two reasons. First, the $c_{200}-M_{200}$ relation in FDM is a positive correlation rather than an anti-correlation (see \citealt{dentler2022}) and for plausible values of $m_{\rm a}$ haloes with $M_{200} \lesssim 10^{11}\, M_\odot$ have $c_{200} \lesssim 5$. Second, abundance-matching stellar-to-halo mass relations within FDM develop a cut-off, which for typical $m_{\rm a}$ values implies a flattening in $M_\ast$ at halo masses below $\sim 5\times10^{10}\,M_\odot$ (e.g. \citealt{banares2023}).
All this makes our best-fitting haloes less extreme in FDM than in CDM. 
For instance, given its $m_{\rm a}$, FDM-Case 1 has a concentration parameter (but consider its poorly constrained posterior) about $1\sigma$ above the expected $c_{200}-M_{200}$ of \citet{dentler2022}, and a halo mass $2\sigma$ below the abundance-matching based FDM stellar-to-halo mass relation (see, e.g. \citealt{banares2023}). For FDM-Case 2, its concentration and halo mass are both within $3\sigma$ (below) the expectations given its $m_{\rm a}$.

All our discussion above suggests that FDM may explain the kinematics of AGC~114905. However, one should also consider the results from \citet{banares2023}. Those authors report that, although they can successfully fit their sample of dwarf galaxies with FDM, their findings do not fully support the FDM model. This is related to the suppression of small-scale structures. \citet{banares2023} show that their best-fitting $m_{\rm a} = 1.9\times10^{-23}\,\rm{eV}$ flattens the stellar-to-halo mass relation at about $M_\ast \sim 1\times10^9\, M_\odot$ for any halo with $M_{200} \lesssim 5\times10^{10}\,M_\odot$, resulting in all of their galaxies being outliers of the relation, having too low $M_\ast$ for their halo mass. Another manifestation of this problem is that their $m_{\rm a}$ would underestimate the number of low-mass galaxies observed in the Local Group. As shown by \citet{banares2023} and references therein, including the effect of the baryons does not change this picture significantly. A potential solution to the excessive structure suppression will be if only a fraction of the dark matter is fuzzy or if there is a whole spectrum of axion masses (i.e. a so-called axiverse, often linked to string theories, e.g. \citealt{arvanitaki2010,visinelli2019}). Therefore, we conclude that FDM is a promising dark matter candidate to explain the kinematics of AGC~114905, but enlarging the sample of dwarf galaxies to test its different predictions is crucial to assess its success or failure.

\subsection{Comparing dark matter models}

In the previous sections, we tested whether the kinematics of AGC~114905 can be explained with dark matter haloes in the CDM, SIDM, and FDM frameworks. We found that within the uncertainties, the three types of dark matter can explain the data under some considerations. In this section, we delve into the comparison between the different models. 

\begin{figure}
    \centering
    \includegraphics[width=0.48\textwidth]{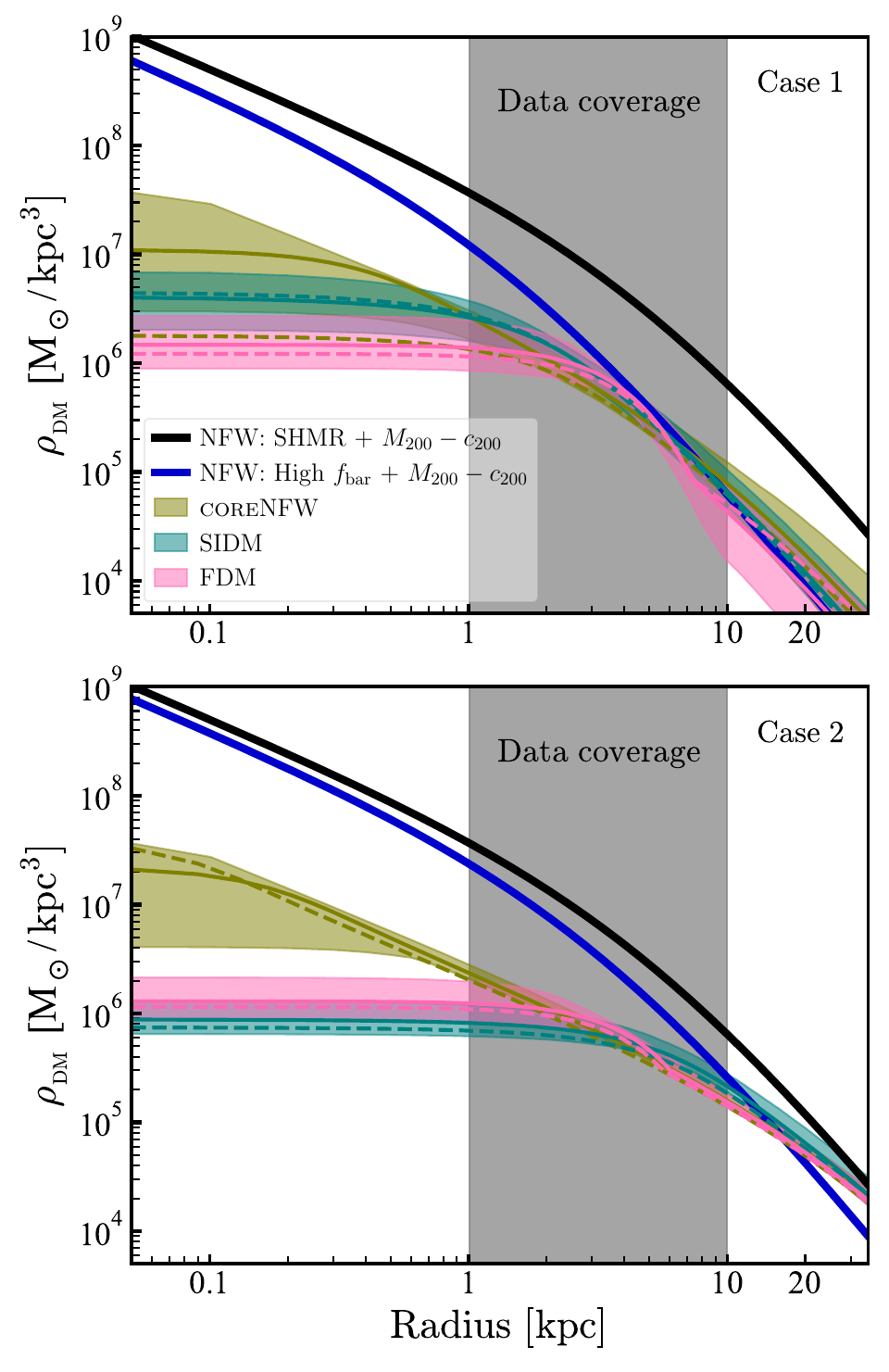}
    \caption{Dark matter density profiles for the dark matter models studied in Secs.~\ref{sec:cdm}-\ref{sec:fdm}. The \emph{top} (\emph{bottom}) shows the Case-1 (Case-2) mass models. On each panel, two reference haloes are plotted. One halo (black dashed line, same for both panels) having a mass expected from an SHMR and a concentration from the $c_{200}-M_{200}$ relation. The other halo (blue curves, different among the panels) instead has a halo mass similar to what we measure for Case 1 and Case 2 and a concentration from the $c_{200}-M_{200}$ relation. The grey region illustrates the approximate extent of our rotation curve.}
    \label{fig:allhaloes}
\end{figure}

In Fig.~\ref{fig:allhaloes}, we plot the density profile of the Case-1 (top) and Case-2 (bottom) dark matter haloes discussed above, together with two reference CDM haloes. The black curve is the same in both panels. It represents a CDM-motivated profile, i.e., an NFW halo with halo mass ($\log(M_{200}/M_\odot) = 10.5$) according to the stellar-to-halo mass relation (SHMR) from \cite{posti2021} and concentration parameter ($c_{200} = 12$) following the $c_{200}-M_{200}$ relation from \cite{duttonmaccio2014}. The blue curve instead is an NFW halo with a halo mass close to our measurements ($\log(M_{200}/M_\odot) = 9.3$ for Case 1 and $\log(M_{200}/M_\odot) = 10$ for Case 2) and a concentration parameter still following the $c_{200}-M_{200}$ relation ($c_{200} = 16$ for Case 1 and $c_{200}=13$ for Case 2); these haloes violate the cosmological baryon fraction.

As can be seen from the top panel of Fig.~\ref{fig:allhaloes}, the three tested Case 1 dark haloes show very similar behaviour, making it hard to distinguish between them -- the most prominent differences are expected in the innermost regions of the galaxy (i.e. within 1~kpc), currently without observational data. Still, it is clear that the models are very different from the ideal more massive and more concentrated NFW halo. The comparison also shows that the models also differ from the NFW halo with similar halo mass but higher concentration. The bottom panel of Fig.~\ref{fig:allhaloes} shows a similar story. The three models resemble each other within the data span, with more noticeable differences expected in the innermost regions. The Case 2 haloes also markedly differ from the two reference NFW haloes. Despite a higher halo mass being enforced with respect to Case 1, these haloes still have a relatively high baryon fraction and, as for Case 1, they need to have low concentration parameters or, equivalently, large $R_{\rm max}$ for their $V_{\rm max}$.

As discussed in the previous sections, despite CDM, SIDM, and FDM producing similar fits to the data, there are further considerations to keep in mind. For instance, at fixed $V_{\rm max}$ or halo mass, CDM N-body cosmological simulations struggle to produce haloes with $R_{\rm max}$ as large and $V_{\rm 2kpc}$ as low as estimated in AGC~114905 and other gas-rich UDGs, while instead such systems are more common in SIDM-based simulations \citep{demao,nadler2023}. Within FDM, the halo mass and concentration parameters are not as extreme as in CDM or SIDM thanks to the different expected SHMR and $c_{200}-M_{200}$ relation. However, it remains unclear whether FDM can simultaneously explain the dynamics of more massive systems (see also \citealt{banares2023}). While our data have enough constraining power to establish the extreme global properties of AGC~114905's halo, it will be crucial to obtain kinematic tracers for the innermost regions to further distinguish between different dark matter models. Besides, enlarging the sample of gas-rich UDGs will help us to set stronger and global constraints on relevant parameters such as $\sigma/m$ for SIDM and $m_{\rm a}$ for FDM. All this will provide key information to compare against large but high-resolution CDM, SIDM and FDM cosmological simulations.

\section{Disc stability and star formation}
\label{sec:stability}
\subsection{The stability of an extreme disc embedded onto an extreme halo}

The extremely low density (both in stars and dark matter) of AGC~114905 makes it a great case to study local and global disc instabilities. In the past, weak instabilities were likely present in the disc of AGC~114905 since the optical morphology shows spiral structures which require at least some mild instabilities \citep{bookFilippo,fiteni2024}. Here, we discuss whether instabilities are still present in this system.

Traditionally, local gravitational instabilities are studied in the context of the Toomre $Q$ parameter \citep{toomre1964}, defined as $Q = \kappa\,\sigma{_{_{\rm HI}}} / (\pi\,G\,\Sigma_{\rm gas})$, where $\kappa$ is the epicycle frequency (which depends on the circular speed, see \citealt{binney1977,bookFilippo}), $\sigma_{_{\rm HI}}$ the gas velocity dispersion, $G$ the gravitational constant, and $\Sigma_{\rm gas}$ the gas surface density. When $Q < 1$, the discs are considered susceptible to local gravitational instabilities, which can likely trigger star formation. While $Q$ was originally derived for razor-thin discs, \citet{romeo2013} have demonstrated that a more appropriate instability diagnostic is $Q_{\rm RF13} = 1.5\, Q$, where the multiplicative factor of 1.5 adapts the traditional $Q$ parameter to a case where discs are thick rather than razor-thin, which helps to stabilise them.

\begin{figure}
    \centering
    \includegraphics[scale=0.52]{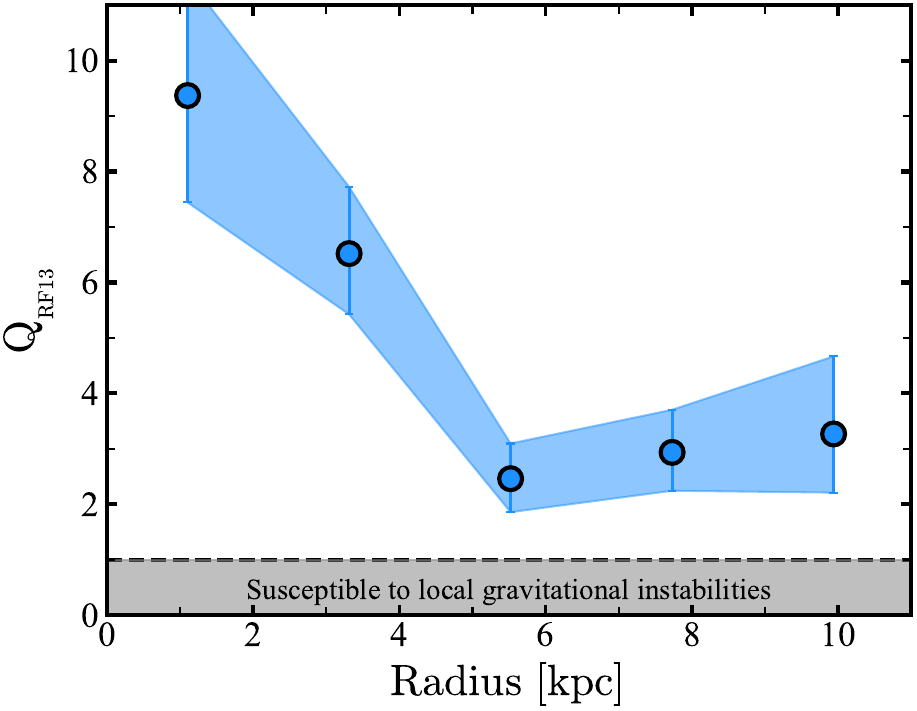}
    \caption{Radial profile of the (thickness-corrected) Toomre parameter ($Q_{\rm RF13}$). $Q_{\rm RF13} > 1$ at all radii suggests that AGC~114905 is not prone to strong local gravitational instabilities.}
    \label{fig:stability}
\end{figure}

We compute the $Q_{\rm RF13}$ parameter, and its radial profile is displayed in Fig.~\ref{fig:stability}. Note that our $Q_{\rm RF13}$ considers only the gas surface density; while the stellar component can play a role in massive spiral galaxies \citep{romeo2017}, this is not the case for AGC~114905, with $\Sigma_{\rm gas} \gg \Sigma_\ast$. 
As can be seen, the evidence suggests that AGC~114905 is stable against local gravitational instabilities, and it has similar values to more massive disc galaxies \citep{romeo2017}. Still, while $Q-$ based diagnostics are simple and useful, it should be noted that both observational and theoretical work has found that they have limitations and should not be over-interpreted \citep{leroy,romeo2013,elmegreen2015,romeo2023}. Taking these results one step further, \citet{ceci_q3d} used the data presented in this manuscript to explore local gravitational instabilities on AGC~114905 using directly a $Q-$like parameter that incorporates the dynamical information of the low-density halo of the galaxy as well as a thick disc ($Q_{\rm 3D}$, see \citealt{nipoti2023}). Those authors found that the disc of AGC~114905 is stable against local gravitational instabilities according to the more appropriate and sophisticated $Q_{\rm 3D}$ criterion, in line with our results in Fig.~\ref{fig:stability}.\\ 

\noindent
AGC~114905 could instead be prone to large-scale global instabilities (e.g. leading to swing amplification or bar formation). Early studies of disc stability based on N-body simulations showed that massive dark matter haloes surrounding galaxy discs are needed to stabilise them against different non-axisymmetric instabilities such as bar formation (e.g. \citealt{ostriker1973,efstathiou1982,mihos1997}). In practice, the picture is more complicated than that, as gas physics processes like feedback and resonances can also stop the propagation of instabilities \citep{binney}, and the thickness of discs also makes them more stable \citep{romeo1992,romeo1994}. These are arguably some of the reasons why simple analytical diagnostics do not succeed at predicting the growth of bar-like instabilities \citep{romeo2023}. Even sophisticated numerical simulations produce systems more prone to instabilities than observed in cosmological hydrodynamical simulations, and the reason for this is not well understood \citep{sellwood2023}. 

\citet{sellwood_agc114905} performed N-body simulations to test the disc stability of AGC~11495 and found that under the assumption of no dark matter, the system develops strong instabilities which disrupt the disc in Gyr-timescales. They report that a stable configuration is reached for either a massive dark matter halo (suggesting that the inclination of AGC~114905 measured by \citet{agc114905} could be underestimated) \emph{or} by a low-concentration halo but considering a mean gas velocity dispersion of about $8~\rm{km/s}$ or higher, above the 6~km/s average reported by \citet{agc114905}. Interestingly, our revised kinematic model (Sec.~\ref{sec:kinematics}) resulted naturally in a velocity dispersion with a mean value of $9~\rm{km/s}$, similar to what is found in other dwarfs \citep{iorio,paperIBFR}. This and the slightly more dominant dark matter halo with respect to that of \citet{agc114905} suggest that AGC~114905 is in global dynamical equilibrium, according to the information reported by \cite{sellwood_agc114905}. In this context, it is also important to highlight that our new optical imaging and the H\,{\sc i} data reveal a fairly regular galaxy without massive companions or clear disturbers, in line with the idea of the system being relaxed and stable. Still, deeper analysis of the stability of AGC~114905 through new N-body simulations and considering their 3D morphology would provide a more definitive view of the long-term stability of the galaxy; efforts to do this are underway and suggest AGC~114905 is dynamically stable over Gyr-scales (Afruni et al. in prep.).

\subsection{Star formation at extreme low densities}

In Fig.~\ref{fig:mass_profs}, we show the stellar ($\Sigma_\ast$) and gas ($\Sigma_{\rm gas}$) mass surface density profiles of AGC~114905. The observed edge of the stellar discs corresponds well with the radius at which $\Sigma_{\rm gas}$ changes slope at about 7.5~kpc. At this radius $\Sigma_{\rm gas}/\Sigma_\ast \sim 30$. As it happens with other low surface brightness galaxies, the low stellar content of AGC~114905 might be related to the gas surface (and volume) density being too low (e.g. \citealt{kennicutt1989, thijs1993,schaye2004,ceci_dwarfs,kado-fong_2022_2}). To test this, we compute the (thickness-corrected) critical surface density (e.g. \citealt{schaye2004})
\begin{equation}
\Sigma_{\rm crit} = 1.5\, Q\, \Sigma_{\rm gas} = 1.5\, \dfrac{\kappa\, \sigma_{_{\rm HI}}}{\pi\, G}~.
\end{equation}

\noindent
Theoretical models predict that star formation should be inefficient whenever $\Sigma_{\rm gas} < \Sigma_{\rm crit}$ (e.g. \citealt{toomre1964,fall1980,zasov1988,kennicutt1989,hunter1998}). Fig.~\ref{fig:mass_profs} shows this is the case at all radii for AGC~114905, especially beyond the truncation of the mass profiles. At the truncation or break radius, $\Sigma_{\rm crit}/\Sigma_{\rm gas} \approx 0.5$, in agreement with observations in nearby spiral galaxies \citep{thijs1993,martin2001,leroy} and with the expectations of models where the star formation threshold in galaxies depends on the thermal instabilities originated from the transition between the warm and cold gas phases \citep{schaye2004}. 

From all of the above, the very low star formation levels observed in AGC~114905 ($\mathrm{SFR}\approx 10^{-2}~M_\odot\,\mathrm{yr}^{-1}$, $\Sigma_{\rm SFR} \approx 1.3\times 10^{-4}~M_\odot\,\mathrm{yr}^{-1}\,\mathrm{kpc}^{-2}$, see \citealt{durbala2020,zhai2024}) seem in line with the observed low gas surface densities (and even lower volume densities), which likely prevent the formation of molecular gas \citep{wang2020_udgs,kado-fong_2022_2}. A more in-depth study of AGC~114905 and other gas-rich UDGs incorporating measured gas and star-formation volume densities (e.g. \citealt{ceci_dwarfs}) would be very insightful to advance our understanding of star formation at very low densities.


\section{Conclusions}
\label{sec:conclusions}

We have conducted ultra-deep optical multi-band ($g$, $r$, $i$) imaging of the gas-rich ultra-diffuse galaxy (UDG) AGC~114905 using the Gran Telescopio Canarias (GTC). This is a puzzling galaxy, as previous studies have suggested it is in tension with the cold dark matter (CDM) model. Yet, such claims relied strongly on the apparent inclination of the gas disc, which was uncertain, making it imperative to obtain an independent measurement. We have assessed this situation, and we summarise our main findings here.

Our imaging, to the best of our knowledge the deepest yet obtained from ground observations (reaching $\mu_{\rm r,lim} \approx 32$~mag/arcsec$^2$ when measuring $3\sigma$ fluctuations in the background of the images in areas equivalent to 10 arcsec $\times$ 10 arcsec boxes), allowed us to characterise the stellar disc of the galaxy with unprecedented detail (Fig.~\ref{fig:images}). AGC~114905 has a startling truncated stellar disc with spiral-arms-like features and an extent and morphology similar to the H\,{\sc i} disc (Figs.~\ref{fig:images}, \ref{fig:sb_profs}, and \ref{fig:mass_profs}). Using isophotal fitting and the Hausdorff distance, we measured an inclination of $i = 31\pm2^\circ$ (Fig.~\ref{fig:isophotes}), in agreement with previous determinations.

We obtained a new kinematic disc model of the H\,{\sc i} emission. We found a rotation curve that reaches its flat part at about 25~km/s and a gas velocity dispersion profile that decreases with radius and has an average value of 9~km/s (Fig.~\ref{fig:kinematics}).

We tested whether the baryons alone (i.e. without dark matter) could reproduce the observed kinematics, finding that their gravitational contribution is not enough to match the data (Fig.~\ref{fig:noDM}). Additionally, we explored whether AGC~114905 conforms to the expectations of Modified Newtonian Dynamics (MOND), finding that this is not the case (Fig.~\ref{fig:mond}).

We estimated the dark matter distribution in AGC~114905 through rotation curve decomposition. We fitted the data with the framework of CDM (Fig.~\ref{fig:coreNFW}), self-interacting dark matter (SIDM, Fig.~\ref{fig:sidm}), and fuzzy dark matter (FDM, Fig.~\ref{fig:fuzzy}). At face value, all the models fit the kinematics of the galaxy, and all of them have low concentration parameters (or in general, a halo with a circular velocity peaking at large distances, i.e. a large $R_{\rm max}$ at fixed $V_{\rm max}$) and high baryon fractions, departing from classical CDM expectations (Fig.~\ref{fig:allhaloes}).
However, the CDM case requires halo structural parameters that are very rarely observed in cosmological simulations. In the case of SIDM, the halo of AGC~114905 is relatively rare, but similar systems appear to be produced in SIDM cosmological simulations. Within FDM, the galaxy is not as extreme as in CDM or SIDM, but it is unclear if FDM can, in general, reproduce the abundance of other galaxy populations. 

Finally, we obtained some simple diagnostics to study the equilibrium of the galaxy and its star formation levels. Our findings suggest AGC~114905 is stable against local (Fig.~\ref{fig:stability}) and global instabilities. We found evidence that the gas density of the galaxy is below the theoretical critical density for star formation, consistent with the low stellar content and current star formation level of the galaxy (Fig.~\ref{fig:mass_profs}).\\

\noindent
Our findings reinforce the idea that the puzzling dynamical properties observed in some gas-rich UDGs, of which AGC~114905 appears to be one of the most extreme cases, are due to atypical dark matter distributions. We expect that as data from deep and large optical and H\,{\sc i} surveys become available, more and more systems like these should be found. Our analysis suggests that the dynamics of this galaxy might be easier to explain in the SIDM or FDM framework than in CDM or MOND. Overall, our study highlights how AGC~114905 and akin galaxies can help us test the extreme conditions at which galaxy formation can occur and obtain new clues on the nature of dark matter. \\

\begin{acknowledgements}
We thank the referee for their valuable feedback, which helped us improve our manuscript. We thank Andrés Bañares-Hernández, Hai-Bo Yu, James Binney, Tom Oosterloo, Cecilia Bacchini, Filippo Fraternali, Alessandro Romeo, Ethan Nadler, Gabriele Pezzulli, Andrea Afruni, Betsey Adams, Justin Read, Jorge Sánchez Almeida, Arianna di Cintio and Chris Brook, for helpful discussions during the development of this work. 

PEMP acknowledges the support from the Dutch Research Council (NWO) through the Veni grant VI.Veni.222.364. PEMP would like to thank the Instituto de Astrofísica de Canarias for the support within its Early-career Visitor Programme. 

IT acknowledges support from the ACIISI, Consejer\'{i}a de Econom\'{i}a, Conocimiento y Empleo del Gobierno de Canarias and the European Regional Development Fund (ERDF) under grant with reference PROID2021010044 and from the State Research Agency (AEI-MCINN) of the Spanish Ministry of Science and Innovation under the grant PID2022-140869NB-I00 and IAC project P/302302, financed by the Ministry of Science and Innovation, through the State Budget and by the Canary Islands Department of Economy, Knowledge and Employment, through the Regional Budget of the Autonomous Community.
MM acknowledges support from the Project PCI2021-122072-2B, financed by MICIN/AEI/10.13039/501100011033, and the European Union “NextGenerationEU”/RTRP. Co-funded by the European Union. Views and opinions expressed are however those of the author(s) only and do not necessarily reflect those of the European Union. Neither the European Union nor the granting authority can be held responsible for them. 
This work is based on observations made with the Gran Telescopio Canarias (GTC), installed at the Spanish Observatorio del Roque de los Muchachos of the Instituto de Astrofísica de Canarias, on the island of La Palma. We also use observations made with the Karl G. Jansky Very Large Array (VLA) of the National Radio Astronomy Observatory (NRAO). NRAO is a facility of the National Science Foundation operated under cooperative agreement by Associated Universities, Inc.

We have used SIMBAD, NED, and ADS services extensively, as well the Python packages NumPy \citep{numpy}, Matplotlib \citep{matplotlib}, SciPy \citep{scipy}, spectral$\_$cube \citep{spectral_cube}, and Astropy \citep{astropy}, for which we are thankful.
\end{acknowledgements}

%
   \bibliographystyle{aa.bst} 
   \bibliography{references,more-references} 
%

\begin{appendix}

\section{Observational strategy and data reduction}
\label{appendix:data_reduction}

Our deep imaging of AGC~114905 was obtained using the instrument OSIRIS and its upgraded version (OSIRIS+) at the Gran Telescopio Canarias (GTC). The field of view of OSIRIS is 7.8~arcmin by 8.5~arcmin (7.8~arcmin by 7.8~arcmin without vignetting). The camera comprised two contiguous CCD (one in the new version) sensors, with a 9.4~arcsec gap between them. The camera pixel scale is 0.254 arcsec/pixel.

We observed the galaxy using the $g$, $r$, and $i$ Sloan filters during seven separate nights in October and November 2022.
Furthermore, we used the OSIRIS+ configuration to retrieve more data in the $g$ Sloan filter on December 12th, 2022. The OSIRIS+ setup represents a comprehensive upgrade to the OSIRIS instrument, which was implemented in 2022. This upgrade includes installing OSIRIS at the Cassegrain focal station and using a new blue-sensitive monolithic detector.
In Table~\ref{tab:pernight} and Table~\ref{tab:entire}, we summarise how the data was retrieved during the different nights of observation, total exposure time per filter, mean seeing and surface brightness limits (3$\sigma$; $10\arcsec \times 10\arcsec$ boxes).

\begin{table}[h]
    \centering
    \caption{Dates, Sloan bands, exposure time per frame, and number of frames taken every night during our observing run. }
    \begin{tabular}{cccc}
         \hline  \noalign{\smallskip}
          Date & Band & $t_{\rm exp,obj}$ & $n_{\rm frames,obj}$  \\
          \noalign{\smallskip}
         \hline  \noalign{\smallskip}
         Oct 23th & $i$ & 180s & 32   \\  \noalign{\smallskip}
         Oct 30th & $g$ & 180s & 16   \\  \noalign{\smallskip}
         Oct 31st & $r$ & 180s & 30   \\  \noalign{\smallskip}
         Nov 1st  & $g$ & 180s & 16   \\  \noalign{\smallskip}
         Nov 25th & $g,r$ & 180s & 16,32 \\  \noalign{\smallskip}
         Nov 28th & $i$ & 180s & 16    \\  \noalign{\smallskip}
         Nov 30th & $i$ & 180s & 26    \\  \noalign{\smallskip}
         Dec 12th & $g$ & 180s & 16    \\  \noalign{\smallskip}
         \hline
    \end{tabular}
    \label{tab:pernight}
\end{table}

\begin{table}
    \centering
    \caption{Details of our observing run. The table lists the total exposure time per filter, the surface brightness limit of the data (3$\sigma$ fluctuations of the background in areas equivalent to boxes of $10$ arcsec $\times$ $10$ arcsec), and the mean seeing.}
    \begin{tabular}{cccc}
         \hline \noalign{\smallskip}
          Band & t$_{\rm exposure}$ & $\mu_{\rm limit}$ [mag/arcsec$^2$] &  Seeing [arcsec] \\ \noalign{\smallskip}
         \hline \noalign{\smallskip}
          $g$ & 3h12min & $\gsblim{}$  & 1.12 \\ \noalign{\smallskip}
          $r$ & 3h06min & $\rsblim{}$  & 1.06\\ \noalign{\smallskip}
          $i$ & 3h45min & $\isblim{}$ & 1.24  \\ \noalign{\smallskip}
        
         \hline
    \end{tabular}
    \label{tab:entire}
\end{table}


\subsection{Dithering pattern}

To achieve our surface brightness level goals (i.e. $\mu_{g}\sim$ 31.0 mag/arcsec$^2$, $3\sigma$; 10$\arcsec$ $\times$ 10$\arcsec$) in our images, we follow an observational strategy similar to that used in previous works using GTC \citep[see, e.g.][]{2016ApJ...823..123T} and other large telescopes such as LBT \citep{LIGHTSs} and Gemini \citep{golini2024}. To accomplish our objectives, we must deal with several observational biases that can affect deep observations, such as scattered light, saturation or ghosts. To do so, we need an accurate estimation of the flat-field correction and careful treatment of the background subtraction.

The twilight flats provided by the observatory are not good enough for our purposes because the variations between the night sky and the twilight illumination result in subtle flat-fielding differences. Therefore, we use the actual science exposures observed with a dithering pattern with large displacements to build our flatfield images. The size of the dithering step has to be similar to the size of the galaxy, so we apply a dithering scheme with 30~arcsec displacement and individual exposure times of 180 seconds. Following such a dithering pattern, the galaxy will be placed at different positions in the CCD, enabling a better determination of the sky background and creating a flat-field mosaic of about $10 \times 12$~arcmin$^2$.

\subsection{Data reduction}
The data were downloaded from the GTC Archive \footnote{\url{https://gtc.sdc.cab.inta-csic.es/gtc/index.jsp} } and reduced using a similar procedure as \citet{LIGHTSs}. Here, we delve into the main steps of the process.
Different pipeline steps rely on the Gnuastro Software \citep{gnuastro} and its tasks, as described below. We also employ publicly available data from the Sloan Digital Sky Survey (SDSS) DR14 \citep{2018sdss} for the photometric calibration of GTC images in $i$ filter, and the Dark Energy Camera Legacy Survey (DECaLs, DR10, brick 0212p072, \citealt{dey2019}) for the $g$ and $r$ filters.


First, for each CCD (two in OSIRIS configuration and one in OSIRIS+) and each frame, we mask pixels with values of 0 that correspond to pixels that do not contribute to the signal due to the detector's readout failure.
Also, we mask all pixels with values greater than $5\times 10^4~\rm{ADUs}$ to account for the sensor's nonlinear response. After this, we produce bias-corrected images.
To do so, we create a masterbias (for each CCD) by combining the individual bias frames with a sigma clipping median using the Gnuastro task \texttt{Arithmetic}. This masterbias is later subtracted from all science images.


\subsubsection{Flat field}

A masterflat is created for each night of observation for each filter separately using the science (bias-corrected) images. This is performed in two steps. First, the science images are normalised as follows. 
The pixel (1052, 1018) of the CCD1 of the OSIRIS camera is used as the centre of a normalisation ring. That ring has an inner radius of 600 pixels and a width of 200 pixels. That ring crosses the two CCDs as it is chosen to cover a region with a similar illumination.
For each CCD and within the corresponding ring region, we calculate the 3$\sigma$-clipped median value of the pixels. 
That value is used to normalise the flux of the CCD for each science image. Then, we combine all the normalised and bias-subtracted science images using a sigma-clipping median stacking. In this way, we get two preliminary masterflats, one per CCD. Science images are then divided by these initial masterflats, which allows us to distinguish the sources in the science images. To further improve our masterflat, we mask all the detected sources using \texttt{NoiseChisel} \citep{noisechisel_segment_2019}, also part of Gnuastro,
and we median-combine all the re-normalised and masked images once again to create a final masterflat (one per CCD). Finally, the individual science images of each filter are divided by their corresponding CCD masterflats. To avoid vignetting problems towards the corner of the detectors, we remove from the images all the pixels where the masterflats have an illumination lower than 95$\%$ of the central pixels.

\subsubsection{Astrometry, photometry, and background determination}

We conduct the following steps to determine the astrometry of our individual science images. We calculate an astrometric solution using Astrometry.net (v0.85; \citealt{Astrometrynet}). We use Gaia eDR3 \citep{Gaiaedr3} as our astrometric reference catalogue. This produces a first astrometric solution, that we improve using SCAMP (v.2.10.0; \citealt{Scamp}). SCAMP reads catalogues that we generate using SExtractor (v.2.25.2; \citealt{Sextractor}) and calculates the distortion coefficient of the images. After this step, we run SWarp \citep{swarp2010} on each individual image to put them into a common grid of 4051$\times$4051 pixels.


Once the astrometry is ready, we subtract the background of the individual CCD exposures by masking the signal of each image using NoiseChisel. For each CCD, we calculate the median of the background image produced as an output by NoiseChisel and remove that value from the image.


Following the background subtraction, we proceed to convert the GTC counts (ADUs) into nanomaggies\footnote{see \url{https://www.sdss3.org/dr8/algorithms/magnitudes.php}}.
To do this, we use SDSS data (for the $i$ band calibration) and DECaLS data (for $g$ and $r$ filters) covering the same field of view around our target. For both the GTC images and the SDSS and DECaLS mosaic, we store a catalogue of the flux of the point-like (PL) sources in the field using circular apertures with a radius equal to the Full Width Half Maximum (FWHM) of the image with Gnuastro \texttt{astmkcatalog}. 
We store only the parameters of the stars that are not saturated in the GTC data (i.e. we use only PL sources with magnitudes $g > 21~\rm{mag}$, $r > 20.5 ~\rm{mag}$ and $i > 19.5~\rm{mag}$) and fainter than $g = 23~\rm{mag}$, $r = 22 ~\rm{mag}$ and $i = 20.5~\rm{mag}$ to minimise errors in the estimation of the flux of the sources. For the selected stars, we compute the flux ratio in the DECaLS (or SDSS) circular aperture to the flux in the GTC aperture for each science exposure. Then, we multiply the GTC image by the resistant median of all these ratios. At the end of the process, we have the pixel values of each pointing in units of nanomaggies.
The error associated with the zero point estimate is $\sim$1$\%$.

\subsubsection{Image coaddition and final stack}
\begin{figure*}
    \centering
    \includegraphics[width=1\textwidth]{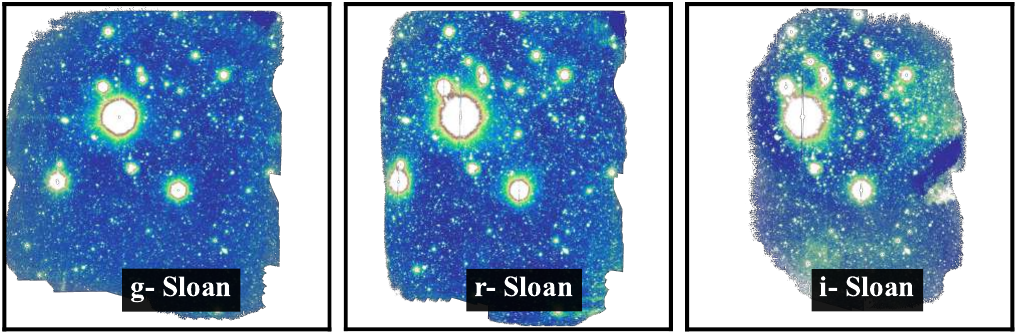}
    \caption{Reduced final mosaics of GTC observations used in this work. A high-resolution version of this figure is available at \href{https://pavel-mancera-pina.github.io/data_to_share/ultradeepA114905_filters.pdf}{this link}.}
    \label{fig:filters}
\end{figure*}

In the final step of data reduction, individual frames are combined to create a final mosaic image. Not all frames are combined with the same weight. This is because the observing conditions change during the night. Several factors, such as the air mass (which varies depending on the position of the target in the sky) and meteorological events such as passing clouds, can affect the sky brightness, degrading its quality and producing larger standard deviations in the sky background pixel values, indicating poorer image quality. 

 The weight assigned to each frame is determined by the ratio of the standard deviation of the background in the best exposure to the standard deviation of the background in the i$^{th}$ frame. Before stacking the data in this way, it is crucial to mask out pixels affected by unwanted signals, such as cosmic rays. These pixels are identified using the following procedure. First, we use \texttt{Swarp} \citep{swarp2010} to put the images into the same astrometric solution. Then, we create a mask to reject the pixels with a value that deviates from the resistant mean (3$\sigma$ rejection is sufficient for our purposes) of all pixels at the same sky location in all frames. We then mask the outliers and stack the data using this sigma-clipped weighted mean. 


The co-added image is significantly deeper than any individual image, and therefore, many very low surface brightness features, previously invisible, emerge from the noise. These features subtly affect the background determination of our individual science images, and it is thus necessary to mask these regions and repeat the entire sky determination process on the individual exposures with an improved mask. In short, we repeat the background estimation (and subsequent reduction steps) described above on the individual images using the improved masks generated by this first data co-addition.  The mosaics for our three photometric bands are derived using the same data reduction pipeline. The final mosaics have a field of view of about $10~\rm{arcmin} \times 12~\rm{arcmin}$ and are shown separately in Fig.~\ref{fig:filters}.

\subsubsection{Scattered light removal}
\label{sec:scatteredlight}
\begin{figure*}
    \centering
    \includegraphics[width=1\textwidth]{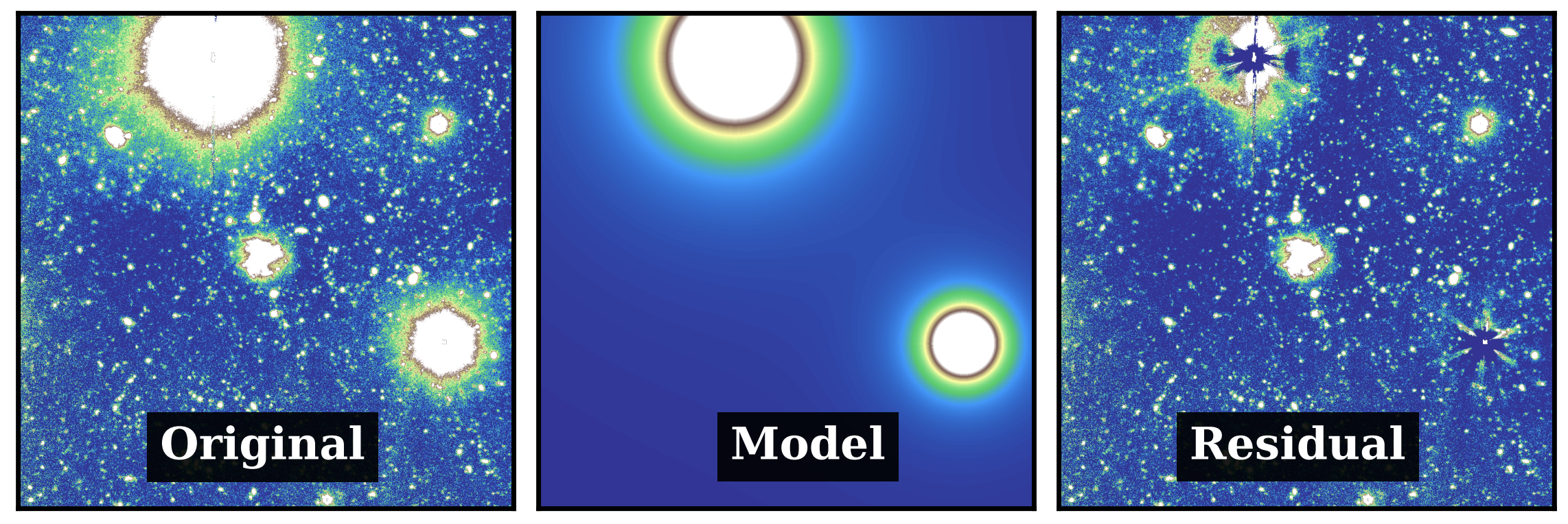}
    \caption{Removal of scattered light contamination produced by bright stars near AGC~114905. The panels illustrate the procedure followed in the $g-$band data. The left and middle panels show the original stacked image and the model of the scattered light, respectively. Our final science images are the result of subtracting the model from the original image, as displayed in the right panel. The field of view of the region is $5~\rm{arcmin} \times 5~\rm{arcmin}$.}
    \label{fig:stcatteredlight}
\end{figure*}

One of the main difficulties in obtaining a high-quality image for low surface brightness science in the field around AGC~114905 is the presence of two bright stars close to the galaxy: TYC 27-1301-1 (01h25m20.7s, +07d23m45.6s) and 2MASS J01251081+0720423 (01h25m10.8s, +07d20m42.6s). Therefore, we must develop a dedicated strategy to remove the light gradient produced by the scattered light of such bright sources on our co-added image.
Fig.~\ref{fig:stcatteredlight} illustrates the three subtraction stages of the scattered light emitted by the stars near AGC~114905.
To remove the light gradient produced by these stars, we use the following strategy.
First, we mask all neighbouring major bright sources in the final mosaic with \texttt{Noisechisel} and \texttt{Segment} around each star.
Second, we obtain the radial profile of the scattered light produced by the stars using circular apertures centred on them (i.e. RA = 21.3362 deg, DEC = 7.3955 deg for the brightest star). 
Such a profile is fitted with a power-law $L(R) = \beta\,  R^{-\alpha}$.
We fit the profile at the radial distance between 0.3' and 1'.
For both stars, we find that the slope of the light due to the scattered light
is well described by a power-law with $\alpha = -2.9$ for the $g$ band, $\alpha = -2.55$ for $r$, and $\alpha = -2.85$ for $i$.
We then convert the power-law radial profile into a 2D circular model centred on the star position. 
We sum the contribution of the two scattered light fields into a single image (middle panel of Fig. \ref{fig:stcatteredlight}) and finally subtract it from the original image. The right panel of Fig. \ref{fig:stcatteredlight} shows the GTC data after removing the scattered light.

\section{Modified$^2$ Hausdorff Distance}
\label{appendix:MMHD}

In this work, we used a two-pronged approach to constrain the position angle and inclination of AGC~114905. Specifically, we used a traditional isophotal analysis and an independent method based on the Modified Hausdorff Distance \citep[MHD,][]{mhd}, following \citet{montes2019}. In this Appendix, we explain how this second method works. 

The MHD measures how far two subsets (in our case contours) are from each other \citep{mhd}. Given two shapes or contours, the MHD quantifies the mean distance between the points of the two datasets. However, a mean value can be biased by outliers or extreme values. Therefore, we decided to modify the MHD to implement a 3$\sigma$-clipped mean (using Astropy's \texttt{sigma_clipped_stats}) of the distribution of distances between the two datasets. We called this updated method M$^2$HD, standing for modified MHD.

\begin{figure*}
    \centering
    \includegraphics[width=\linewidth]{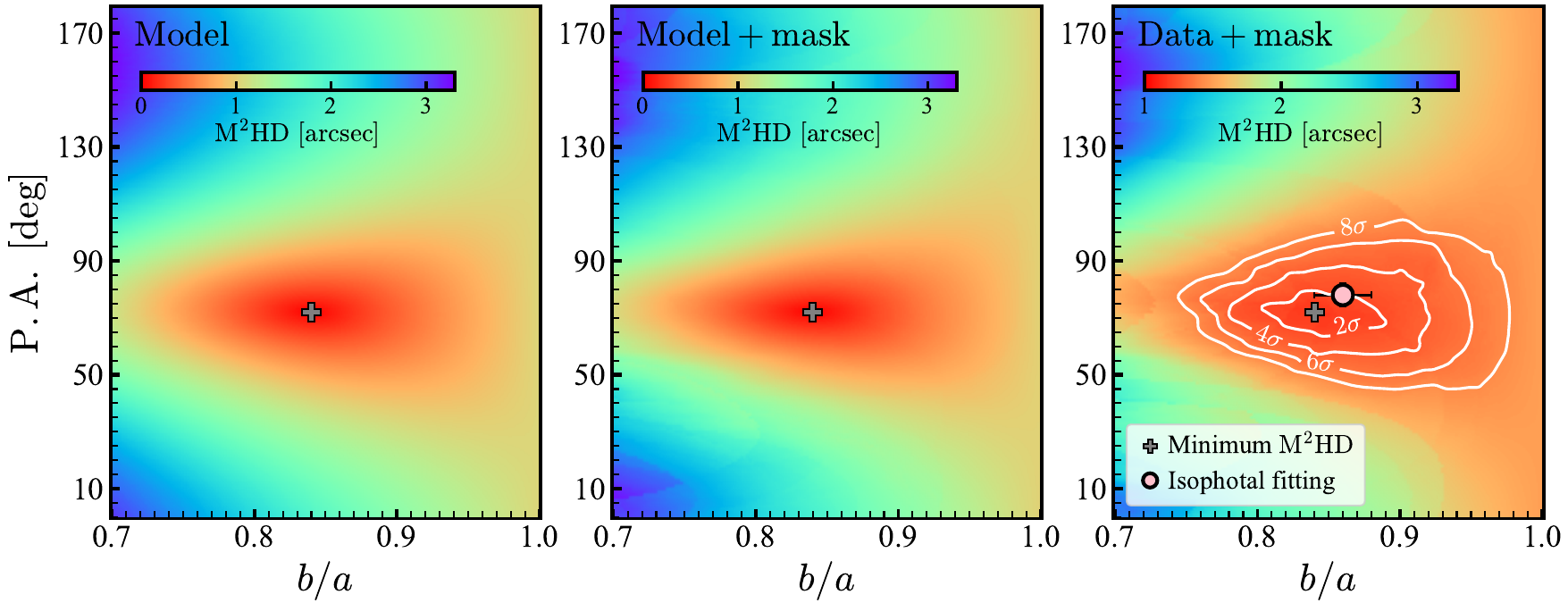}
    \caption{Modified MHD (M$^2$HD) maps. The \emph{left} panel presents the idealised M$^2$HD when comparing a set of elliptical isophotes against a particular ellipse with $\mathrm{P.A.} = 72^\circ$ and $b/a = 0.84$. The \emph{middle} panel shows the same comparison but using the mask applied in this paper, which introduces some uncertainty. The \emph{right} panel instead shows the comparison between the set of ellipses and the real (masked) isophote at the edge of the galaxy. All the panels compare isophotes at 19 arcsec, i.e. right before the location of the edge radius of  AGC~114905 at about 20 arcsec. In all the panels, the grey cross represents the minimum M$^2$HD value. For the data, the global minimum of the map (white cross) is consistent within $2\sigma$ with the estimates of $\mathrm{P.A.}$ and $b/a$ derived through isophotal fitting (see Sec.~\ref{sec:optical}). Note that the first two panels have the same colorbar, but the last panel shows a narrower range to better appreciate the structure of the map.}
    \label{fig:mhd_comp}
\end{figure*}

We are interested in deriving the galaxy's position angle and axis ratio close to the truncation radius, which is a good proxy for the disc's edge and representative of its shape. For that, we compare the shape of an isophote immediately before the galaxy edge to a large set of ellipses of different position angles ($\mathrm{P.A.} = 0^\circ-180^\circ$) and axis ratios ($b/a = 0.7-1$). Fig. \ref{fig:mhd_comp} shows the M$^2$HD map when comparing our set of ellipses with a particular ellipse shape (left), the same ellipse shape with the same mask applied (middle), and the actual isophote of the galaxy (right). The left panel uses an ellipse with $\mathrm{P.A.} = 72^\circ$ and $b/a = 0.84$ as the basis for the comparison to show how the map is structured in an ideal (i.e. perfect elliptical isophote) case. The middle panel shows the effect of adding a mask (the mask used is described in Sec.~\ref{sec:optical}) to the elliptical isophote: the M$^2$HD values present a slight increase and some substructure due to the mask removing some of the points of the isophotes, increasing the uncertainty. Finally, in the right panel, we show the map comparing pure elliptical shapes with the real isophote of AGC~114905 at 19~arcsec, just before the edge of the disc. In this case, we see that the M$^2$HD values increase with respect to the previous idealised maps due to the deviation of the shape of the real isophote of the galaxy compared to perfect elliptical shapes. The minimum of the map (white marker) is $\rm{M^2HD} = 1.15$, at $\mathrm{P.A.} = 72^\circ$, $b/a = 0.84$. The associated uncertainty (see also Fig.~\ref{fig:isophotes}) is the standard deviation from 1000 jackknife realisations, taking half of the M$^2$HD values and computing the M$^2$HD for each subsample. The pink marker shows the values calculated in Sec.~\ref{sec:optical} from the isophotal fitting, consistent within $2\sigma$ (dashed contour) with the values derived here.

\section{CDM mass models using a double power-law halo}
\label{appendix:dpl}

In the main text, we used the \textsc{coreNFW} profile (Eq.~\ref{eq:corenfw}) to obtain CDM mass models. In this Appendix, we show an analogous analysis under a different functional form, namely a double power law (DPL) of the form

\begin{equation}
    \rho_{\rm DPL}(r) = \rho_s \left(  \dfrac{r}{r_s} \right)^{-\gamma} \left( 
1 + \dfrac{r}{r_s} \right)^{\gamma - \beta}~,
\end{equation}
where $r_s$ is again a scale radius and $\gamma$ and $\beta$ are free parameters related to the inner and outer shape of the profile. The concentration parameter in this case is also given by $c_{200} = r_{200}/r_{-2}$, but $r_{\rm s} \neq r_{-2}$, with $r_{\rm s} = (\beta - 2)\, r_{-2} / (2-\gamma)$. Note that \emph{only} for $\gamma = 1$ and $\beta = 3$ the NFW profile is recovered  and $r_{\rm s} = r_{-2}$. This DPL profile has been shown to fit well the kinematics of some gas-rich ultra-diffuse galaxies \citep{dicintio2014,demao}.

\begin{figure}
    \centering
    \includegraphics[scale=0.6]{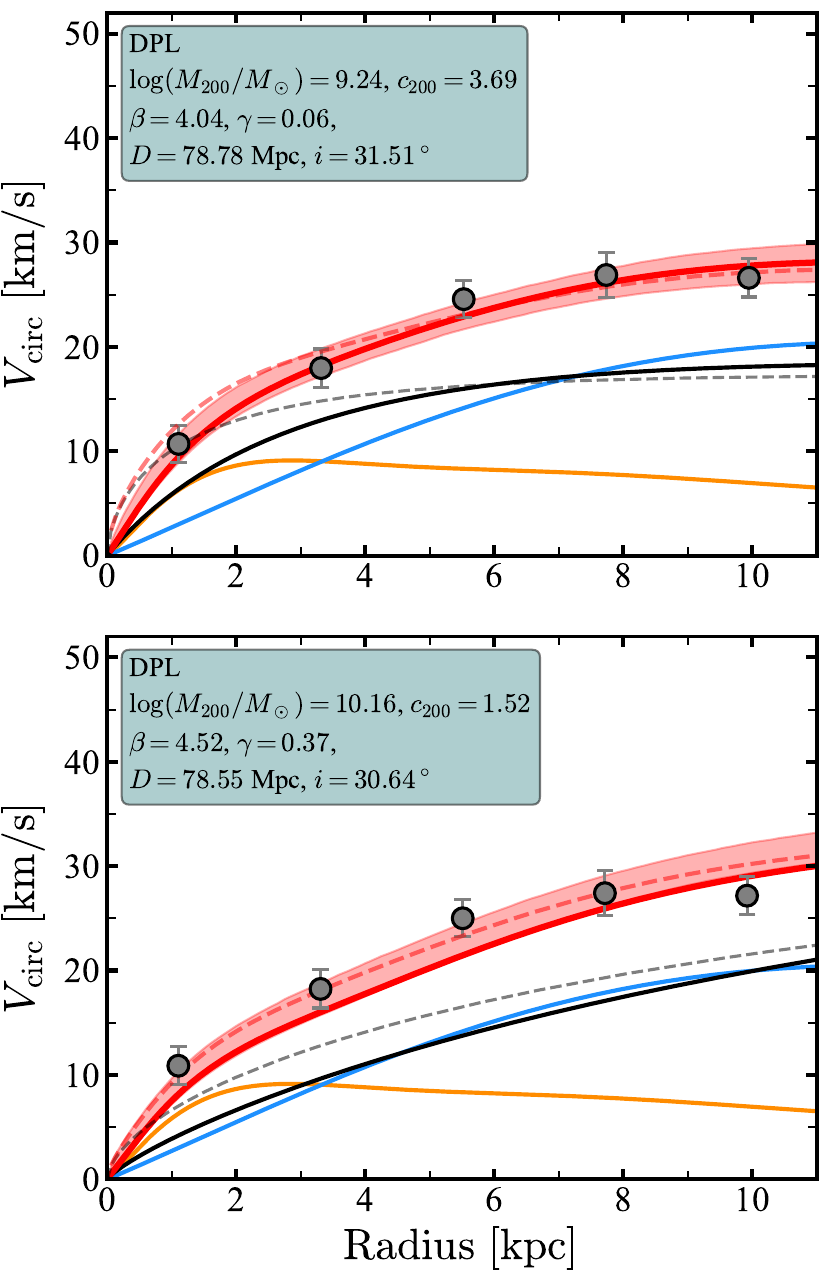}
    \caption{CDM mass models for a DPL halo. The top panel shows the results when $M_{200}$ is completely free, while the bottom panel imposes a lower bound in $M_{\rm 200,min}$. Curves are as in Fig.~\ref{fig:coreNFW}.}
    \label{fig:dpl}
\end{figure}

When doing our MCMC fitting, we consider $\log(M_{200})$, $\gamma$, $\beta$, $i$, and $D$ as free parameters. We adopted the same priors as for the \textsc{coreNFW} on $\log(M_{200})$, $\log(c_{200})$, $D$, and $i$, and the additional flat priors for $\gamma$ and $\beta$ in the ranges $\log(2) < \log(\beta) \leq \log(10)$ and $\log(0.001) \leq \log(\gamma) < \log(2)$. As in the main text, we explore two models, one where $M_{200}$ can be as low as $10^7~M_\odot$, and one where $M_{200}$ ensures a baryon fraction lower or equal to the cosmological baryon fraction. The DPL mass models can be seen in Fig.~\ref{fig:dpl}, and their posterior distributions can be found in the next section (Figs.~\ref{fig:post_cdm_dpl_nominM200} and \ref{fig:post_cdm_dpl_minM200}).
These mass models depict the same picture as the \textsc{coreNFW} models. If the limit on the cosmological baryon fraction is not imposed, the best-fitting halo has $\log(M_{200}/M_\odot) = 9.24^{+0.35}_{-0.29}$, and the galaxy would have a baryon fraction about 5 times higher than the cosmological average. When a lower limit on $M_{200}$ is imposed, we find $\log(M_{200}/M_\odot) = 10.16^{+0.39}_{-0.15}$, which results in a more moderate (but still high) $f_{\rm bar} \approx 0.6$.

Finally, we would like to emphasise that even when fitting DPL haloes, low concentration parameters (defined as $c_{200} = R_{200}/r_{-2}$) are needed. Since the definition of concentration can change depending on the functional form of the halo (see discussion in \citealt{demao}), it is also helpful to think of these haloes in terms of their $V_{\rm max}$ (maximum of $V_{\rm DM}$, a proxy of halo mass) and $R_{\rm max}$ (radius at which $V_{\rm DM}=V_{\rm max}$) values. As shown in Table~\ref{tab:v2kpc}, regardless of the exact functional form used to describe the haloes, at fixed halo mass AGC~114905 \citep[and other UDGs, see ][]{demao,brook2021} has a shallow and extended halo, different from those frequently seen in CDM-based simulations \citep{demao,nadler2023}.

\section{Posterior distribution of fiducial mass models}
\label{appendix:posteriors}

This Appendix provides the MCMC posterior distributions for our fiducial mass models. The main text discusses the main features of these plots.

\begin{figure}
    \centering
    \includegraphics[scale=0.5]{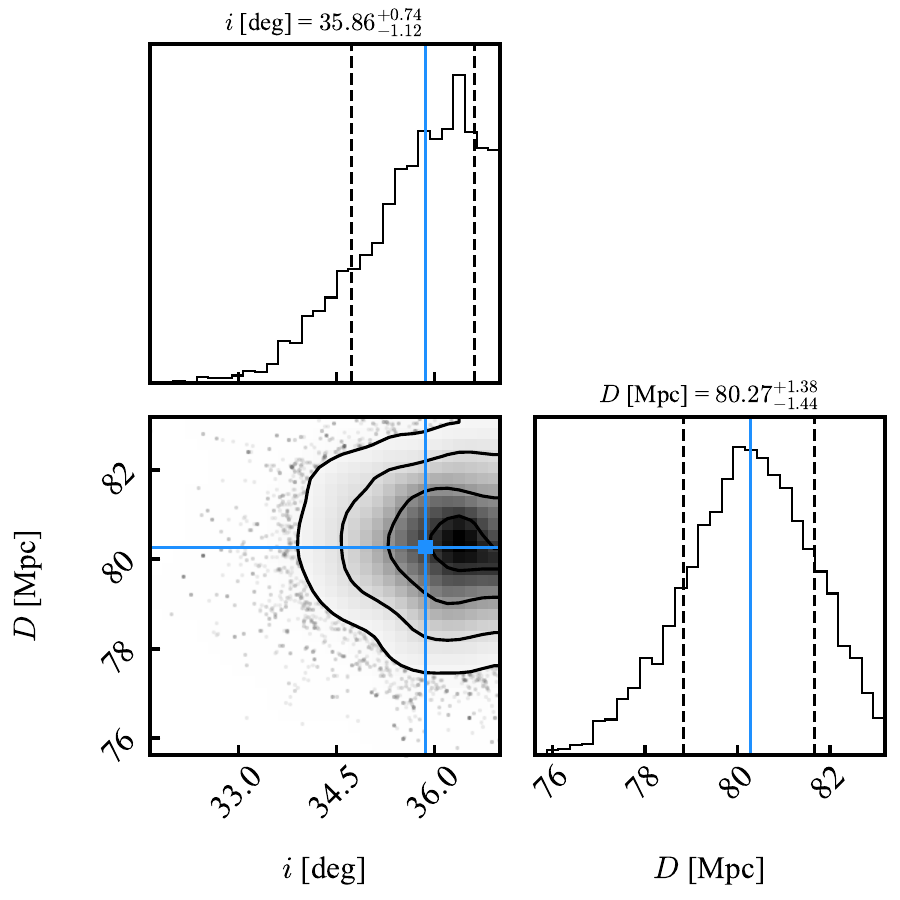}
    \caption{Posterior distribution of the mass model without dark matter shown in Fig.~\ref{fig:noDM}.}
    \label{fig:post_noDM}
\end{figure}

\begin{figure*}[h]
    \centering
    \includegraphics[scale=0.44]{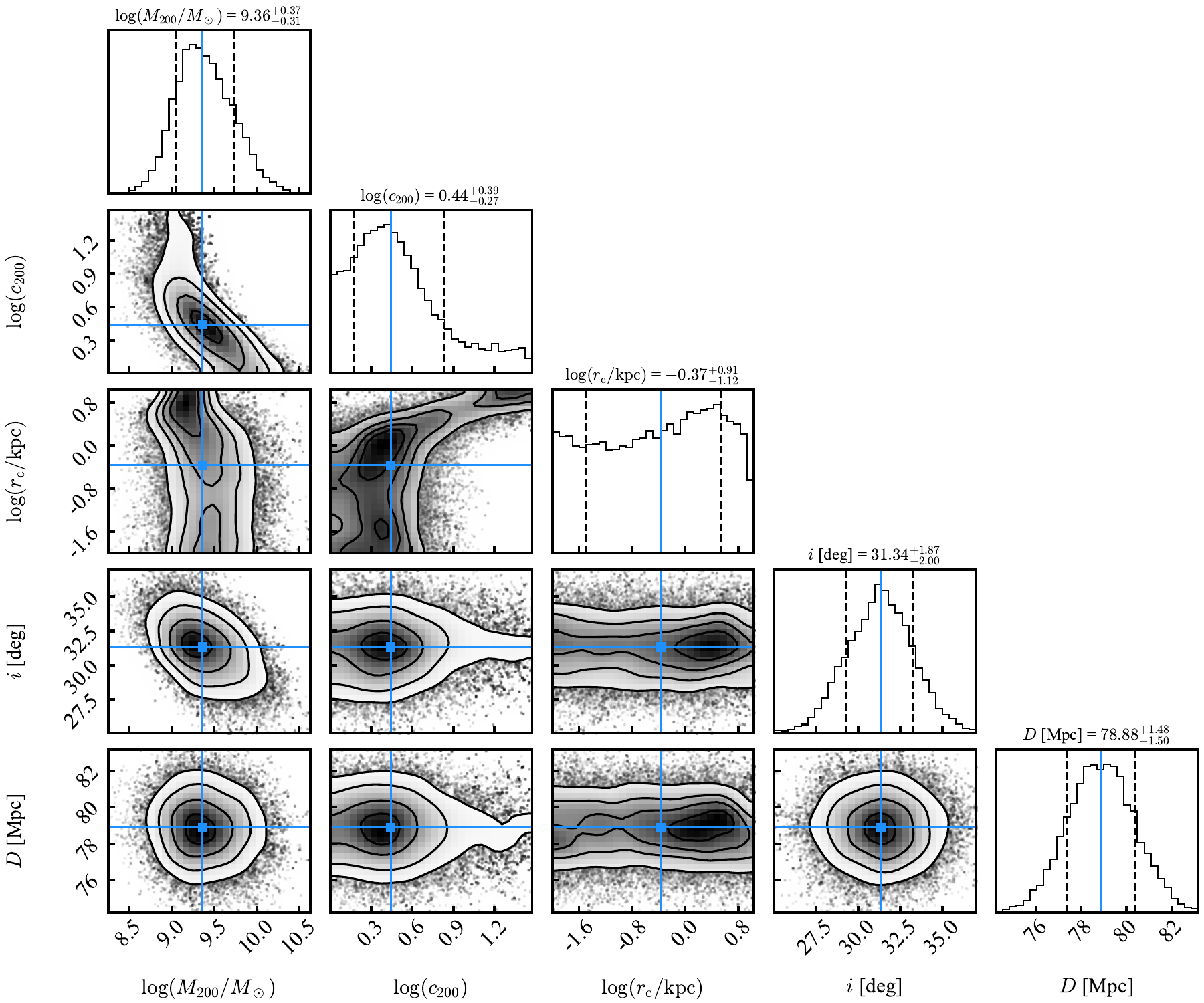}
    \caption{Posterior distributions of the \textsc{coreNFW} mass model shown in the top panel of Fig.~\ref{fig:coreNFW}. The $50^{\rm th}$ percentiles of the distributions, corresponding to our adopted values, are shown in blue. The $16^{\rm th}$ and $84^{\rm th}$percentiles, corresponding to our uncertainties, are shown with black dashed curves.}
\label{fig:post_cdm_corenfw_nominM200}
\end{figure*}

\begin{figure*}
    \centering
    \includegraphics[scale=0.44]{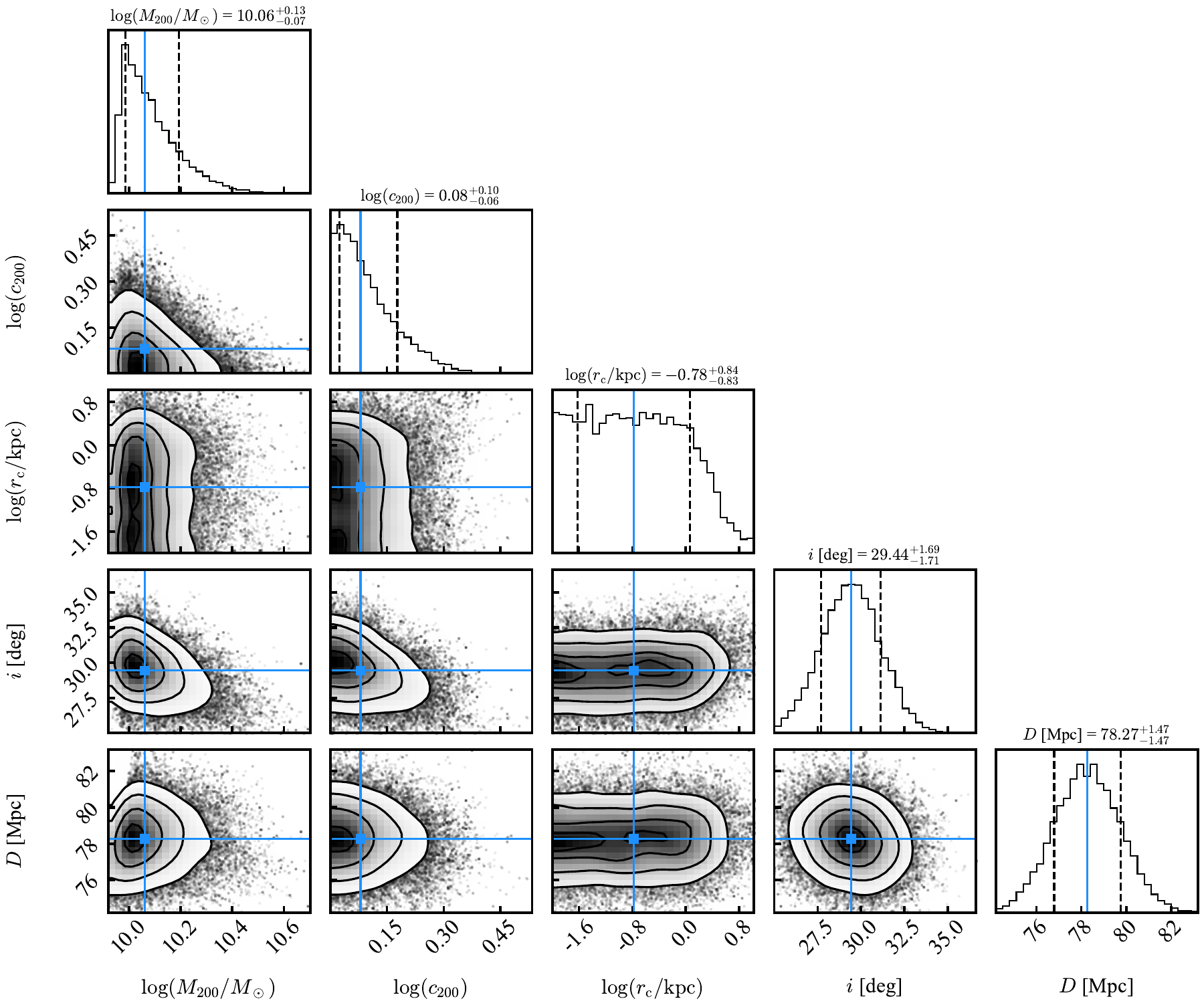}
    \caption{Posterior distributions of the \textsc{coreNFW} mass model shown in the bottom panel of Fig.~\ref{fig:coreNFW}. }
    \label{fig:post_cdm_corenfw_minM200}
\end{figure*}

\begin{figure*}
    \centering
    \includegraphics[scale=0.44]{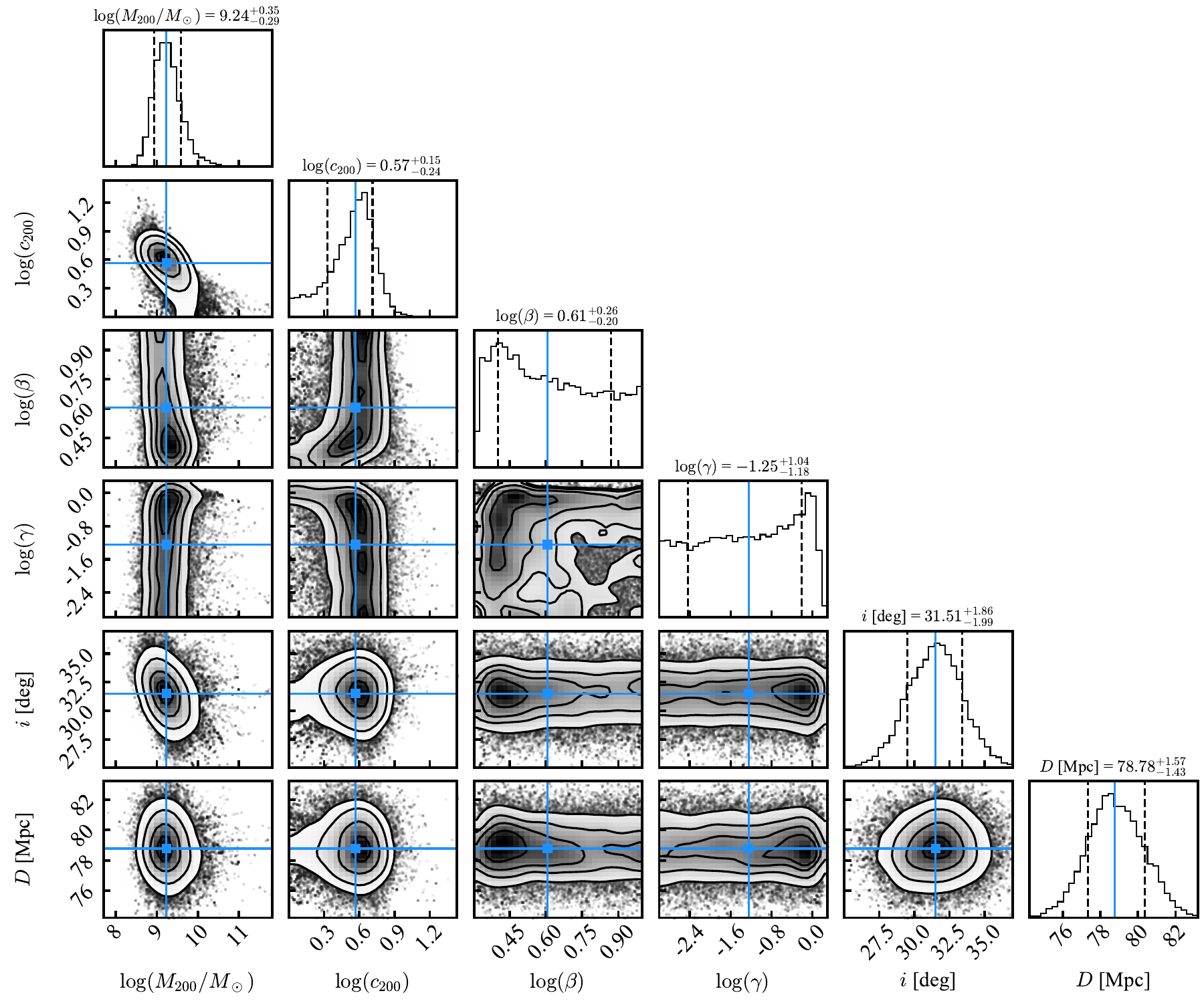}
    \caption{Posterior distributions of the DPL mass model shown in the top panel of Fig.~\ref{fig:dpl}.}
    \label{fig:post_cdm_dpl_nominM200}
\end{figure*}

\begin{figure*}
    \centering
    \includegraphics[scale=0.44]{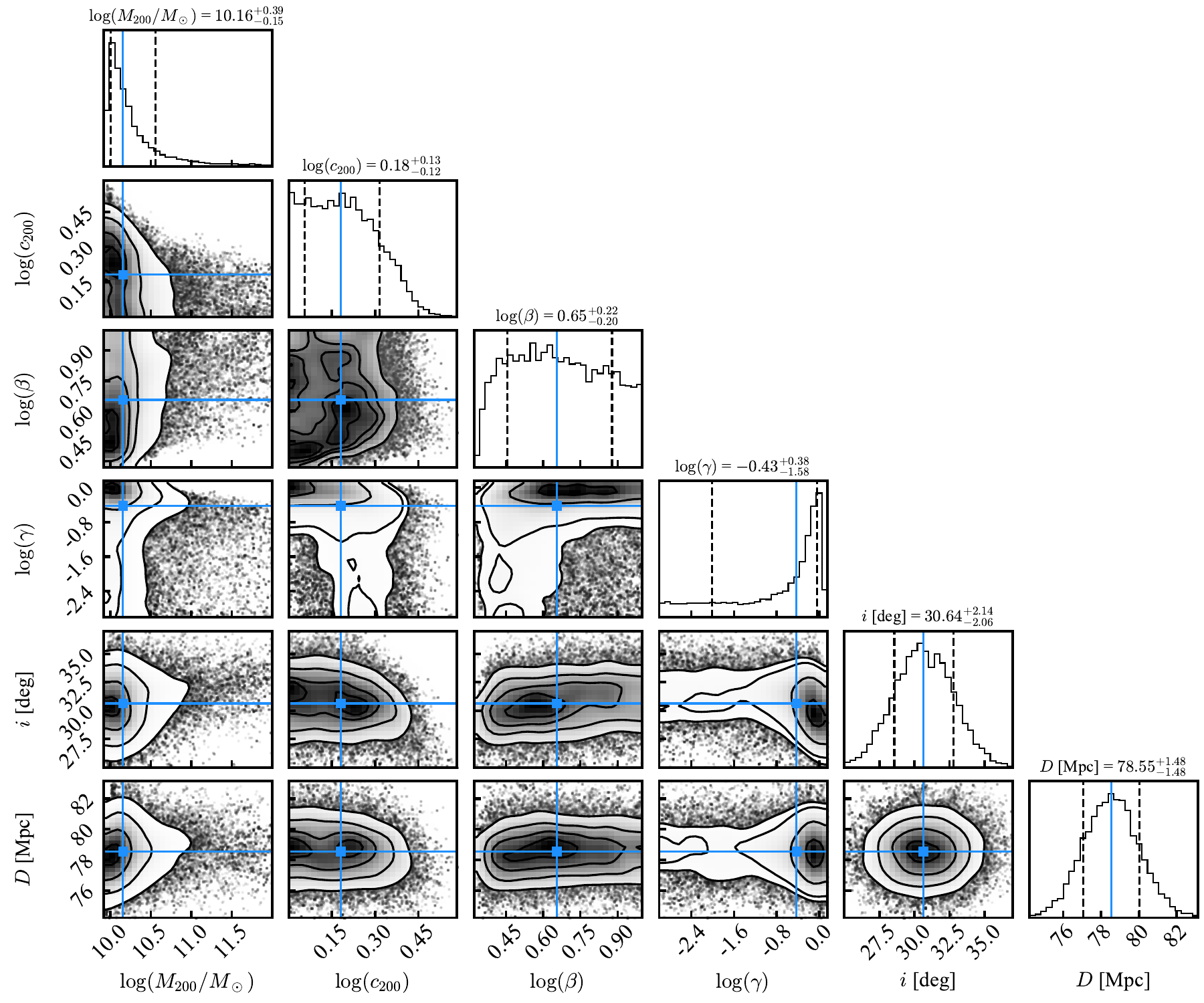}
    \caption{Posterior distributions of the DPL mass model shown in the bottom panel of Fig.~\ref{fig:dpl}. }
    \label{fig:post_cdm_dpl_minM200}
\end{figure*}

\begin{figure*}
    \centering
    \includegraphics[scale=0.44]{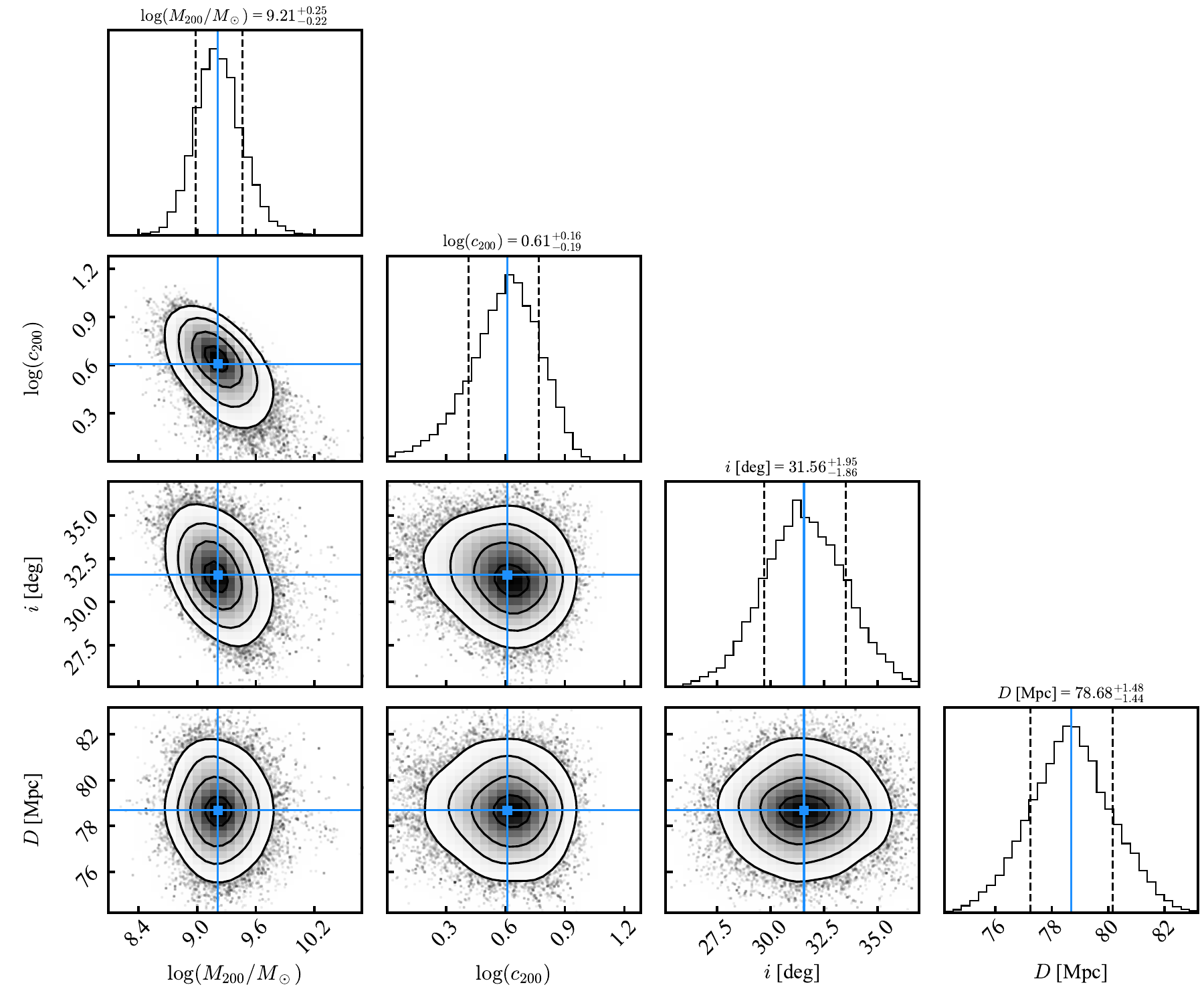}
    \caption{Posterior distributions of the SIDM mass model shown in the top panel of Fig.~\ref{fig:sidm}.}
    \label{fig:post_sidm_nominM200}
\end{figure*}

\begin{figure*}
    \centering
    \includegraphics[scale=0.44]{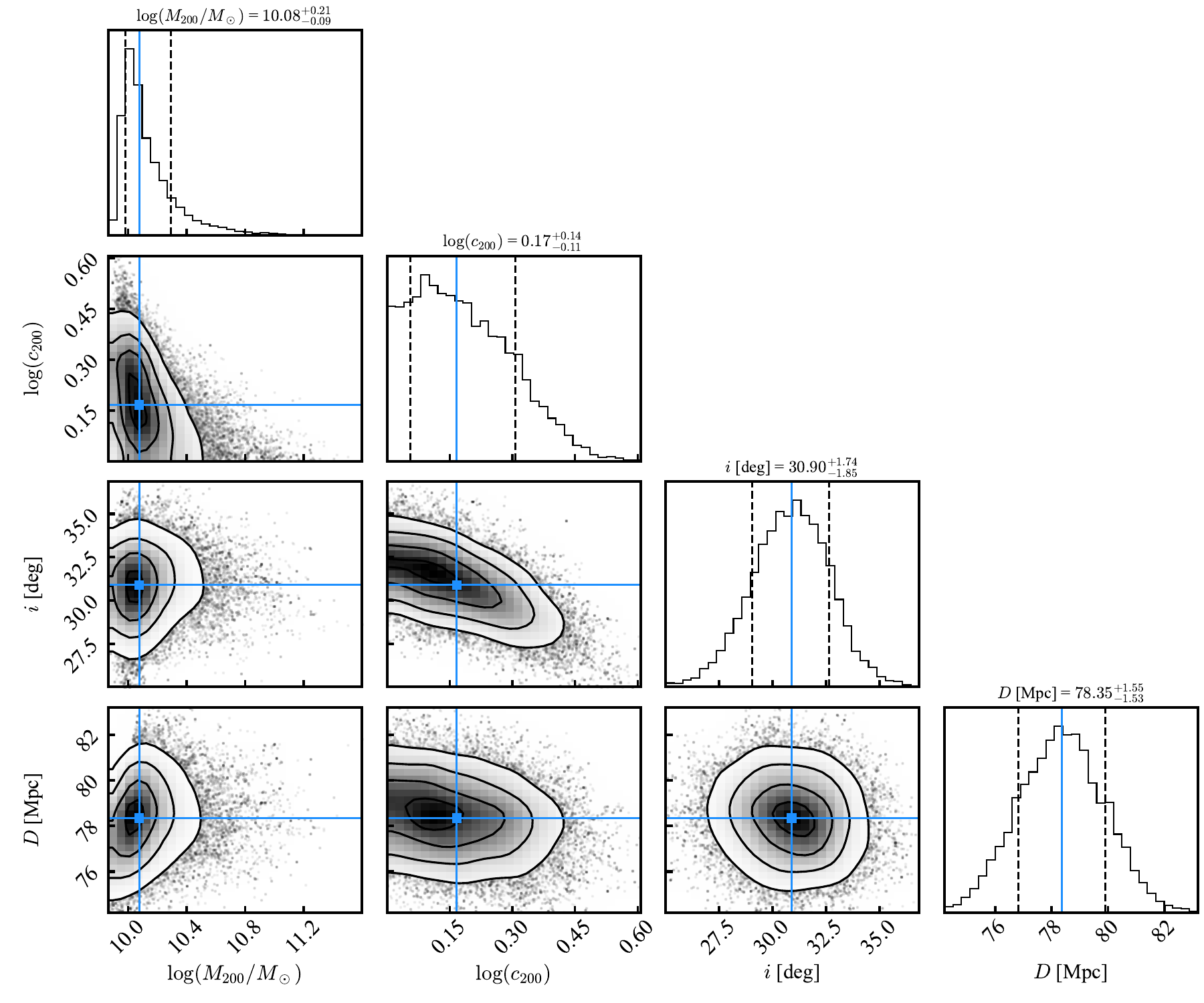}
    \caption{Posterior distributions of the SIDM mass model shown in the bottom panel of Fig.~\ref{fig:sidm}.}
    \label{fig:post_sidm_minM200}
\end{figure*}

\begin{figure*}
    \centering
    \includegraphics[scale=0.44]{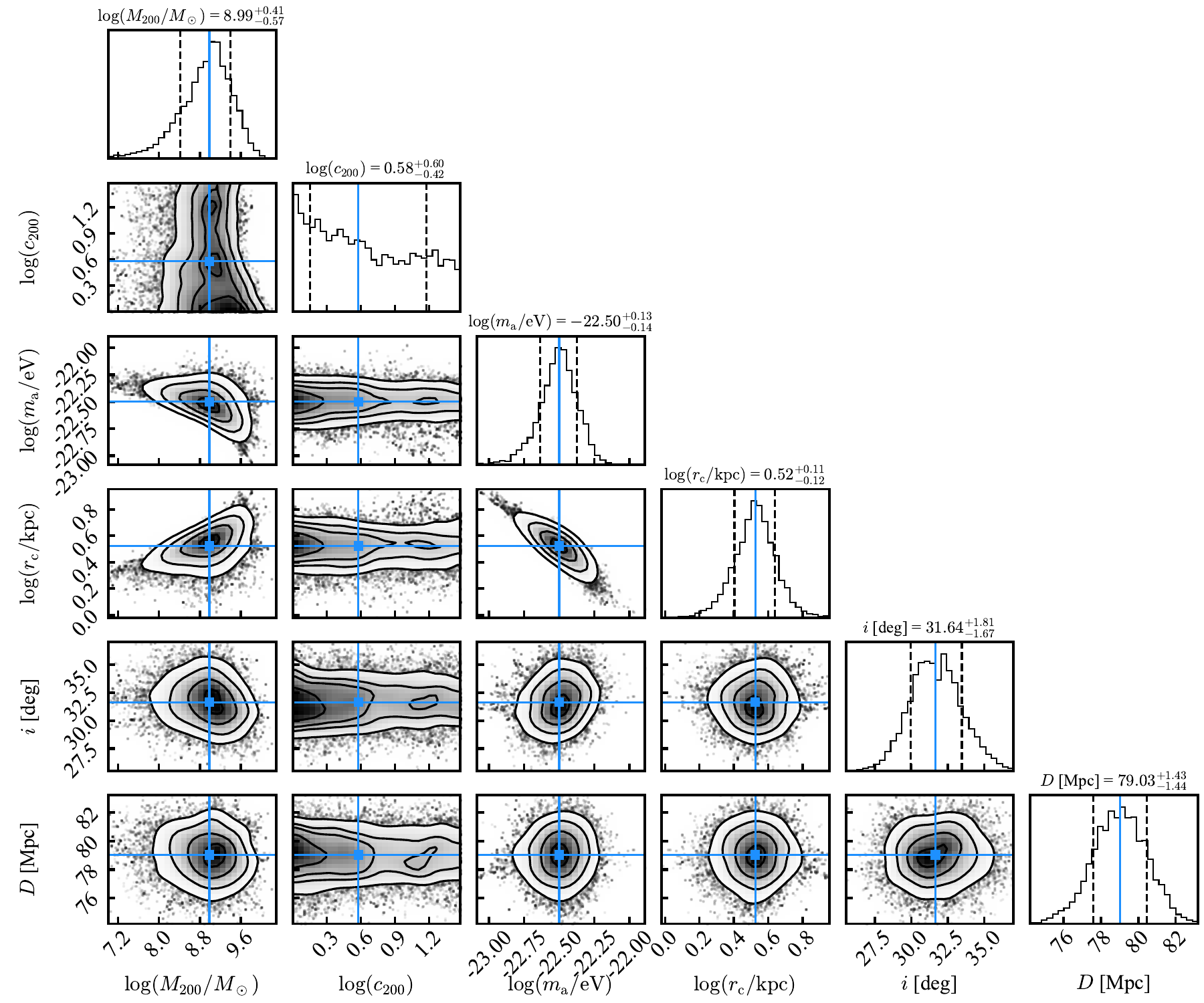}
    \caption{Posterior distributions of the Fuzzy dark matter mass model shown in the top panel of Fig.~\ref{fig:fuzzy}. }
    \label{fig:post_fuzzy_nominM200}
\end{figure*}

\begin{figure*}
    \centering
    \includegraphics[scale=0.44]{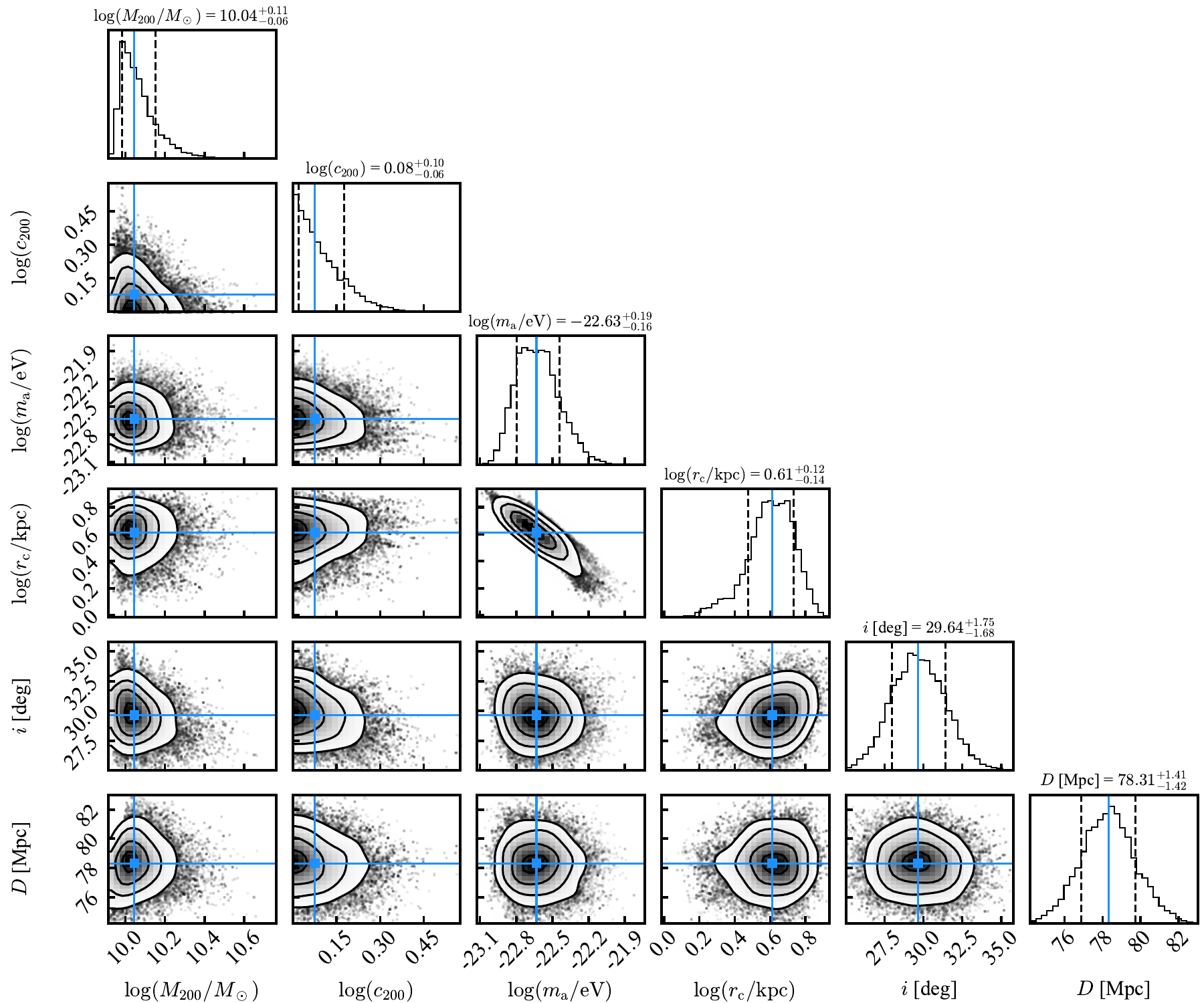}
    \caption{Posterior distributions of the Fuzzy dark matter mass model shown in the bottom panel of Fig.~\ref{fig:fuzzy}. }
    \label{fig:post_fuzzy_minM200}
\end{figure*}

In Fig.~\ref{fig:post_noDM}, we show the posterior distribution of the mass model with no dark matter of Fig.~\ref{fig:noDM}.
In Fig.~\ref{fig:post_cdm_corenfw_nominM200} (\ref{fig:post_cdm_corenfw_minM200}), we show the posteriors for the \textsc{coreNFW} fit in the top (bottom) panel in Fig.~\ref{fig:coreNFW}. 
In Fig.~\ref{fig:post_cdm_dpl_nominM200} (\ref{fig:post_cdm_dpl_minM200}), we show the posteriors for the DPL fit in the top (bottom) panel in Fig.~\ref{fig:dpl}. 
In Fig.~\ref{fig:post_sidm_nominM200} (\ref{fig:post_sidm_minM200}), we show the posteriors for the SIDM fit in the top (bottom) panel in Fig.~\ref{fig:sidm}.
In Fig.~\ref{fig:post_fuzzy_nominM200} (\ref{fig:post_fuzzy_minM200}), we show the posteriors for the Fuzzy dark matter fit in the top (bottom) panel in Fig.~\ref{fig:fuzzy}.

\end{appendix}

\end{document}